# From Traditional Adaptive Data Caching to Adaptive Context Caching: A Survey

Shakthi Weerasinghe[1], Arkady Zaslavsky, Seng W. Loke, Alireza Hassani, Amin Abken, Alexey Medvedev

School of Information Technology, Deakin University, Burwood, Australia

*Abstract*—Context information is in demand more than ever with the rapid increase in the number of context-aware Internet of Things applications developed worldwide. Research in context and context-awareness is being conducted to broaden its applicability in light of many practical and technical challenges. One of the challenges is improving performance when responding to a large number of context queries. Context Management Platforms that infer and deliver context to applications measure this problem using Quality of Service (QoS) parameters. Although caching is a proven way to improve QoS, transiency of context and features such as variability and heterogeneity of context queries pose an additional real-time cost management problem. This paper presents a critical survey of the state-of-the-art in adaptive data caching with the objective of developing a body of knowledge in cost- and performance-efficient adaptive caching strategies. We comprehensively survey a large number of research publications and evaluate, compare, and contrast different techniques, policies, approaches, and schemes in adaptive caching. Our critical analysis is motivated by the focus on adaptively caching context as a core research problem. A formal definition for adaptive context caching is then proposed, followed by identified features and requirements of a well-designed, objective optimal adaptive context caching strategy.

## I. INTRODUCTION

THE adoption of Internet-of-Things (IoT) has been on the rise over recent years ("Australia's IoT Opportunity: Driving Future Growth," 2018). It is perceived that IoT will contribute to an average 2% improvement in productivity each year over a period of 8-18 years as a result. With 75 billion IoT devices expected by 2025 (Ruggeri et al., 2021), it consequently increases the demand for derived knowledge to make smarter and more informed decisions. From small-scale monitoring devices, e.g., smartwatches, to complex industrial IoT, e.g., Integrated System Health Management, derived knowledge can provide valuable insight about entities and the environments with which they interact (ur Rehman et al., 2019). We refer to this knowledge as context information. Big IoT data which is foundational to context, is characterized by volume, velocity, and variety (Perera et al., 2014; ur Rehman et al., 2019). They are still major challenges to efficiently responding to queries for context. Distributed systems play a major role in this aspect according to literature in related areas owing to massive scalability and low cost (Perera et al., 2014).

According to the definition given by Dey, context is "any information that can be used to characterize the situation of an entity. An entity is a person, place, or object that is considered relevant to the interaction between a user and an application, including the user and applications themselves" (Abowd et al., 1999). Context is unique characteristically to traditional data shared on the World Wide Web (WWW). It is a derived piece of information - an output of a reasoning process, which involves incorporating data from many entities that can exceed even the size of data about an entity. For example, the model, colour, and speed of a vehicle could be data. Information on whether the vehicle is driving "hazardously" by reasoning over the speed of the vehicle, speed limits, condition of the roads, and status of the components (e.g., brakes) is context (note that what is context is subjective to the use case or scenario). Applications that use context in business logic are context-aware (Baldauf et al., 2007; Bauer et al., 2002; Perera et al., 2014). Smart health and smart cities are such applications.

Context Management Platforms (CMP) (de Matos et al., 2020; Hassani et al., 2018; Li et al., 2015) deliver context to context-aware applications called Context Consumers (CC) by retrieving and processing data from IoT data providers. It abstracts connections to IoT data providers and the process of inferring context (Figure 1). According to (Li et al., 2015), there are four challenges when responding to a large number of context queries. Firstly, context information is derived from continuous streams of IoT data streams. IoT data are ephemeral in lifetime and context information in extension is transient as well (Weerasinghe et al., 2022b; Zhu et al., 2019), e.g., traffic condition close to a car parking facility is a temporal construct. Authors measure transiency using different metrics such as lifetime (Weerasinghe et al., 2022b), time-to-live (TTL) (Schwefel et al., 2007), and age-of-information (AoI) (Abd-Elmagid et al., 2019). Secondly, context is both time and quality critical. For instance, a driver needs to be notified well before driving "too fast" to avoid being penalized. Thirdly, Big IoT data used to derive context is heterogeneous, e.g., in size and format (a temperature sensor reading versus a traffic video stream). Fourthly, the enormity of data required to infer context information. Consider the challenges of Big IoT data (Perera et al., 2014; ur Rehman et al., 2019) and the many different contextual information that can be requested and derived. They require an enormous amount of processing, memory, and storage. Instead, storing the smaller pieces of contextual information is more efficient than storing large volumes of real time streamed data.

However, it is extremely expensive to optimize for all four challenges in near-real time. Related work (Chabridon et al.,

[1] Corresponding author. School in Information Technology, Deakin University, Burwood, Victoria, Australia 3125.
   Email address: syweerasinghe@deakin.edu.au

2013; *Context Information Management (CIM): Application Programming Interface (API)*, 2020; "FIWARE-Orion," n.d.; Li et al., 2018) suffer either from scalability, complexity, and/or cost of responding to queries as a result.

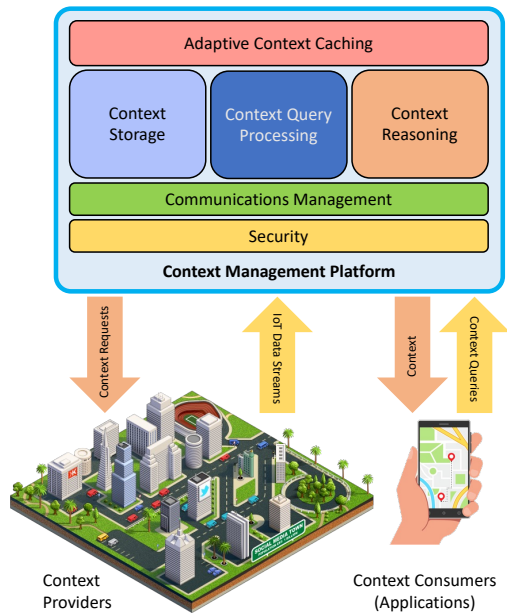

**Fig.1.** Overview of a Context Management Platform with Adaptive Context Caching.

Caching is a popular, and time proven strategy to manage scale and costs by reusing and repurposing data. Efficient caching has a significant impact on performance compared to accessing data via an expensive backend (Choi et al., 2020), or Context Providers in our case. Hail et al. (Hail et al., 2015) indicate that caching significantly improves the performance of Information Centric Networks (ICN) for transient IoT data. One of the critical features of a query load including for context is dynamism. The popularity of data objects (Zhou et al., 2015), and request patterns (Y. Wang et al., 2019) varies temporally. Previous work investigate adapting to dynamic queries based on data lifetime (Zhang et al., 2018), properties of network queues (Sun et al., 2016; Xu and Wang, 2019), popularity (Nasehzadeh and Wang, 2020; Sadeghi et al., 2019b; Sheng et al., 2020; Zhu et al., 2019), and cost of caching (Sadeghi et al., 2019b; Somuyiwa et al., 2018; Zhu et al., 2019). However, these are data caching solutions and we distinguished data from context information above.

Concepts applied in data caching are not always applicable when caching context (Weerasinghe et al., 2022c). For instance, authors pre-define a file library (Sadeghi et al., 2019b) or a data library, e.g., each sensor produces distinct data items (Nasehzadeh and Wang, 2020), in data caching. A context library cannot be defined similarly owing to the dynamic emergence of novel context. Secondly, the efficiency objectives are achieved as a function of evictions. Consider a limited sized cache using the Least Frequently Used eviction policy. Under eviction-based optimization, an item ($i_1$) that is requested only once could evict an item ($i_0$) that was cached immediately before due to lack of access frequency at the time $i_1$ is retrieved. $i_1$ can still be a frequently accessed item resulting in at least one more eviction. Eviction-based optimization is hence both cost and space inefficient due to redundant operations. Blanco et al. (Blasco and Gunduz, 2014), Sadeghi et al. (Sadeghi et al., 2019b), and Chen et al. (Chen et al., 2020) selectively cache to minimize network calls and maximize the utility of cache occupation. It allows to maintain a consistent Quality of Service (QoS) which is unachievable with frequent evictions. Further, managing context derived from transient Big IoT data in the cache emphasizes the need to measure and address the balance of quality and cost of refreshing context in a timely manner.

There are several further key differences between caching context from data caching (Weerasinghe et al., 2022c). We can discuss three differences relating to the scope of this paper. Firstly, the primary objective of data caching is to minimize data retrieval latency from servers. Data caches are fast, proximally located servers that temporarily store copies of data. But there are several features of the cached context in contrast. Derived context (Medvedev et al., 2018; Perera et al., 2014) inferred by applying (a) aggregation, or (b) inferencing, e.g., based on Context Space Theory (Boytsov and Zaslavsky, 2010), may not be in the same format or representation as the providers' data (Henricksen et al., 2002). For instance, the situational description of a room derived from a video stream is a JSON object. The cached context is not a fast-access copy of a piece of data generated elsewhere. Context Consumers may be interested in different aspects, and presentations of the same contextual information based on their requirements as well. Secondly, cached context is logically hierarchical as we will indicate in Figure 3 compared to data objects, e.g., a flat file such as a JPEG. IoT data (we refer to as low-level context) are shared among the higher-level cached contexts. They refresh reactively (Medvedev, 2020) to changes in the value of the retrieved IoT data. Thirdly, caching any IoT data from Context Providers may not be useful or would be cost-efficient. For instance, compare caching the geolocation of a moving vehicle (low-level context) versus caching the estimated trajectory (high-level context) to produce proactively derive context such as for altering purposes. Another unique challenge to context caching is the variety of context queries and heterogeneity of IoT data that lack clear patterns (Medvedev et al., 2018). Therefore, adaptive context caching (abbreviated as ACC) is beyond the current state-of-the-art for CMPs based on the limited research being conducted in the area.

The main aim of this paper is to distinguish ACC mechanism from traditional adaptive data caching, as would be carried out within a CMP (our notion of adaptive context caching is broader and can be considered beyond CMPs, but we consider context caching within a CMP in our paper to make the ideas more concrete). The objectives to achieve this aim are twofold: (a) evaluate and compare different approaches to adaptation in related areas to identify research gaps from a CMP perspective, and (b) highlight the significant challenges and future direction for adaptive context caching.

To the best of the authors' knowledge, this paper presents the first comprehensive survey on adaptive context caching which is different from traditional data caching. We reviewed the literature on adaptive data caching as far as from 1998 to 2022. Our primary aim of this exercise is to conceptualize a solution for adaptive context caching for CMPs. We focused on the literature that views cache adaptation from the system-wide perspective among other viewpoints. The main contributions of this paper are as follows:

- We summarize a large collection of related work in traditional adaptive data caching,
- We conceptually differentiate between context and data caching,
- We critically evaluate and compare different techniques, and strategies to achieve different objectives in adaptive context caching,
- We develop a formal definition for "Adaptive Context Caching" (ACC), and
- We identify requirements for adaptive context caching based on our reviewed work.

The rest of the paper is structured as follows. Section II describes a motivating scenario. Then in Section III, we compare context caching against context-aware caching followed by establishing our goals of adaptation for ACC in Section IV which we will use in this survey to focus our critical review. Then in Section V, we provide an overview of adaptive data caching and in Section VI, we discuss approaches to adaptive data caching. Section VII broadly categorizes approaches to cache efficiency followed by a discussion of techniques used in literature to achieve efficiency goals in Section VIII. We evaluate scalable Cloud cache memory as a suitable technology for ACC in Section IX. Next, in Section X, we summarize all the parameters considered for adaptive caching and isolate the set relevant to ACC. Then, in Section XI, we evaluate the current state-of-the-art in context caching comparing our benchmark CMPs. In Section XII, we differentiate ACC from adaptive data caching and propose a formal definition. It is followed by a discussion about the functional requirements of ACC in Section XIII. Finally, Section XIV concludes and highlights the direction for future work in this area.

## II. BUSY CITY CIRCLE – A MOTIVATING SCENARIO

Let us consider a real-world scenario surrounding a city circle with multiple intersections as illustrated in Figure 2. Considering context queries to the CMP are either push- or pull-based, TABLE I lists several context queries written in Context Definition and Query Language (CDQL) (Hassani et al., 2019, 2016). A large sample of these context queries can be found in GitHub[1]. In this scenario, we identify three types of context-aware application users: drivers, riders, and joggers. Context Providers (CP) can also be CCs, e.g., cars, and could be mobile, e.g., bikes, or stationery, e.g., car parks (its sensors providing occupancy data).

Context queries are parsed into multiple sub-queries called context requests (CR) with respect to context entities. The Context Query Coordinator in the Context Query Engine (CQE) (Hassani et al., 2018) coordinates the execution of the CRs to produce the context query response. Figure 4 illustrates the query breakdown for $Q_4$ and $Q_5$ respectively. The blue colour is used to indicate the context query, orange colour for CR, green for entities, and yellow for context attributes. The arrows indicate the logical relationship between these context information. Immediate child nodes are sub-components of the parent nodes which are required to derive context, e.g., "*get car parks*" is required to generate context information for $Q_4$, or describe the parent node, e.g., address and location attributes describe the building entity.

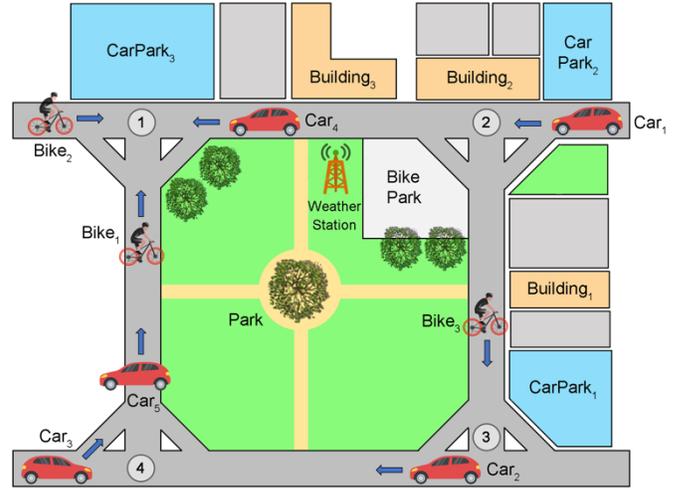

**Fig.2.** An intersection with multiple Context Providers and Consumers.

TABLE I
CONTEXT QUERIES IN THE MOTIVATING SCENARIO

| Notation | Query Description | Type |
|---|---|---|
| $Q_1$ | Search for car parks with available slots in the area. | Pull |
| $Q_2$ | Search for car parks with available slots in the area that charge less than a given amount per hour. | Pull |
| $Q_3$ | Search for car parks with available slots that are less than a certain distance from the target location. | Pull |
| $Q_4$ | Search for car parks with available slots given they are *good for walking* to the target location. | Pull |
| $Q_5$ | Check whether it is good for jogging at the park. | Pull |
| $Q_6$ | Search for bike parking spots in the near vicinity of the rider. | Push |
| $Q_7$ | Is the rider vulnerable to crashing into an object at the next closest intersection? | Push |

There are several challenges and features that could be observed from these queries. They are as follows:

- **C1:** Notice the CR - "*goodForJogging*", share the same entities and context attributes as the "*goodForWalking*" in $Q_4$, e.g., temperature and humidity are attributes of the weather entity. It is prudent to cache such items for reuse to avoid redundant retrievals.
- **C2:** The context attributes of certain entities such as the of the *weather* entity, could be very slow to vary and might be applicable uniformly across a large area. Consider the response for "*goodForWalking*". It is less likely to change in the scenario irrespective of the target location (e.g., a building, or a park in the area) because the weather is, (a) relatively static for a foreseeable period, such as on a sunny summer day, and (b) relevant for the entire region depicted in the scenario. But on the contrary, contextual information such as *"distance"* and *"isAvailable"* frequently vary, and they require refreshing in the cache to maintain the stored context is consistent with the real world.

---
[1] Available at: https://bit.ly/sample-context-queries

```
prefix mv:http://schema.mobivoc.org, schema:http://schema.org,
pull (targetCarparks.*)
define
entity targetLocation is from schema:place
    where targetLocation.address="221 Burwood Highway Burwood VIC 3125"
entity cosumerCar is from schema:car
    where consumerCar.vin="CDC24R5"
entity targetWeather is from schema:weather
    where targetWeather.location=targetLocation
entity targetCarpark is from mv:carpark
    where (
    ((distance(targetCarpark.location, targetLocation, "walking")
        <{"value":500, "unit":"m"} and
    goodForWalking(targetWeather)>=0.8) or
    (distance(targetCarpark.location, targetLocation, "walking")
        <{"value":1, "unit":"km"} and
    goodForWalking(targetWeather)>=0.5)) and
    isAvailable(targetCarparks.availability, {"start_time":now(),
        "end_time":{"value":"2022-03-07T17:30:00", "unit":"datetime"})
)
```

(a)

```
prefix schema:http://schema.org,
pull goodForJogging(targetWeather)
define
entity targetPark is from schema:place
    where targetPark.address="Bennetswood Reserve"
entity cosumerCar is from schema:car where consumerCar.vin="CDC24R5"
entity targetWeather is from schema:weather
    where targetWeather.location= targetPark
```

(b)

**Fig.3.** CDQL query for (a) $Q_4$, and (b) $Q_5$.

- **C3:** The other function of context data caching is repurposing. For instance, assume the $car_5$ executes $Q_2$ which is travelling near $car_3$ that executes $Q_1$. Then, the retrieved data from nearby car parks for $Q_1$ could be repurposed for $Q_2$ using only a filter.
- **C4:** Consider C2, C2, and C3. Each of them focuses on a logical level of context. In C1, the is decision related to the context attributes. In C2, the decision is made relevant to entities and context requests. In C3, the decision relates to the highest level of context – the query response. The effectiveness of caching for reuse and repurposing in each of these logical context levels depends on different factors.
- **C5:** In Figure 5, we indicate the relationships between the entities in the scenario. It illustrates relationships between (a) co-occurring entities, such as those declared in relational functions (Hassani et al., 2019, 2016), e.g., the car is owned by a person, and (b) correlated entities, e.g., condition of an access road

(entity) to a car park (entity), when making context caching decisions.

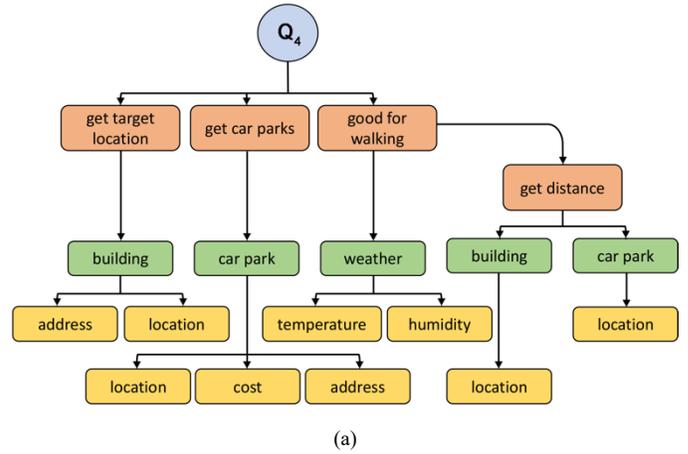

(a)

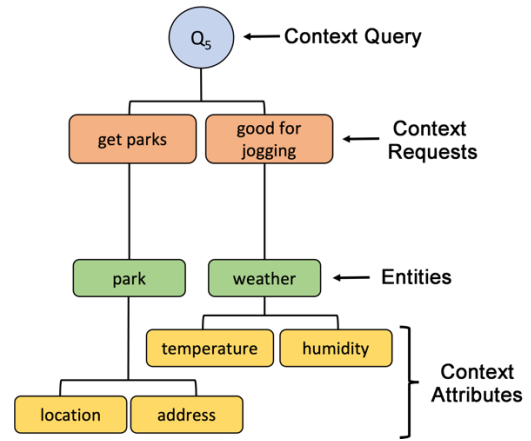

(b)

**Fig.4.** Context query breakdown for (a) $Q_4$, and (b) $Q_5$.

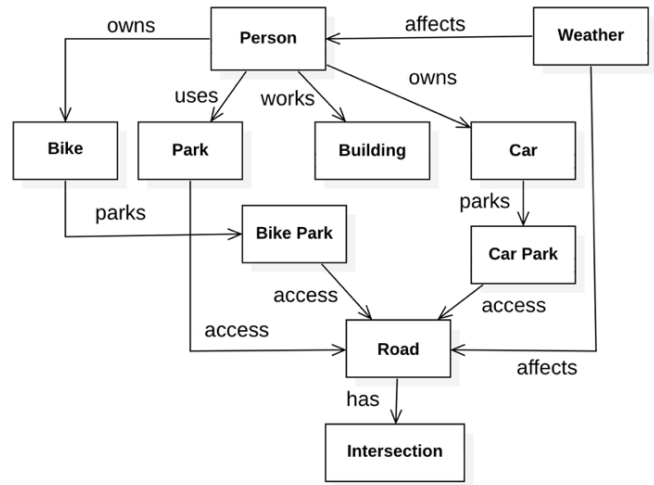

**Fig.5.** Relationships among entities in the scenario.

- **C6:** Consider the relationship between the *Car Park* entity with the *Road* entity in Figure 5. Access to a car park may be hindered via an access road for vehicles approaching from a specific direction. Then the road condition, e.g., traffic level is a vital context information for responding to the context query. The accumulated size of relevant context to respond to a context query is larger than it is implicit in the query statement.

- **C7:** Further to the above, when executing the $Q_1$-$Q_4$, all the relevant car parks (after context service selection) will be retrieved. When the retrieved candidate car parks are large in number, so would be the number of caching decisions for each candidate entity.
- **C8:** The heterogeneity of context query loads depends on time. It is the primary motive in investigating an adaptive solution to make context caching decisions. By observing recurring patterns and features in query loads over a period of time, the application of an optimal caching model could improve the quality of service (QoS). For example, the request rate ($\lambda$) for $Q_1$ is $2.5 \pm 0.01$ per second from 7:30-8:30 AM during the weekdays. We refer to such a period as *planning periods* (or *PlnPrd* for short) (Medvedev, 2020).

A fundamental challenge here is what we refer to as the *lack of prior knowledge*. Consider the sudden occurrence of an accident (context) in the scenario. Assuming this situation was not previously observed, there is no information to base the decisions to cache for related context queries, e.g., no previous experience leading to a *cold start*. This is also a good example to indicate the practical infeasibility to predefine the transitional probabilities of all possible cache state–action pairs. We also attribute the imperfections in context data (Henricksen et al., 2002; Krause and Hochstatter, 2005) to the lack of priors since it induces uncertainty in statistically derived priors.

The interesting problem here is the reliable delivery of context information. Consider a Context Consumer that demands context (e.g., the status of the car park) no older than 30 seconds, with completeness, e.g., provide all car parks that meet the contextual criteria. Then, the decision to cache a car park entity could be based on: (a) the expensiveness of the context retrievals compared to the cost of caching and retrieving from the cache, (b) congestions in the mobile telephone network in the area via which the IoT devices are connected; so that it is difficult to receive data within an acceptable period of time from all the car parks, and/or (c) popularity by preference. There exists no definite set of rules applicable across any scenario to make selective caching decisions. The applicable parameters and weight of them in the decision criteria are subject to the *current circumstance* (which is also *context information*).

### III. CONTEXT CACHING AND CONTEXT AWARE CACHING

In this section, we distinguish context caching from context-aware caching due to taxonomical similarity.

#### A. Context Caching

Context caching is the process of temporarily storing context information in a fast access memory, facilitating device and location transparency when retrieving. In other words, context caching (e.g., within a CMP) facilitates accessing context data from a temporary location without having to obtain it each time from the Context Providers, or re-executing the (a) aggregation, and/or (b) inferencing processes.

#### B. Context-aware Caching

Context-awareness (CA) has several definitions, including that by Dey (Abowd et al., 1999). Huebscher et al. (Huebscher and McCann, 2004) defined CA as the ability of an application to adapt itself to the context of its user(s). Baldauf et al. (Baldauf et al., 2007) define context-awareness of a system as the ability to adapt operations to the current context without explicit user intervention and aiming at increasing usability and effectiveness by taking environmental context into account.

Accordingly, we define context-aware caching as the process of temporarily caching data based on context (Weerasinghe et al., 2022a). Context could be of the data query load, users, or about the situation/scenario which triggers the data queries. Context-aware caching can be proactive (e.g., prefetch data to cache during network off-peak hours) or reactive (e.g., caching the items that are most popularly accessed). Context-awareness in selective caching is aimed at either maximizing the Quality of Service (QoS) such as by increasing the hit rate, or the Quality of Experience (QoE) such as maintaining a consistent data query response latency.

To the best of the authors' knowledge, this survey critically reviews existing data caching techniques and their applicability for adaptively caching contextual information. Although adaptive data caching has been widely investigated in the areas of Mobile and Internet Edge Computing (MEC/IMC) (Orsini et al., 2016), Mobile Networks (MN) (Blasco and Gunduz, 2014), Fog Radio Access Networks (FRAN) (R. Wang et al., 2019), IoT Networks (IoTN) (Xu and Wang, 2019), Wireless Sensor Networks (WSN) (Chatterjee and Misra, 2016, 2014; Zhou et al., 2015), Content Delivery Networks (CDN) (Kangasharju et al., 2002; Kiani et al., 2012; Tadrous et al., 2013), and Information Centric Networks (ICN) (Meddeb et al., 2017; Zhang et al., 2018), adaptively caching context information is not yet been studied. Data shared in the areas above are often stationary files hosted centrally, which researchers refer to as content libraries. In such a setup, work into adaptive data caching at network edges (H. Wu et al., 2021) and CDNs have benefited from prior or posteriorly learnt knowledge such as file popularity distributions. So, Bayesian techniques have been implicitly used. On the other hand, continuously streamed IoT data produced by Context Providers (CPs) are characteristically transient (Weerasinghe et al., 2022b; Zhang et al., 2018). Limited-sized caching systems that are used in systems classified under the said areas do not have the luxury of storing large volumes of streamed data. We contend that storing the smaller-sized, interpreted piece of information – called context, is more efficient, such that Adaptive Context Caching is a crucial area to study to enhance the cutting-edge CMP technology.

In the next section, we identify what are the goals and objectives of an adaptive context caching algorithm.

## IV. GOAL OF ADAPTIVE CONTEXT CACHING

In this section, we describe several concepts relevant to ACC, identify the relationships to our problem, setting our criteria for reviewing literature in related areas, and identify ACC as a multi-objective optimization problem.

### A. Taxonomy in Context Management

**Cost of Context (CoC)** is a related terminology for ACC which can be defined as the cost of acquisition of context, often relating to the *acquisition* phase in the context life cycle (Perera et al., 2014). Villalonga et al. (Villalonga et al., 2009) however define CoC as also encompassing non-monetary aspects, e.g., resource consumption. In either case, they ignore caching. We can view CoC as a component of the Cost of Context Caching (CoCa) which involves both monetary, e.g., cost of retrieval for refreshing cached context, and non-monetary costs, e.g., CPU cycles utilized for processing the context caching decisions. Jagarlamudi et al. (Jagarlamudi et al., 2021) define CoC as the perceived value of context query responses. The perceived value is a function of Quality of Context (QoC) - any information that describes the quality of information that is used as context information (Buchholz et al., 2016). In the case of context caches as a source for context acquisition, QoC relies on, (a) context caching resource allocation, e.g., availability of context cache memory by size, and (b) appropriate processes, e.g., efficient context refreshing to maintain a high degree of cached context freshness. Both of these impacts the Quality of Service (QoS) of the CMP.

**Quality of Service (QoS)** is a metric that indicates a platform's performance in delivering the respective services to the user. Response time (RT), availability, and reliability of the CMP – parameters of QoS define its capability to manage context efficiently. We use Service Level Agreements (SLAs) to define acceptable QoS parameters such as the maximum acceptable RT, non-compliance to which can incur penalties (Weerasinghe et al., 2022b). For example, inadequate cache space (assuming the cache size is scalable) may delay responding to a context query (resulting in a penalty cost) because of accumulated retrieval latencies from CPs (CoC). It then violates the CCs requirements stated in the SLA. But adequately meeting the QoS and QoC are constrained by how much costs the CMP is willing to incur to remain or improve its cost-efficiency. Cost constraints are set by the Service Level Agreements a CMP has with CPs, CCs, and resource providers (e.g., Cloud cache services). We indicate this relationship in Figure 6. For instance, freshness of context (a QoC metric) can be maximized by frequent refreshing, but it is both network and process intensive. QoS can be maximized by using an unlimited amount of processing units or cache memory. These options are infeasible due to cost. Therefore, how well the CMP meets the QoS and QoC depends on its budget. We use the term "cost" to refer to the cost of context caching (CoCa) in this work because of this reason. CoCa is an abstract cost comprising of both monetary, e.g., cost of cache memory, and non-monetary costs, e.g., cache seek time to incorporate all relevant expenses.

### B. Perspectives on Efficiency

We can broadly categorize previous literature in adaptive caching for cost-efficiency (Dasgupta et al., 2017; Menache and Singh, 2015; Venkataramani et al., 2002) into two main perspectives as indicated in Figure 7: (a) optimizing for the total cost (minimizing) or returns (maximizing) for all data queries (Dasgupta et al., 2017; Venkataramani et al., 2002; Zhu et al., 2019) – *Perspective A*, and (b) maximizing the number of context query responses that are beyond an acceptable Quality of Service (QoS) metric that significantly contribute to the cost or return, e.g., hit rate, response latency – *Perspective B*. A similar notion can be found in (Narasayya et al., 2015) in a related area.

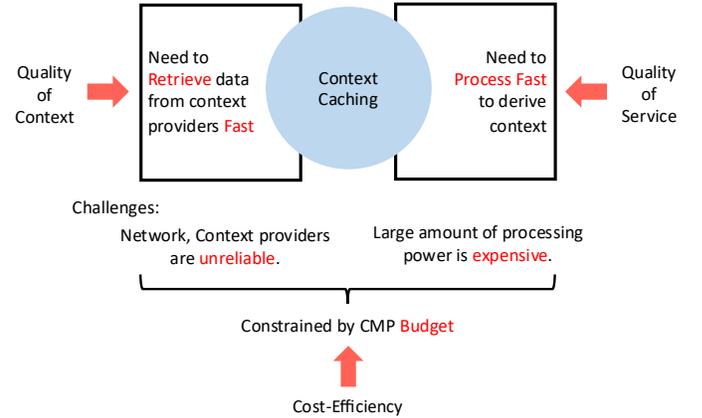

**Fig.6.** Challenges and opportunities in Adaptive Context Caching problem.

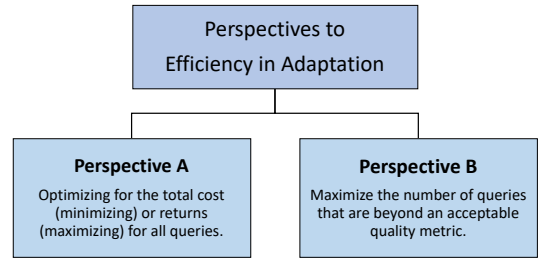

**Fig.7.** Perspectives on efficiency in adaptive caching.

There are several reasons against Perspective B in our scope. First, acceptable parameter values depend on the use cases and certain classes of context queries. We use the SLAs that are attributed to consumers to overcome this issue (Medvedev, 2020; Weerasinghe et al., 2022b). Therefore, it is infeasible to predefine thresholds applicable for all or the majority of the queries similar to optimizing database or search engine transactions. A CMP is a middleware serving a multitude of context queries that need to perform "objectively" – unbiased towards optimizing only a subset of context queries. Secondly, the thresholds are time variant – one of the features of adaptation. Thirdly, we need to consider acceptable costs for each distinct query. Consider $Q_1$ and $Q_7$. $Q_7$ is qualitatively a matter of whether the rider faces significant injury whereas $Q_1$ is about missing a spot for which an alternative can be suggested. The acceptable cost for $Q_7$ cannot be easily defined and is much more critical than $Q_1$. Taking these reasons into account, we resort to *Perspective A* in formulating a solution for cost-efficient adaptive context caching (ACC) in our work.

### C. Objectives of Efficiency

In the previous sub-sections, we described the relationships between CoC, QoC, QoS, and CoCa in achieving cost- and performance-efficiency for a CMP (e.g., cache memory utilization is crucial for maximizing cost efficiency (Dasgupta et al., 2017; Menache and Singh, 2015; Venkataramani et al., 2002; Wu and Kshemkalyani, 2006), and perceived utility). Then we opt for a system-wide optimization goal measured in total costs incurred or earned by the CMP. So, we focus our critical reviewing process on evaluating the cost-efficiency of the solutions against the ability to maximize the QoS of the CMPs. Overall, our adaptive context caching problem is at least a tri-objective optimization problem, i.e., minimizing cost while maximizing QoS (or in other words, the performance of the CMP for simplicity) and QoC.

V. OVERVIEW OF ADAPTIVE DATA CACHING

In this section and several sections to follow, we critically analyse and survey previous work in adaptive data caching. But first, we define adaptive data caching, then discuss the practical implications of IoT data caching using scenarios to guide the discussions in later sections and reflect on the applicability of previous work to adaptive context caching.

*A. Adaptiveness and Adaptive Caching*

We first discuss the keyword "adaptation". Gramacy et al. (Gramacy et al., 2002) denoted adaptation as "an online algorithm is called adaptive if it performs well when measured up against off-line comparators". Kouame et al. (Kouame and Mcheick, 2018) define adaptation as a process of four activities: (a) trigger, (b) planification and decision, (c) realization of the adaptation, and (d) reconfiguration. Raibulet et al. (Raibulet and Masciadri, 2009) define three metrics to measure the performance of adaptation: (a) performance latency, (b) quality of response (which we can refer to by QoC in our scope), and (c) performance influence on adaptation, e.g., loss of QoC in the cache by minimizing refresh rate – which aligns with our tri-objective problem defined in Section IV-C. We also identify two types of adaptation.

- **Rule-based adaptation** – adapting based on a predefined set of condition-action pairs, e.g., (Muller et al., 2017).
- **Context-aware adaptation** – adapting to the logically inferred context of an entity, i.e., context of the data such as the cost, e.g., (Dasgupta et al., 2017; Menache and Singh, 2015; Scouarnec et al., 2014; Venkataramani et al., 2002) and Quality of Experience (QoE), e.g., (Chockler et al., 2010; Choi et al., 2020).

Accordingly, adaptive caching is changing the state of the cache memory (i.e., what is cached and not cached) to better suit the features of data, user requests, and sometimes the users. By the classification above, context-aware caching is a form of adaptive caching among others. We discuss related literature in adaptive caching in general in this paper, as a result.

*B. Practical implications of adaptive IoT data caching*

We mainly focused on the literature involving ICNs, CDNs, IoT Networks (IoTN), and Wireless Sensor Networks (WSN) since information is uniquely recognized as a piece of content, similar to context information. Caching strategies in these areas can be broadly categorized into three kinds as indicated in Figure 8, and we describe below.

Firstly, consider the objectives of minimizing response latency, and distance to data. The authors in (Pahl et al., 2019) use under-utilized memory in networking devices for *on-path caching*. This could be further broken down into two sub-objectives:

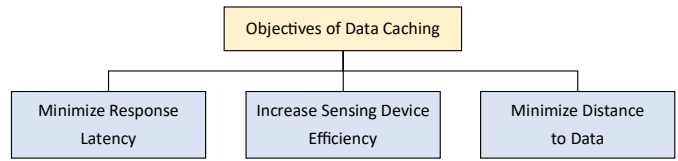

**Fig.8.** Objectives of caching in ICNs, CDNs, IoTN, and WSN.

- **Minimizing asynchrony** - IoT devices are connected through a network. There are several challenges observed in the literature to streamlining data ingestion for processing context: (a) difference in clock speeds, (b) unsynchronized network queues (Kaul et al., 2012; Sun et al., 2016), insufficient provider availability (Pahl et al., 2019), and (c) availability in bandwidth (Zhou et al., 2015). These can create network congestion. Cache strategies developed for ICNs aims to reduce stress, and load on back-haul networks, e.g., cooperative caching (Psaras et al., 2012; Y. Wang et al., 2019; Zhang et al., 2018; Zhou et al., 2015), effectively reducing operational costs originating from network delays.
- **Balancing caching with network processing** - Cache memories in networking devices are small, and shared, i.e., only the available memory after data forwarding is used for caching in a router. Cache replacement has been thoroughly investigated in this view to improve efficient occupancy. For instances, Psaras et al. (Psaras et al., 2012) investigate replacing at *line speed* – the number of data items received per second.

Secondly, consider the other objective of improving device efficiency (Pahl et al., 2019; Xu and Wang, 2019; Zhang et al., 2018). The problem to optimize device efficiency arises from IoT often being small, low powered devices with no or limited computational capacity, e.g., narrowband IoT (nb-IoT). These are predominantly *read-only* and used to measure natural properties, e.g., temperature. Caching sensor readings has two efficiency objectives: (a) improve battery life (Sheng et al., 2020; Zhang et al., 2018) and (b) overall performance efficiency, e.g., the number of triggered readings (Zhang et al., 2018). Both aspects are central to CMPs and Adaptive Context Caching since the lifetime of a sensor can underpin the QoC of context information. For instance, consider the following extended scenario for $Q_7$, assuming the hazard is an open door of a parked car in Figure 9. The door sensors on vehicles are CPs, and the CC is a cyclist. The status of the door is sensed by a battery powered retrofitted sensor, and the battery life is inversely proportional ($1/\propto$) to the number of requests to execute sensing. Here, $vehicle_1$ is closer to the rider, but the rider is suggested a hazard further ahead, e.g., the open door of *vehicle₂* because of the obliviousness of the context retrieval and refreshing process to power management.

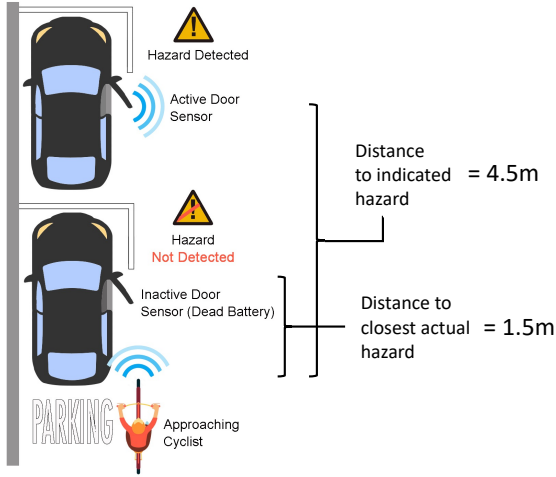

**Fig.9.** Extended scenario for $Q_7$ – open car doors as a hazard.

A similar case is managing precision measurement sensors that could be defined in an SLA, e.g., the location accuracy of a car to execute $Q_4$ in Figure 3(a) should be less than 5m. As in the scenario, they can be crucial in meeting the QoC requirements of prefetched context into the cache. The criticality is the same when sensors are distributed sparsely (e.g., not having sensors at each intersection to execute $Q_7$) especially when redundancy is not enforced (e.g., the context about the intersection₁ cannot be applied to intersection₂, if the latter is not equipped to produce context information).

An overlooked area of research in this respect is the round-trip latency arising from errors, or unavailability in the network, or CPs. Round trip latency for a CMP refers to the sum of time spent on context service resolution, uplink, and downlink time to retrieve context and processing time. Noise, cross talks, fading links, and impairments in the transmission media could cause errors in the received data (Song et al., 2020), e.g., Baccelli et al. (Baccelli et al., 2014) test the impact of transmission losses on the freshness of data, while intermittent loss or permeant loss (i.e., packet loss as a result of a system failure) of entity of a resolved service could trigger re-attempts culminating in additional latencies. Although this area is widely discussed in signal processing, the issue is of paramount importance to IoT based systems alike. So, there are two risks to (a) QoS – e.g., extra latency to retrieve context during a reactive refreshing (Medvedev, 2020; Weerasinghe et al., 2022b) could violate the SLA for response time, and (b) QoC. We tested the impact of noise on the estimations for proactive context refreshing, and cost in (Weerasinghe et al., 2022b), where it was found that noise reduces the cost-efficiency of ACC. We also found that (a) the reliability of context information (e.g., in (Weerasinghe et al., 2022b), the cached values for available slots were incorrect due to noise), and (b) the completeness of context information, (e.g., context response missed hazard closer in Figure 9) were also affected by noise.

Considering the discussion above, measuring network, and IoT device reliability can improve efficiency in ACC. It allows to select and balance the trade-off between cost and quality as early as at context service selection, e.g., selecting the most reliable (e.g., guaranteed to retrieve the "right" context in a single round trip to cache) yet the cheapest Context Provider meeting the minimum constraints in SLA. These are also examples to illustrate that ACC is a cohesive system of several functions.

## VI. APPROACHES TO ADAPTIVE DATA CACHING

In this section, we present, compare, and contrast different methods of data and context cache adaptation in the previous literature. They can be broadly categorized into eight methods: (a) adaptive cache replacement, (b) caching policy shifting, (c) hierarchical caching, (d) proactive caching, (e) elastic caching, (f) adapting cache space allocation, (g) adapting by locality, and (h) adapting for data refreshing, each of which is distinct in terms of the strategy, factors considered for adaptive decision making, and/or the focal resources used to adapt in a distributed system. These are summarized in Figure 10 and discussed below. Approaches coloured in yellow are strictly reactive, whereas those coloured in blue, indicate evidence of being a proactive technique.

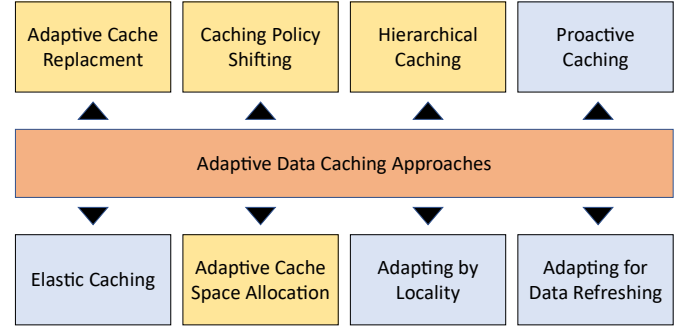

**Fig.10.** Approaches to Adaptive Data Caching.

### A. Adaptive Cache Replacement

Cache replacement is a form of adaptation that replace cached items with those which are more relevant in responding to contemporary queries. We refer to this method as "optimizing by eviction" because swapped items are selected based on achieving an objective function, e.g., maximizing hit rate by replacing with items expected to have a higher hit rate (HR) than the items evicted. But the pre-condition is the limited size of the cache memory.

Adaptive cache replacement underlies the infrastructure which dictates the different optimization objectives. For instance, consider the following work:

- The remaining memory of network devices, e.g., routers are used as caching nodes in ICNs. Cache size is extremely small and variable due to this reason. As a result, in (Anandarajah et al., 2006; Pahl et al., 2019; Psaras et al., 2012; Sheikh and Kharbutli, 2010; Vural et al., 2017), authors are concerned with replacing based on, (a) data lifetime and (b) data popularity, to maximize reusability of cached data items.
- In mobile networks, base stations are relatively large cache memories, sometimes of which are dedicated. In (Blasco and Gunduz, 2014; Zhou et al., 2015), the
8

authors replace based on, (a) popularity, (b) lifetime, and (c) size, for each locality. The objective is to maximize the mobile service experience in each locality by caching items most accessed by the users in the area.
- Edge caches are dedicated large cache memories. They could afford to cache for each application. In (H. Wu et al., 2021), application user behaviours, and access pattern is used to make replacement decisions.

Although a large body of research uses adaptive cache replacement, we focus more on the strategies applicable in a scalable cache environment given our objective for consistent QoS. Then, only a limited work such as that of Wang et al. (Y. Wang et al., 2019) could be found. It attempts to merge the benefits of collaborative caching (refer to Section VI-G), and Cloud caches to perform cost-effective cache replacement using a cost, and popularity-based model. But the homogeneous cost modelling, e.g., uniform transfer cost between servers, contradicts with price models offered by commercial vendors, e.g., AWS Elasticache price varies by availability zones, and node type ("Amazon ElastiCache - In-memory datastore and cache," n.d.).

*B. Caching Policy Shifting*

First-in-first-out (FIFO), last-in-first-out (LIFO), least recently used (LRU), least frequently used (LFU), and random are frequently used cache replacement policies used in computing systems. Each could be identified with advantages and disadvantages, e.g., complexity when implementing LFU and cache pollution of LRU. For instance, LFU is a suitable policy when the popularity distribution of data items follows a power tail distribution (Choi et al., 2020). However, the policy performs poorly under uniform popularity distributions because recently cached items are prone to get evicted due to lack of access during the short cache residence. So, LRU could be more appropriate. Caching policy shifting (*PoliShft*), therefore, refers to a form of cache adaptation, in which the cache replacement policy is switched among a predefined set of eviction policies (*PoliSet* ⊆ *Eviction Policies*) in relation to observed metric, e.g., miss rate in (Choi et al., 2020).

*PoliShft* is significantly different from other caching mechanisms. But there are two fundamental drawbacks to it. Firstly, the runtime complexity is high, i.e., fetch, evict, re-fetch, and policy changes are handled at the physical level. Secondly, it is non-trivial to design because: (a) it is impossible for a cache designer to design for all possible scenarios of the system beforehand in order to decide on the cache replacement policy set; so, a rigid rule-based adaptation policy is difficult to design, and (b) selective caching decision of data items can differ in relation to the selected policy; therefore, *PoliShft* requires to track a limited history of cache replacement to trigger to re-fetch data items that may need to be cached under the new policy, e.g., an item evicted under LRU when switching from LFU may need to be cached again if the policy is switched back to LFU.

Finally, consider the work of Gramecy et al. (Gramacy et al., 2002), where re-fetching is defined as the retrieval of a previously evicted object preferred by a previous policy, e.g., item$_1$ ($i_1$) is evicted by the LRU policy, although it was preferred for caching under LFU; then re-cached after shifting again to LFU. *PoliShft* can be cost-inefficient (e.g., due to re-fetching costs) under volatile query load circumstances, e.g., that of a CMP. As a result, none of the policies would stabilize for an adequate time to converge to sufficient efficiency. Considering that consistency in delivered QoS is one of the primary objectives for ACC (refer to Section IV-C), this feature can be detrimental to the perceived QoS.

*C. Hierarchical Caching*

This strategy refers to managing cache data between several distinct levels of physical caches, e.g., slow- and fast-cache (Chatterjee and Misra, 2016; Cheng et al., 2015). A logical cache level (i.e., logical context cache levels in (Medvedev, 2020)), may utilize up to all the physical levels which could technologically be different. One of the drawbacks of a unitary cache organization, i.e., single physical cache, is the impact that non-transient data (e.g., Vehicle Identification Number of a car, address of a location) have on transient data such as speed of the car, number of parking slots available, from being cached considering the work in (Chatterjee and Misra, 2016, 2014; Schwefel et al., 2007; Vural et al., 2017; Zhang et al., 2018). Previous work points out polyglot persistency ("FIWARE-Orion," n.d.; Jayaraman et al., 2016) to allow unique cost and performance benefits for different features of the same application. In data caching, Kabir et al. (Kabir and Chiu, 2012) illustrate a two-level hierarchical cache memory, e.g., memory as a cache in an Elastic Cloud Computing (EC2) instance – the fast, in-memory cache, and a Simple Storage Service (S3) bucket – the slow, persistent cache, which achieves near 100% hit rate (HR). The primary feature here is the balance between seek time, cache size (which are QoS parameters), and cost of cache memory – fast caches are expensive but small, whereas slow caches are cheap but large. It is advantageous to adopt multiple technologies of cache memory that are orderly organized considering all of the costs, seek time, and size.

An alternative hierarchical cache structure can be found in the work of Chatterjee et al. (Chatterjee and Misra, 2016, 2014) which investigates "external" and "internal" caches in a WSN. This is a logical hierarchy compared to the physical hierarchies we discussed so far. We identify this as an alternative structure to the Adaptive Cache Space Allocation strategy in Section VI-F due to the similarity in objective.

One of the drawbacks of hierarchical caching is the complex data management procedures such as (a) managing data between different cache technologies, e.g., different file systems, and (b) unifying access to cache memories. Cache virtualization in (Chatterjee and Misra, 2016, 2014), and cache broker virtualization in (Kiani et al., 2012) share the advantage of a single interface to the caching module since it abstracts retrieval, processing, and distribution of cached content. In a distributed system, virtualization improves location transparency that is proportional to QoS (e.g., by providing unified access to a large, distributed cache memory space), hiding complexity (e.g., inter-communication between cache memory instances), and load balancing. This points to Cloud-based caching services such as ElastiCache ("Amazon ElastiCache - In-memory datastore and cache," n.d.) to be useful in ACC due to these reasons.

*D. Proactive Caching*

Proactive caching (or prefetching) is to retrieve data into the cache memory prior to them being accessed by consumers. Proactive caching anticipates demand for particular data in the foreseeable future using many different techniques. Surveys on performance-efficient web data prefetching can be found in (Ali et al., 2011; Domènech et al., 2007, 2006). Although the strategy is commonly used to reduce response latency at peak-hours by maximizing network utility during off-peak hours (Sadeghi et al., 2018), it is a form of adaptation that change the state of the cache to the anticipated context of queries, entities, and/or data, e.g., a situation (a high-level context) such as a hazard to a rider (entity) in $Q_7$. However, proactive caching requires continuous monitoring (Hassani et al., 2018), and prior knowledge of temporal dependencies of the queries, and data. So, it is a continuous learning task that is computationally intensive, and scalability by design is a critical requirement for ACC.

Bogsted et al. (Bøgsted et al., 2010) define models for two main classes of the proactive strategy, which are (a) event-driven, e.g., rider hazard, and (b) periodic, e.g., prefetching car parks for $Q_1$-$Q_4$ before the rush hour. Both these classes apply to ACC. We identify two alternate classes of proactive caching in the literature:

- **Prefetching using a cost-function** - prefetch content into a cache based on an estimated cost calculated using popularity (Sadeghi et al., 2018; Tadrous et al., 2013; Wu et al., 2019; Xin Chen and Xiaodong Zhang, 2003; Zhong et al., 2018), data or process lifetime (Papaefstathiou et al., 2013), or its combination (Sadeghi et al., 2018; Somuyiwa et al., 2018; Zhu et al., 2019). There are three rationales against this approach for a CMP: (a) tracking and estimating cost for each context is expensive, (b) cost-functions are myopic (ignores long-term features, such as re-occurring situations/context and sequentially accessed dependent context information), and (c) considering one, or few parameters to estimate the cost can be inadequate, e.g., cost as a function of remaining life (Somuyiwa et al., 2018) ignoring reliability of the connectivity to the provider can result in costs borne by lack of QoC.
- **Objective-greedy prefetching** (Wu and Kshemkalyani, 2006) – maximizing, e.g., Hit Rate (HR) (Venkataramani et al., 2002), or minimizing, e.g., network utility (Sadeghi et al., 2018), perceived latency and cost (Faticanti et al., 2021), a chosen metric by prefetching into the cache. Previous work employing Machine Learning techniques (Pahl et al., 2019; Scouarnec et al., 2014; H. Wu et al., 2021; Zameel et al., 2019) can be categorized in this class.

There are other forms of functions used for prefetching, related to the former. Venkataramani et al. (Venkataramani et al., 2002) propose a long-term prefetching approach using a calculated parameter called *GoodFetch* – a method for balancing the bandwidth usage and access frequency of an item. The common drawback of function-based prefetching lies with predefined thresholds. In a dynamic system, no items may get cached or cached context would generate negative returns when the load is fairly distributed among context queries and/or information, e.g., prefetching context of an empty intersection for $Q_7$ (this is an example of *over context caching*).

Using context-awareness for proactive caching is proposed in (Zameel et al., 2019). But, measuring the micro-variations in context about the sensed environment is an existing problem, e.g., changes after a roadside accident. These can be identified as residuals to the current context depending on the time scale. Exceptions could be observed for well-established planning periods (PlnPrd) (Medvedev, 2020; Weerasinghe et al., 2022b). Trends and seasonalities can be fundamental for prefetching, e.g., perfecting for $Q_1$-$Q_7$ before the rush hour (seasonality) except during the holiday season (seasonality). Further, utilizing slower cache instances when the request rate by trend is estimated below a certain threshold in a year.

There are several distinct ways of implementing proactive caching in the literature, some of which are implicit. The methods provided, so far, explicitly prefetch context into the cache. PRESTO (Ming Li et al., 2009) implements a query-based proactive data retrieval adaptation for WSN. The main aim of PRESTO is to minimize the number of requests to sensors by using a predictive model. The sensors push data to the system only when the measurement is different from the prediction, and the query processor may respond to queries without retrieving from the providers, e.g., sensors. Now consider the notion of caching the estimated path of a "hazardous" (which is context about, e.g., a car, that could be derived based on the Context Space Theory (Boytsov and Zaslavsky, 2011a) using, e.g., ECSTRA (Boytsov and Zaslavsky, 2011b) or Bayesian techniques (Ullah et al., 2022)) vehicle rather than the transient location for $Q_7$. This is an form of proactive context caching since caching the model is cost-efficient compared to retrieving location and then compute.

Predictive modelling has been comprehensively investigated in the literature under prefetching. Schwefel et al. (Schwefel et al., 2007) developed models for calculating the (a) mismatch probability, (b) mean access delay, and (c) network overhead. Based on the provided mathematical models, the authors formulated two scenarios for maximizing cache lifetime: (a) minimizing average access latency and network usage, while keeping the probability of mismatch under a certain threshold, and (b) minimizing mismatch probability, while keeping the latency of access and network usage under a defined threshold. Bogsted et al. (Bøgsted et al., 2010) extended the probabilistic models to optimize access to transient data items in both reactive and proactive setups. But, mismatches in (Bøgsted et al., 2010; Olsen et al., 2006; Schwefel et al., 2007) are calculated using absolute measurements which are distant to the practical environment. Aging of transient data items, e.g., temperature reading, should rather be modelled using a decay function (Medvedev, 2020; Weerasinghe et al., 2022b). Therefore, further improvements to the models are necessary prior to be useful for prefetching in ACC.

*E. Elastic Caching*

The majority of the literature considered in this survey uses a limited sized cache memory considering in-network nodes that share the memory with the cache. But Infrastructure-as-a-Service (IaaS), or Platform-as-a-Service (PaaS) in the Cloud provides horizontal scalability to cache memories that can virtually scale on-demand allowing to theoretically access an unlimited amount of cache space. However, the pay-as-you-go model (e.g., Cloud SLA) requires applications to incur costs for the resources used. Then it is cost-inefficient to acquire a large number of cache resources that may be underutilized, e.g., caching all context (another example of *over context caching*), some of which are accessed only once or intermittently. So, Elastic Caching (DyScl) for context caching refers to acquiring and relinquishing caching resources (cache memory, and processing units required to make caching decisions) dynamically in relation to the current context query load.

We identify two forms of elasticity in adaptive caching, primarily distinguished by the size of the cache memory used.

- **Elastic cache allocation** – dynamically allocating a proportion of the limited physical cache memory space based on the criteria, e.g., tenant-based SLA (Chockler et al., 2011), caching policies (Gramacy et al., 2002), features of data (Kiani et al., 2012), etc. Due to the considerable large body of work in this area, we discuss this separately as another method of cache adaptation in the next sub-section, Section VI-F.
- **Caching in dynamically sized cache memory** – caching in a virtualized cache memory that dynamically varies its physical size to, (a) accommodate for large, cached data volume (Bibal Benifa and Dejey, 2019; Cidon et al., 2015), (b) minimize competition among competing context information for cache space (Guo et al., 2013), e.g., queries that are likely to generate similar earnings for the CMP by accessing data from the cache rather than retrieving from the provider, and (c) optimize a performance metric that is a function of cache performance, e.g., QoS measured in HR, average response latency, and cost per hour in (Kabir and Chiu, 2012). As we described earlier, our focus is on this form of elastic caching. We further elaborate our rationale and compare ACC using limited and scalable cache memories in Section IX.

Minimizing competition for cache is one of the objectives of eviction policies too in limited cache memory. But consider time-aware eviction policies – TLRU (Bilal and Kang, 2014), TTL, or Volatile TTL ("Redis," n.d.), as examples. These policies are biased for items with longer lifetimes (e.g., low transient). For instance, consider caching weather entity for $Q_4$ and $Q_5$ on a sunny summer day. The cache decisioning is advantageous for the non-critical, but long-lived weather entity when the cache space needs to be shared with short-lived, but more critical contexts, e.g., the location of a hazardous vehicle for $Q_7$. It is a necessity that all relevant contexts are cached for CoaaS to be time-critical (a QoC parameter), for which *DyScl* could be viewed as a potential solution.

Implementing *DyScl* is simple considering the existence of a number of commercial PaaS providers such as AWS as indicated in (Scouarnec et al., 2014). But IaaS frameworks provide more flexibility in solution development for elastic computing (Bibal Benifa and Dejey, 2019; Choi et al., 2020). Resources are allocated at a price per unit, however. Therefore, a novel cost optimization problem could be realized.

*F. Adaptive Cache Space Allocation*

Adaptive Cache Space Allocation or Space optimization (SpcOpt) refers to dynamically partitioning a fixed cache space using a predefined criterion. For example, setting the space in the cache memory that different tenants can access (Chockler et al., 2010). The work in (Kiani et al., 2011) adapts the virtual cache allocation proportional to the ratio of data with long and short lifetimes. Hence, it is strictly optimal-greedy for cache resource utility.

We can identify two classifications for work in this area:

- **By scaling unit of cache** – refers to allocating virtual cache pages, e.g., as in Gramacy et al. (Gramacy et al., 2002), or physical cache instances e.g., as in Hafeez et al. (Hafeez et al., 2018).
- **By the allocatee of the virtual cache space** – refers to the parameter by which the allocations are made to. We identify three types of allocatees:
  - **Tenants** – allocating cache space guaranteed by SLA to a tenant. The objective of this strategy is to ensure QoE. The guarantees are either (a) minimum space (Chockler et al., 2010), or (b) minimum performance, e.g., HR in (Chockler et al., 2011).
  - **Features of Data or Queries** – optimally separating the cache space based on a feature such as (a) size classes of objects (Bruce M Maggs and Sitaraman, 2015), or (b) lifetime of data (Kiani et al., 2011) so that a performance metric is maximized.
  - **Caching Policies** – optimally separating the cache space among *PoliSet* to maximize the HR by applying the relevant eviction policy suitable by the nature of the data access, e.g., the work in (Gramacy et al., 2002).

Kiani et al. (Kiani et al., 2011) investigate Bipartite Caching – a two-layer segregated cache design, each accommodating data items exhibiting short, and long lifetimes. The authors identify two areas for adaptation: (a) accommodating emerging scenarios and (b) meeting heterogeneous requirements. Consider a fixed ratio between long-short term spaces, e.g., 2:3, and a total context information size equal to cache size. It creates a competition for cache space, e.g., when 80% of the data are highly transient, and under-utility, e.g., the remaining 20% occupy only half of the allocated space. Accordingly, dynamic partitioning is advantageous. But the results produced by the authors are not significant. We identify the reason being the competition for space between the two types causing evictions from either to adjust space, e.g., "pre-mature eviction" – evicting a large data item by size that is expected to be accessed over a long period, to accommodate scaling needs of the short-term cache. We indicate this graphically in Figure 11. We later discuss in Section IX the usefulness of the notion of "slab classes" in Memcached ("Memcached - a distributed memory object caching system," n.d.) to overcome this problem for adaptive context caching.

Similarly, the Greedy-Dual-Size (GDS), and GDS with Frequency (GDSF) [83] algorithms assign a value to each object, by computing a function as a function cost of retrieving, size, age, and popularity, for each object. They are examples of function-based policies (which include policies such as LVF (Bruce M Maggs and Sitaraman, 2015) and Need-to-Refresh (Medvedev, 2020) referred to in our previous work as we discuss in Section VIII-A). These overcome a lot of drawbacks of traditional policies, e.g., LRU and LFU. But formulating the "right" function, involving "right" parameters, and weights is a complex task. So, attributing a priority value when adjusting cache space (e.g., as a multi-variate function) is an important challenge to investigate in this area.

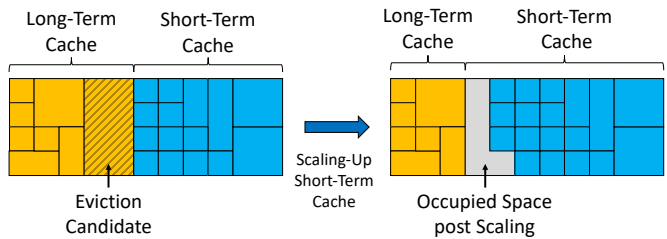

**Fig.11.** Drawback of adaptive bi-partite caching.

Tenant-based adaptive caching (Chockler et al., 2010; Choi et al., 2020) guarantees a minimum cache space for individual applications (i.e., Context Consumers in our scope), and services by the SLA. During our extensive literature search, however, we do not find any of the existing CMPs to provide this guarantee.

Consider a niche set of highly active consumers dominating the context management middleware in terms of query load, e.g., 50% of the context queries in the above scenario are generated by a parking assistant application. It is logical to maximize returns by adapting to these consumers by allocating more resources. On the other hand, this has a negative impact on less dominant consumers for whom, meeting the SLA could be more challenging with the remaining resources. Tenant-based caching, therefore, is a resource optimization problem that addresses (a) how to improve resource utility under a minimum allocation scheme, and (b) how to fairly allocate resources among dominant and non-dominant tenants.

Chockler et al. (Chockler et al., 2011) define the *space utility model* which factorizes HR for a given tenant over a specific period of time as a function of the dedicated cache size to solve the aforementioned problem. Constraints to the objective function in (Choi et al., 2020) are defined using an estimated HR as well. The strategy indicates good adaptiveness but is computationally intensive requiring re-computation of the hit-histogram for each request – a common drawback of function-based policies.

Implementing *SpcOpt* is straightforward. However, there are two critical aspects not evident in the reviewed literature. Firstly, how physical cache memory (i.e., pages) is handled in dynamic virtual allocation. For instance, consider facilitating uniform access to all levels of a logically related data structure (e.g., accessing all context levels in Figure 4 with the latency). Secondly, how to selectively evict cached items in a receding segment (e.g., long-term cache as in Figure 11) of a virtualized cache memory to minimize post-scaling degradation.

*G. Adapting by Locality*

Adapting by locality (AdLoc) refers to a form of adaptation at the infrastructural level in which caching is eventually focused on the features of the query load generated by the locality. The objective is to cache in close proximity to the consumer. One of the key challenges in context-aware systems is to provision contextual information about anything, anytime, and anywhere, according to Weiser (Weiser, 1999). This is achieved using edge caching. Hence, this technique applies only to distributed CMPs with cache-enabled and/or dedicated caching nodes. Adaptiveness in *AdLoc* is a consequent result of edge nodes adapting to cache content based on a set of metrics, e.g., the popularity of topics in each network cell (Muller et al., 2017; Zhou et al., 2015).

*AdLoc* can be discussed under two methods: (a) localizing data cache (Guo and Yang, 2019), and (b) localizing context cache processing (Faticanti et al., 2021; Orsini et al., 2016; Sheikh and Kharbutli, 2010), as in Figure 12. In the figure, the centralized node is often a controller hosted in the Cloud, but there can exist edge nodes with adequate processing power that localize ACC decision processes for a set of neighbouring caching nodes (i.e., nearest processing nodes). We refer to such a cluster of nodes that manage context information as a Context Management Cluster in later work. These processing nodes at the edge can also be caching context or dedicated only to executing ACC agents.

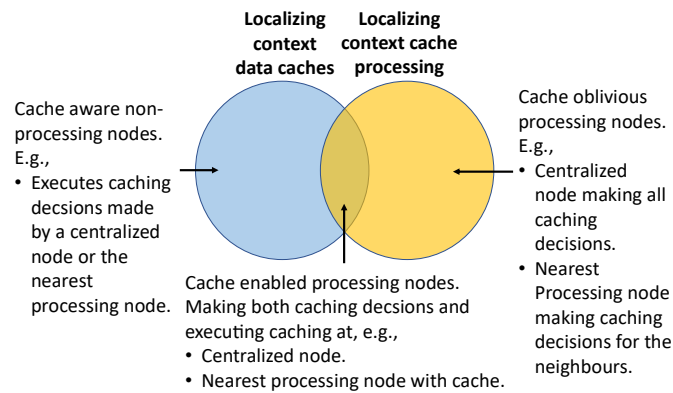

**Fig.12.** Adaptive data caching by locality.

Locality Aware Cost Sensitive (LACA) cache (Sheikh and Kharbutli, 2010) performs equally well against other locality-based policies such as the *deadlock predictor* that manages complex distributions, and wireless cell selection (Zhou et al., 2015). These fall under the category of *Optimization by Cache Organization* which we will discuss in Section VII. The major advantage of these techniques is the high HR due to the local relevancy of cached data suggesting that smart placement of data items in close proximity to the consumer is highly efficient. Sheikh et al. (Sheikh and Kharbutli, 2010) refer to an early concept by Sirinivasan et al. (Srinivasan and Lebeck, 1998), in which *critical cache blocks* are preserved in a special critical cache, or used to initiate prefetching for the locality. Criticality was estimated by keeping track of a load's *dependency chain* – which is also an interesting concept for CoaaS. For instance, the cached context in our motivating scenario is not relevant at a different city circle (different locality). So, ACC could benefit in QoS, QoC, and cost efficiency by adopting a similar approach by virtue of separating and prioritizing interesting context queries by locality.

Fanelli et al. (Fanelli et al., 2011) merge locality and QoS optimization in a peer-to-peer system that develops quality profiles for neighbouring consumers. This approach has several drawbacks which disqualify it as a strategy for a CMP. Firstly, in order for this approach to be useful and effective, all consumers should essentially be operating in coherence with a shared objective as in the scenario discussed in the paper. For instance, consider $Q_1$-$Q_7$. It is unnecessary to share the same quality profile for two or more adjacent cars, one of which may be searching for parking spaces close by, where the available parking spots are being competed for, in contrast to searching in a suburban area where parking is abundant. Secondly, to minimize cost inefficiencies, the disparity of quality requirements should be minimal among the neighbours. Extending the example discussed before, consider the additional refreshing cost incurred assuming proactive caching at the end of *ExpPrd* for two pairs of cars where the freshness thresholds are ($car_3$=0.9, $car_5$=0.8), and ($car_3$=0.9, $car_2$=0.4) respectively. In the latter pair, $car_2$ could be seen to incur additional costs to the CMP in-fetching compared to $car_5$ in the former pair because it is now subject to a 0.9 freshness threshold due to its proximity to the neighbour. Thirdly, when the neighbourhood is random and moving, quality profiles would need regular updates. Whilst it is computationally intensive, the definition of boundaries to a neighbourhood is abstract and difficult to define in a practical environment compared to the single hop criteria used by the authors in (Fanelli et al., 2011; Hou et al., 2019; Zhang et al., 2020).

We like to draw attention to *cooperative caching* (Psaras et al., 2012; Y. Wang et al., 2019; Zhang et al., 2018; Zhou et al., 2015) although it is not the most suitable in a cost-efficient ACC problem. We observe that this strategy results in redundantly caching the same data items across neighbouring nodes in the downlink. This is expensive especially if nodes are leased from a 3rd party IaaS vendor. Further, for the benefit of discussion, we point out that some of the work assumed low transiency for some data types (e.g., temperature) as a counter argument to compromise on quality (as a result of the reduced number of refreshing that is expensive in cooperative caching since multiple caches need to be updated). The assumption facilitates propagation of smaller-sized *deltas,* i.e., changes in value compared to the last reading, e.g., (Venkataramani et al., 2002), across the network. But the assumption is not generalizable because of three reasons. Firstly, it is applicable only at the context attribute level because, derived context can be complex objects, e.g., JSON-LD. Only a part of the JSON-LD document may not have any meaning. Secondly, certain context information (e.g., the response to $Q_7$) such as those derived from the location are also transient – an update of a context attribute results in a chain reaction that updates high-level context (Medvedev, 2020). Finally, the assumption is invalid because of inconsistencies in propagation delays. For example, different values of the same data in multiple cached nodes when *sampling_interval < network_traversal_time*. ProbCache (Psaras et al., 2012) probabilistically cache data that is expected to be accessed by (a) minimizing redundant copies to update and (b) leaving cache space for other data flows. The authors, however, do not discuss the costs related to cache misses or invalid responses. Zhang et al. (Zhang et al., 2018) propose and evaluate a cooperative caching scheme that dynamically adapts by auto-configuring according to IoT data lifetime and request rate. This underpins the concept of *ExpPrd* and freshness is considered when making the caching decision. Fatale et al. (Fatale et al., 2020) indicate an improvement in hit rate when items are non-redundant selectively cached and refreshed using the least useful policy in a cooperative caching setup, but ignoring delays of transmission among caches.

*H. Adapting for data refreshing*

Assuming a transient data item is already cached, one of the important decisions is – when to acquire a fresh snapshot of the data, to ensure the validity of the cached item. Different metrics to trigger refresh can be found in literature such as (a) need to refresh (NTR) (Medvedev, 2020), (b) age penalty (Sun et al., 2016), (c) cost of update delay (Kosta et al., 2017), to name a few. Adaptive refreshing changes the refresh rate according to the variability of the data item's lifetime, e.g., the time between the temperature sensor reading $31^0$ C to the time of the previous reading of $30^0$ C (Weerasinghe et al., 2022b). We identify this method to be proactive considering that an optimal refresh rate could fetch an item prior to a fresh copy being requested. Proactive cache strategies (e.g., reuse with and without shift, no ruse) in (Medvedev, 2020), and full-coverage retrieval in (Weerasinghe et al., 2022b) use this rationale.

An imminent advantage is the reduced network utility compared to a simple periodic refresh tested in (Weerasinghe et al., 2022b). The frequency of variation of sensed data (e.g., temperature) is subject to randomness and can be complicated to accurately estimate. It is why quantifying the Value of Information of Update (VoIu) – the importance of refresh data defined by how much it improves the accuracy of the predicted refresh rate (Abd-Elmagid et al., 2019) is a vital definition. Systems using cache refreshing for transient data are most likely to operate sub-optimally, e.g., in either accuracy of data or cost-efficiency in refreshing, unless the intervals between two different measured values can be accurately estimated (e.g., the probability of a parking slot being occupied in the next minute is 95%) or observed to be consistent over a longer period (e.g., temperature changes every 2 hours).

In TABLE II, we summarize the pros and cons of all the adaptive caching strategies discussed in this sub-section. In the next section, we will critically evaluate the literature considering the different approaches used to achieve the optimization goals discussed above.

VII. APPROACHES TO ACHIEVING EFFICIENCY GOALS IN ADAPTIVE DATA CACHING

In the previous section, we broadly discussed multiple methods of adapting. But they too can be categorized into two distinct approaches to cache adaptation: (a) optimization by cache organization (OpByOrg) and (b) adaptation for optimality guarantee (or learning an optimal adaptive caching policy). We indicate this categorization using a Venn diagram in Figure 13.

TABLE II
COMPARISON OF ADAPTIVE CACHING STRATEGIES

| Adaptation scheme | Pros | Cons |
|---|---|---|
| Adaptive cache replacement | <ul><li>Extensively researched forms of adaptive caching approach.</li><li>Most have simple definitions and hence, easy to implement.</li></ul> | <ul><li>Needs extensive monitoring on each data item and computation to decide items to replace, e.g., relative popularity (Pahl et al., 2019).</li></ul> |
| Caching policy shifting | <ul><li>Context-aware.</li><li>High cache hit rate when cached data access follows a consistent behaviour (Choi et al., 2020).</li></ul> | <ul><li>Probability of frequent policy shifts when no dominant context exists.</li><li>Additional cost to re-fetching to replace items evicted by the previous policy.</li></ul> |
| Hierarchical Caching | <ul><li>Logically and physically separates types of resource competing items (Chatterjee and Misra, 2016).</li><li>Virtualized. The view of the cache is uniform and unitarily accessed (Kiani et al., 2012).</li><li>Cost optimal to place data/context in differently expensive cache memories when consumers latency requirements varies or changes with time.</li><li>Higher hit rate because of longer, and larger cache data retention (Kabir and Chiu, 2012).</li></ul> | <ul><li>Demands for duplicate or redundant cache memories.</li><li>Complex processes to manage data among many levels, and technologies of cache.</li></ul> |
| Proactive caching | <ul><li>Can utilizes idling resources – cache, network, and processing (Sadeghi et al., 2018).</li><li>Caches data prior to requests are realized. No or minimal delay to reach optimal minimal response latency (Venkataramani et al., 2002).</li></ul> | <ul><li>Estimating items to proactively cache is non-trivial. Especially for long term cost minimization goals.</li><li>Hit rate depends on prediction accuracy.</li><li>Suffers from *cold start problem*. Optimal performance is reached after a delay.</li><li>Requires priori knowledge for prediction, which in some scenarios and data could be unavailable. Therefore, should resort to complex filtering algorithms.</li><li>Computationally expensive given the model requires to re-learn iteratively to adapt.</li></ul> |
| Elastic Caching | <ul><li>Accessibility to resources is theoretically unconstrained.</li><li>Enables optimizing for infrastructural, and resource costs (Bilal Benifa and Dejey, 2019).</li><li>Easily implementable using existing cloud-based Infrastructure-as-a-Service (IaaS), e.g., Choi et al., 2020), or with Platform-as-a-Service (PaaS), e.g., Scouarnec et al., 2014.</li></ul> | <ul><li>Applicable only in a Cloud environment.</li><li>Not economically viable if cache memory or scaling it is costly.</li><li>The decision to scale need to be proactive to fully yield the cost-efficiency benefit. But it is non-trivial when priori information is unavailable or uncertain.</li><li>Scaling up and down are complex operations at run-time (See above), e.g., due to managing post scaling degradation.</li><li>Research and experimentation on a commercial Cloud can be costly. Using simulators is more viable.</li></ul> |
| Adaptive Cache Space Allocation | <ul><li>Maximizes space utility of resources. Marginal utility of cache memory is maintained high.</li><li>Could reach optimal hit rates when the factor(s) used to decide the space allocation is/are highly skewed in values (Chockler et al., 2011).</li></ul> | <ul><li>Applicable only to limited sized cache systems.</li><li>Physical cache page management for dynamic space variation is non-trivial.</li></ul> |
| Adapting by Locality | <ul><li>Traffic is confined locally. Therefore, backhaul congestion is limited, consequently resulting fair allocation of backhaul network allocation among localities (Faticanti et al., 2021).</li><li>Shorter latencies serving from local caches (shorter hop distance), e.g., Guo and Yang, 2019.</li><li>Round trip to retrieve from cache is shortened when approaches like collaborative caching is utilized (Y. Wang et al., 2019).</li></ul> | <ul><li>Applicable only to when multiple cache memories are independently deployed in the network.</li><li>More suitable when cached data values are monotonic or static.</li><li>It is essential there exist similarity in requests for data/context among users in the locality to achieve efficiency objectives.</li></ul> |
| Adapting for Data Refreshing | <ul><li>Minimizes cost on context /data retrieval (CoC), e.g., Weerasinghe et al., 2022b; Medvedev, 2020.</li><li>Reduces load on backhaul networks.</li></ul> | <ul><li>Data lifetime of natural data are random. Therefore, non-trivial to accurately learn.</li><li>Invalid data can be served unless the refresh interval is accurate against the actual lifetimes.</li><li>Decision on optimal refresh interval is complex when multiple SLAs and dependencies among data items exists.</li></ul> |

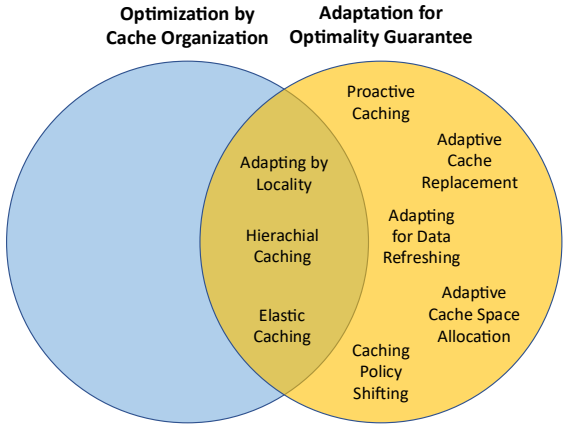

**Fig.13.** Classification of adaptive caching by approaches.

We refer to *OpByOrg* as the approach that employs infrastructure, cache memory organization, and/or data structures along with relevant algorithms to implicitly induce efficiency in to caching. Consider adapting by locality, e.g., cooperative caching (Y. Wang et al., 2019), or on-path caching (Pahl et al., 2019) which confederate data between different levels of proximally placed distributed caches to minimize the distance between a consumer and data even though the provider may be distantly located. The precondition here is the availability of a network of data caching nodes that could be selected at will to execute caching and internetworking that transfer data from any node to another.

Adaptation for optimality guarantee (AdOpGu) refers to methods ensuring the adaptive actions result in higher cost-efficiency, or QoS. In other words, *AdOpGu* is being objective greedy. Consider, *PoliShft* as an example. The switches (adaptive action) are aimed at maximizing the HR which are made by a process that observes, learns, and executes accordingly, e.g., using machine learning (ML) (Pahl et al., 2019). Two main features distinguish these approaches: (a) *OpByDes* is based on the assumption of perpetual connectivity (i.e., which is a feature of ICN networks) whereas *AdOpGu* is a constraint problem, i.e., managing the limited cache space, (b) *OpByDes* is coordinated, e.g., collaborative caching, whereas *AdOpGu* can be executed independently, e.g., selective caching agent in the Cloud (Blasco and Gunduz, 2014; Kirilin et al., 2019; Xu and Wang, 2019), and (c) *OpByDes* is infrastructure based, i.e., need many connected cache-enabled nodes to implement hierarchical caching (Chatterjee and Misra, 2016, 2014), whereas *AdOpGu* is software-based, e.g., cache replacement using an ML (Pahl et al., 2019; Xu and Wang, 2019), or reinforcement learning (RL) (Nasehzadeh and Wang, 2020; Sheng et al., 2020) models.

In Figure 13, we did not indicate any adaptive strategy that is purely *OpByDes* since none of the reviewed literature used the pure form. For instance, a simple rule-based adaptation by locality method could forward and cache a data item in the next node(s) closer to the consumer if its access frequency is more than a set threshold. This rule is unaware of optimal cache placement to maximize HR since the migrated data item could evict the item in the receiving node without guaranteeing an increase in the HR, e.g., replacing using LRU.

We have discussed the literature on different strategies of adaptation, which we also classified into two approaches. In the next section, we survey the different technical solutions that these strategies used to achieve their optimization goals. As per our classification in Figure 13, we refer to them as techniques to guarantee optimality.

## VIII. TECHNIQUES TO GUARANTEE EFFICIENCY IN ADAPTIVE DATA CACHING

Cache efficiency is a subjective construct, which depends on the goal of using cache memory. We indicated several such goals, e.g., minimizing the cost of responding to queries (Scouarnec et al., 2014), maximizing HR (Schwefel et al., 2007), etc. Despite the objectives, the strategies for adaptation are limited as we discussed in Section VI. Further, Mehrizi et al. (Mehrizi et al., 2020) state that there exist only two problems that adaptive caching needs to solve (a) data selection problem – identifying "*what*" data items to cache, and (b) cache placement problem – identifying "*where*" the data item should be cached. In this section, we survey the different technical solutions used to solve the adaptation problems. They are indicated in Figure 14 and discussed below.

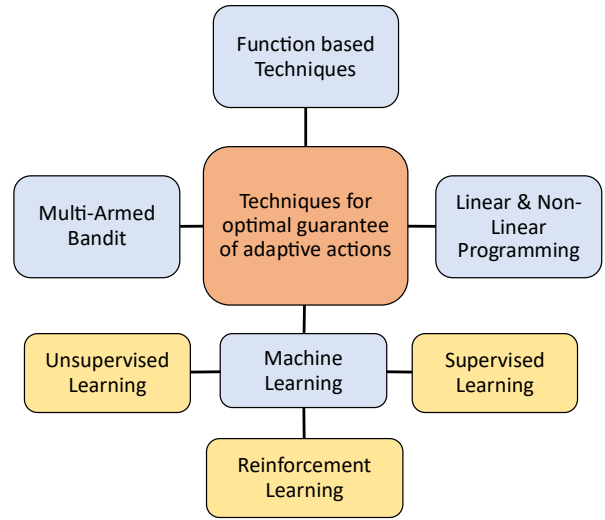

**Fig.14.** Techniques used for optimality guarantee of adaptive actions.

### A. Function-based techniques

The first technique involves defining a decision metric as a function of sever parameters (Al-Turjman et al., 2013; Anandarajah et al., 2006; Chatterjee and Misra, 2016, 2014; Schwefel et al., 2007). Decision criteria can be static, e.g., (Anandarajah et al., 2006), adaptive, e.g., (Venkataramani et al., 2002), or threshold(s) based. The decision metric can be any of:

- **Monetary cost** – e.g., in (Scouarnec et al., 2014). We previously discussed of (Sheikh and Kharbutli, 2010; Y. Wang et al., 2019) as cost models under ICN cache policies.
- **Performance metric** – e.g., (Schwefel et al., 2007) adopts *mismatch probability* rather than the hit probability considering it as a direct indicator for cache eviction and TTL using mathematical proof.

- **Abstract value** – e.g., in the Least Value First (LVF) policy (Al-Turjman et al., 2013), Least Useful policy (Fatale et al., 2020), Need to Refresh (NTR) value in (Medvedev, 2020), and cost of a data item at a router in (Vural et al., 2014).

But there are several drawbacks to this technique. First, functions are attributed to each data item, which requires to iteratively update for each read or write event. Unless a reliable estimation technique that could calculate for all data items using only metrics of a sample of data items, e.g., estimating based on similarity, formula-based approaches are computationally expensive at scale.

Secondly, weights defined for a formula are translatable from SLAs, e.g., in (Menache and Singh, 2015), authors use a weighted caching model in which cache misses are translated to fixed costs defined per user. In a scenario that involves multiple, complex SLAs, it is non-trivial to define formulas.

*B. Linear and Non-Linear Programming*

There exists a significant body of research applying linear and non-linear programming approaches to solve multivariate caching problems. Rugegeri et al. (Ruggeri et al., 2021) investigate the optimal content placement of selected items for caching using Integer Linear Programming. Authors consider the popularity and IoT data lifetime to formulate the decision criteria. The problem however is the NP-hardness of solving the problem since the expected accumulated latency gain is calculated for each data item-cache memory pair in discrete time intervals. Menache et al. (Menache and Singh, 2015), and Dasgupta et al. (Dasgupta et al., 2017) described applying convex optimization to online caching to reduce infrastructure costs in Cloud environments. The linear programming approach is applied to the costs of various types of Cloud compute instances, Cloud storage costs, access latencies to data, and different SLAs. This solves one of the questions we discussed in Section VI - optimizing the number of purchased Cloud cache instances. This approach can also be applied to efficiently allocate tasks, e.g. localizing cache context processing (Sheikh and Kharbutli, 2010) (refer Section VI), and localizing data caching (Malawski et al., 2013; Wyrzykowski et al., 2014) (refer Section VI).

We observe that a significant number of Machine Learning (ML) techniques including Reinforcement Learning (RL) and Multi-Armed Bandit (MAB) involve non-linear optimizers, e.g., policy gradient methods implement a gradient decent – a non-linear programming model (Sutton and Barto, 1998). So, we break our survey into three sub-sections as follows.

*C. Machine Learning*

Machine Learning (ML) techniques are commonly used in prediction tasks. Techniques that involve pattern recognition, anomaly detection, feature learning, regression analysis, etc., are trained using large representative historical data sets. ML is broadly categorized into three paradigms: (a) supervised – features tagged with the class labels, (b) unsupervised – classes inferred during training, and (c) reinforcement learning. Previous work that utilized Reinforcement Learning is discussed in the next sub-section. Shuja et al. (Shuja et al., 2021) have conducted an extensive study on applying ML for adaptive data caching, however, not discuss their application for caching contextual information.

Supervised learning for ACC is infeasible due to two main reasons: (a) volatility of context resulting in lack of knowledge, e.g., new context information, variation of context lifetime (Sheng et al., 2020; Weerasinghe et al., 2022b) and (b) dimensionality problem – the impossibility to define all features and caching actions relevant to the ACC design at a given time.

Unsupervised learning techniques such as online clustering are attractive techniques to make selective caching decisions when prior knowledge is not available. Khargaria et al. (Khargharia et al., 2022) implement an ACC mechanism that estimates the popularity of a cluster of context entity types. While this approach is completely oblivious to the QoC and cost efficiency of the CMP, the popularity of a cluster cannot practically be homogenous to all its context entity instances. So, one of the limitations of clustering context information is defining a valid clustering criterion. Secondly, it is difficult to verify or validate clustered context information as it can be highly subjective to the entity or the scenario. For instance, consider a contextual model that represents the stress level of a driver in our scenario. Apart from the context of being stressed being highly subjective to the individual even under the same conditions, the lifetime can similarly vary. Clustering such context information based on representation or definition can be inaccurate. Any estimations made using these context clusters, e.g., estimated popularity, are hence not generalizable, nor accurate. So, there exists no ground truth to clustering context information.

Zameel et al. (Zameel et al., 2019) trains an time-series model using ARIMA. The authors showed an 81.4% increase in HR for a 50% increase in cache size. $HR \propto CacheSize$ is a feature of LRU. There is no compelling evidence to suggest an additional advantage of applying time-series predictions, which could be reasoned out using the volatility of the context information that does not conform to trends or seasonalities.

Phal et al. (Pahl et al., 2019) propose a self-organizing machine learning (ML) model to prefetch sensor measurements purely on a theoretical basis. It is however interesting to compare and contrast this model against (H. Wu et al., 2021) and (Scouarnec et al., 2014) in which ML is utilized the same way but for edge caching. In order to apply ML, the items should be uniquely identifiable which is a challenge with context. As in the case of (Pahl et al., 2019), the authors adopted the Virtual State Layer (VSL) middleware for tagging content using VSL ID. A similar approach to virtualizing Sensor Cloud is found in (Chatterjee and Misra, 2014). In either case, it enforces an abstraction of the sensing devices as a service – which is similar to the notion of context service in CoaaS (Hassani et al., 2018).

A key drawback of the works mentioned in this sub-section is the lack of details in the presentation on the processes, e.g., how Extract-Transform-Load (ETL) is performed, and how data is localized in making caching decisions. This is a limitation when critically analysing this work.

A promising work using ML technique for IoT data caching can be found in (Ale et al., 2019). Authors employ a three-block architecture consisting of a 1-D convolutional neural network (1-D CNN) for feature selection, followed by an LSTM network purposed for predicting requests as a time-series and a fully connected neural network (FCNN) to filter the most likely items to access (for prefetching). The solution attempts to solve the problem of temporal association of queries, which we have discussed in the previous sections. Authors assume there exist similarities among different user requests (as in the case of *AdLoc* by locality) which is acceptable. The drawbacks however are, (a) the imminent process heaviness incorporating three large neural networks which do not scale well with very large amounts of IoT data; the authors do not indicate any QoS measurements, (b) the accuracy of prediction is only 60%, which indicates that ~40% of the cache occupancy generate a *hold-up* and additional retrieval costs; it is not cost-efficient, therefore.

*D. Reinforcement Learning*

There is a significant body of research in adaptive data caching that utilizes Reinforcement Learning (RL). Wang et al. (Wang and Friderikos, 2020) provide a comprehensive survey on different approaches of RL for edge caching. We find the main reason for this popularity is the model-freeness of developing an RL solution. There are four key advantages to ACC using RL: (a) learns a model without prior knowledge, i.e., as we discussed the nature of context and its priors in Section 1, (b) solution design is focused on developing the "right" objective and reward functions, and (c) RL models evolve over time, e.g., decisions adapt with experience (Zhu et al., 2019), and (d) relaxation of dimensionality concerns at design time. We discuss each of these using the related literature below.

In TABLE III, we first summarize all the RL architectures and algorithms used in previous literature and indicate their abbreviations for brevity in our later discussions. Authors in (X. Wu et al., 2021) use a technique called *Soft Policy Improvement* (SPI) for policy optimization, which is a form of *Proximal Policy Optimization* (PPO). SPI is hence indicated under PPO. Then in TABLE VI, we indicate the solution design by architecture and optimizer for each relevant work.

TABLE III
RL ARCHITECTURES AND OPTIMIZERS USED IN LITERATURE

| | Abbrev. | Description |
|---|---|---|
| **RL Architecture** | AC | Actor-Critic Network (He et al., 2019) |
| | A2C | Advantage Actor-Critic Network (Sheng et al., 2020) |
| | A3C | Asynchronous Actor Critic (Zhu et al., 2019) |
| | MA-AC | Multi-Agent Actor Critic Network (X. Wu et al., 2021) |
| | MDP | Markov Decision Process (Somuyiwa et al., 2018) |
| | MDP-SI | Markov Decision Process with Side Information |
| | QN | Q-Network (He et al., 2019) |
| | DQN | Deep Q-Network (Sheng et al., 2020) |
| | EMRQN | External Memory based Recurrent Q-Network (Wu et al., 2019) |
| | FTN | Fixed Target Network (Zameel et al., 2019) |
| **Optimizer** | OGD | Online Gradient Decent Method (Sheng et al., 2020) |
| | PG | Policy Gradient Method (Zhu et al., 2019) |
| | PPO | Proximal Policy Optimization (X. Wu et al., 2021) |
| | FDM | Finite Difference Method (Somuyiwa et al., 2018) |
| | IDSG | Iterative Dual Sub Gradient Method (Sadeghi et al., 2018) |
| | DDPG | Deep Deterministic Policy Gradient (Zhong et al., 2020) |

Firstly, the ability to learn without prior knowledge, e.g., unknown transition probabilities (Sadeghi et al., 2019b), is a necessity for ACC due to the nature of context as we discussed earlier. Secondly, we agree with Wang et al. (Wang and Friderikos, 2020) that the reward scheme should be appropriately designed to ensure that the RL agent does not converge at short-term rewards (local optima) for a long-running dynamic system. For instance, consider a reward > 0 at each step for correct action (e.g., caching in the short run). The agent, in turn, could improve the accumulated short-term benefits although not yielding long-term rewards, e.g., due to over context caching. So, the definition of an optimal policy ($\pi^*$) and reward scheme to reflect long-term orientation is a design question. But there are three major concerns: (a) identifying the constraining parameters, e.g., the budget of the CMP for ACC, and (b) measurement of parameters, e.g., how to measure and represent co-occurring context queries and similar complex dependencies among context information, and (c) storage of measured parameters, e.g., in Context Storage and

TABLE IV
COMPARISON OF ADAPTIVE DATA CACHING STRATEGIES

| Literature | RL Architecture | | | | | | | | | | Optimizer | | | | | |
|---|---|---|---|---|---|---|---|---|---|---|---|---|---|---|---|---|
| | AC | A2C | A3C | MA-AC | MDP | MDP-SI | QN | DQN | EMRQN | FTN | OGD | PG | DDPG | FDM | IDSG | PPO |
| He et al., 2019 | | | | | | | ✓ | | | | | | | | | |
| Nasehzadeh and Wang, 2020 | ✓ | | | | | | | | | | | | | | | ✓ |
| Sadeghi et al., 2019a | | | | | | | ✓ | | | | | | | | | |
| Sadeghi et al., 2019b | ✓ | | | | | | | | | | | | | | | |
| Sadeghi et al., 2018 | | | | | | | ✓ | | | | | | | | ✓ | |
| Sheng et al., 2020 | | ✓ | | | | | | ✓ | | | ✓ | | ✓ | | | |
| Somuyiwa et al., 2018 | | | | | ✓ | ✓ | | | | | | | | ✓ | | |
| Wu et al., 2019 | | | | | | | | ✓ | ✓ | | | | | | | |
| X. Wu et al., 2021 | | | | ✓ | | | | | | | | | | | | ✓ |
| Zameel et al., 2019 | ✓ | | | | | | | ✓ | | ✓ | | | | | | |
| Zhong et al., 2018 | | | | ✓ | | | | ✓ | | | | | ✓ | | | |
| Zhong et al., 2020 | ✓ | | | | | | | | | | | | ✓ | | | |
| Zhu et al., 2019 | | | ✓ | | | | | | | | | ✓ | | | | |

Management System (CSMS) (Medvedev, 2020; Medvedev et al., 2017). This is non-trivial for a large near real-time system. Sheng et al. (Sheng et al., 2020) use a rolling window strategy to measure cache performance parameters and execute learning based on them. We find two advantages to this method: (a) prevents the system from compromising to a sub-optimal cache state in attempting to optimize for a very long term, e.g., optimal policy during 8:00-9:00 am is not applicable during 1:00am-4:00 am for $Q_1$-$Q_4$ in Melbourne (Auditor General of Victoria, 2013) and (b) the notion of windows is similar to the notion of *PlnPrd*s in which we have already established to prune the parameter space to a manageable set.

There are three categories into which the reward functions can be classified by objective:

- **Maximizing reward for the CMP** – reward function designed using QoS parameters - e.g., Wu et al. (Wu et al., 2019) demonstrate a higher HR compared to the benchmarks under dynamic data and request popularity and in (X. Wu et al., 2021), authors minimize the average age-of-information and cost of cache updating. We focus more on this approach in our discussion since it aligns with *Perspective A* of efficiency.
- **Maximizing rewards for the consumer** – reward function designed using QoE parameters, e.g., He et al. (He et al., 2019).
- Maximizing reward for the CMP using consumer modelling, which is a hybrid of the above functions in which the individual user behaviours, e.g., access patterns, are modelled to estimate the QoS of the CMP.

Zhu et al. (Zhu et al., 2019), Sheng et al. (Sheng et al., 2020), and Nasehzadeh et al. (Nasehzadeh and Wang, 2020) pursue the objective of maximizing the overall reward for the system. The goal is to maximize HR (the reward) by estimating the popularity of data items. The authors of (Zhu et al., 2019) estimate the IoT data popularity using a collaborative filtering technique based on user request patterns. The solution has a simple construct compared to the latter two because of two reasons. Firstly, the model adapts episodically and converges to local optima per user. Therefore, there is a mutual exclusion among consumers on sub-optimal adaptive performance during their respective learning periods. Models are therefore subjective and easier to train using user preferences as a *Knapsack problem*. Secondly, user models are less likely to change over shorter periods of time. In contrast, consider models developed for maximizing reward for the CMP, e.g., system-wide cost-efficiency in (Sadeghi et al., 2019b). In the worst-case scenario of complete randomness, it is bound not to converge. Assume a set of dominant context information, e.g., consider $Q_1$-$Q_7$ to be dominant by popularity among other context queries in the area. The problem would rather be to find a caching model most appropriate for the dominant queries. *Greedy policy selection,* e.g., defining clipped functions (Nasehzadeh and Wang, 2020; X. Wu et al., 2021), would be more effective to direct the policy search (which is a matter of QoS, since shorter convergence time results in higher accumulated perceived value to the CMP in the long term) compared to randomly searching for optimal policies among the policy set.

In Section IV, we defined the Cost of Context Caching (CoCa) as an abstract construct. We find several ways this concept is realized in literature. Sadeghi et al. (Sadeghi et al., 2019a) align the reward function to minimize CoCa. But the optimal value function is approximated. It maximizes process efficiency at the expense of long-term cache performance-efficiency. In (Sadeghi et al., 2018) cost is defined as the difference between what is cached now, and what is decided to be cached. The more different they are, higher the cost. This is a de-facto reward function. For a minimizing objective, the model can converge to a state that is least probable to incur the costs in the immediate decision window. So, the reward function encourages short-term optimality by design. Similarly, Wu et al. (Wu et al., 2019) define the reward function by the popularity of current cache items, and the weighted popularity of cache items retained from the previous window. It maximizes the popular data that stays in the cache for a longer period while a negative reward is assigned to cache evictions. We identify that it balances myopic behaviour that the agent may exhibit only if the positive reward was considered.

Thirdly, RL models evolve with experience which we discussed as a need for ACC based on the volatility of context query loads. Adapting either by proactive caching (Sadeghi et al., 2018; Somuyiwa et al., 2018; Wu et al., 2019; Zhong et al., 2018; Zhu et al., 2019), selective caching (Kirilin et al., 2019), or cache replacement (Nasehzadeh and Wang, 2020; Sadeghi et al., 2019b; Sheng et al., 2020; H. Wu et al., 2021), the convergence occurs over multiple iterations. This contrasts with the drawbacks discussed for literature using ML in Section VIII-C.

It is important to discuss why a greater number of works in adaptive caching that implements DDPG methods. We make this observation using TABLE IV. The actor-critic architecture shows promise in the applicability for the adaptive context caching problem because the critic network could refine the actor-network by feedback, e.g., against a loss function (Nasehzadeh and Wang, 2020; Zameel et al., 2019). According to (Wang and Friderikos, 2020), RL can be categorized into two types: (a) value function based, e.g., Q-Learning in (Sadeghi et al., 2019a; Wu et al., 2019) and (b) policy search methods, e.g., PG in (Zhu et al., 2019), PPO in (Nasehzadeh and Wang, 2020), FDM in (Somuyiwa et al., 2018), DDPG in (Sheng et al., 2020; Zhong et al., 2018), and IDSG in (Sadeghi et al., 2019a). But we note that there is (a) a lack of prior knowledge - e.g., in (Sadeghi et al., 2019b), the authors' state file requests have unknown transition probabilities, and unknown time-varying popularity profiles considered in (Liu et al., 2021) and (b) complexity to define all relevant dimensions for input in ACC because of the multitude of parameters that affect various aspects of the adaptive caching decision. It is infeasible to adopt value-functions-based approaches since it is essential to have knowledge of at least the transition probabilities to accurately estimate either value or Q-functions (Wang and Friderikos, 2020). The rationality to adopt a policy search method can be attributed to four reasons:

- Faster convergence (Wang and Friderikos, 2020) – which is advantageous to minimize the negative

impact of (a) the *cold start problem*, (b) temporally varying characteristics of the query load, e.g., the composition of context queries (Medvedev et al., 2018), and (c) emergence of new context and queries, on the objective function.
- Policy search methods handle stochastic processes better than the function approximation techniques, e.g., in (Zhong et al., 2018), request frequency is derived as a statistical process. Note that internally computed metrics to ACC are stochastically derived such as the co-occurrence probability of a pair of context information.
- Parametrizes the policy ($\pi_\theta$) by known parameters ($\theta$), which are metrics measured internally, e.g., HR, CoCa, or externally provisioned, e.g., penalty cost in SLA. Further examples are (Sadeghi et al., 2019a; Zhu et al., 2019) where the policy is parametrized in a multivariate manner by, (a) file popularity, (b) cost of caching, and (c) loss of freshness.
- Tunes the parameters of the policy function in search of an optimal policy using the gradients of the performance metrics, e.g., if the performance metric is Hit Rate (HR), the search follows the policy that maximizes the gradient of the HR variation. Similarly, we can develop to search policy that minimizes the gradient of Cost of Context Caching.

It is evident from the discussion above that, although improving cache decisions by experience is a feature of RL, not all techniques apply to Adaptive Context Caching.

Fourthly, there is work to discuss the advantage of the reduction in dimensionality concerns when developing adaptive caching solutions. There are two aspects to it: (a) identifying feature space and (b) identifying action space. All the literature listed in TABLE IV is defined using multivariate feature vectors, which is different from previous work in the areas that are more focused on a single feature. The curse of dimensionality is handled by tuning weights which are learnt iteratively (Sutton and Barto, 1998; Wang and Friderikos, 2020).

In the literature we have already discussed, the possible caching actions are predefined, e.g., all candidate files to cache (Nasehzadeh and Wang, 2020; R. Wang et al., 2019; Zhang et al., 2020). But we identified defining all actions for context caching as non-trivial. The number of all possible actions is also a concern when considering RL for ACC. Nasehzadeh et al. (Nasehzadeh and Wang, 2020) propose the Wolpertinger architecture (Dulac-Arnold et al., 2015) to heuristically estimate actions which otherwise would result in exponential growth in cost and time-complexity as a result of a large action space. Using PPO for policy optimization, the authors attempt to improve the efficiency and accuracy of caching by using k-Nearest Neighbour (kNN) clustering greedy selection. A DRL agent is used for prefetching and making caching decisions to store the frequently requested contents at local storage as a *replay memory*. Similar is the work in (X. Wu et al., 2021).

But there are three key issues in applying RL as a technique for ACC are summarized in Figure 15 and discussed below.

Firstly, when using gradient decent methods for cache optimizing (Nasehzadeh and Wang, 2020; Sadeghi et al., 2019a, 2019b; Sheng et al., 2020; Somuyiwa et al., 2018; Zhu et al., 2019), *vanishing gradient problem (VanGrdPro)* – the problem of the gradients of loss functions getting smaller, eventually leaving weights of some layers in the neural network to unchanged, is a concern. This is induced by an enlarging action space, e.g., in relation to novel context information. The difference between any two caching actions can get indifferent, e.g., caching context of $Q_1$ versus $Q_7$. As a result, the gradient decent may never converge to an optimum. There are two solutions investigated in the literature. Wu et al. (Wu et al., 2019) propose to use Long Short Term Memory (LSTM) to alleviate this problem. They predict the outcome of a state-action pair based on gradients of recent actions. Perseverance of recently learnt information and the ability to use them in the current time slot is a feature of LSTM (Ale et al., 2019). The other solution is to generate differentiable actions accepting the increase in the number of actions. Wu et al. (X. Wu et al., 2021) customize the Gumbler-SoftMax sampler for this purpose. The authors also propose to improve exploration of the action space as a countermeasure to prevent premature convergence. It is however not clear the sustainability of this solution in the long-term because of (a) the infinite number (theoretically) of context caching actions that are hardly differentiable and (b) the effectiveness of exploration when the probability of selecting a context caching action rapidly decrease.

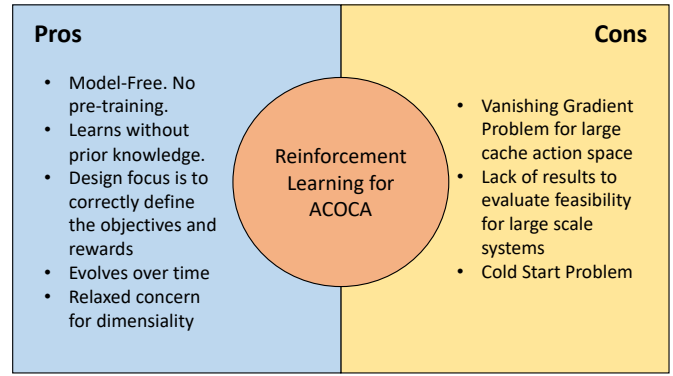

**Fig.15.** Pros and Cons of applying RL for ACC.

Secondly, experimental setups in related work are small to evaluate their scalability. We summarize several experimental setups in TABLE V. Further investigations are needed (a) when there are multiple caching nodes, and (b) for heterogenous data, types, queries, popularity distributions, and request rates.

TABLE V
SUMMARY OF TEST SETUPS USED IN RL APPROACHES TO ADAPTIVE CACHING. IS? – IS SIMULATED, NF – NOT CLEARLY INDICATED, N – VARIABLE, BS – BASE STATION

| Research | Sample Size | Caching Nodes | IS? |
|---|---|---|---|
| He et al., 2019 | 1001 chunks | 1 BS | Yes |
| Kirilin et al., 2019 | 49-436 million requests 7-56 million unique objects | 1 | No |
| Nasehzadeh and Wang, 2020 | NF | 1 | Yes |
| Sadeghi et al., 2019a | ~ (< 10,000) files | 1 | Yes |
| Sadeghi et al., 2019b | 1,000 leaf nodes | N | Yes |

|  | 1,000 files |  |  |
|---|---|---|---|
| Sadeghi et al., 2018 | 10 contents | 1 | Yes |
| Sheng et al., 2020 | NF | 1 | Yes |
| Somuyiwa et al., 2018 | NF | 1 | Yes |
| Wu et al., 2019 | 20 users | 1 | Yes |
| X. Wu et al., 2021 | 10 sensors<br>100 users per edge node | 4 | Yes |
| He et al., 2019 | 1001 chunks | 1 | Yes |
| Zhong et al., 2020 | 10,000 requests<br>5,000 files | N | Yes |
| Zhong et al., 2018 | 10,000 requests<br>5,000 files | 1 | Yes |
| Zhu et al., 2019 | 50 sensors | 1 | Yes |

Thirdly, the model-free nature of RL imposes a QoS problem of concern to near real-time systems. Chugh et al. (Chugh and Hybinette, 2004) describes it and is observable in the results of many work (Nasehzadeh and Wang, 2020; Zhu et al., 2019). We referred to it by the *cold start problem* – sub-optimal cost- and performance efficiency until convergence. This is where prior knowledge is useful, as in (Nasehzadeh and Wang, 2020), in order to pre-train and minimize convergence latency. This notion is clear using sampling techniques such as *Thompson sampling* (Zhu et al., 2020) where central tendency measures stabilize for a distribution of data items after a number of iterations. Let us assume that an RL model is trained for each *PlnPrd* and applied accordingly for ACC. Two reasons would not allow ACC to perform efficiently right away in the current *PlnPrd*: (a) difference in parameters and metrics observed in the previous *PlnPrd*; so certain metrics need to gather an adequate amount of data in the current *PlnPrd*, and (b) new context information due to possible lack of pattern in the query load (Medvedev et al., 2018), e.g., context about the $car_6$ not previously observed in the scenario in Figure 2.

*E. Multi-Armed Bandit methods*

Multi-Armed Bandits (MAB) is a sub-class in Reinforcement Learning where the action space is strictly confined to a predefined set among which a single action is chosen at the decision time. The goal state of MAB is concrete and assumes the convergence is guaranteed over a finite number of iterations. Previous literature often assumes minimum or no state transitions in the interacting environment, e.g., variation in preference for certain car parks, to ensure convergence. MAB is, therefore, suitable for well-defined optimization problems where the observed entity or entities are unlikely to change frequently.

A considerable body of research optimizing for QoE of individual users (Mehrizi et al., 2021; Yu et al., 2020; Zhang et al., 2020), such as training recommendation models to proactively cache for individual users could be found in the literature. For instance, Zhang et al. (Zhang et al., 2020) used non-negative matrix factorization to estimate the probability to decide to cache new data items using the user preference matrix of known items. Mehrizi et al. (Mehrizi et al., 2020) investigate a similar proactive caching strategy using tensor train decomposition, and Kalman filtering, while Yu et al. (Yu et al., 2020) adopt a variational auto-encoder. The focus of these work is to mitigate the lack of prior knowledge, which is a challenge in ACC as well. MAB approaches to proactively caching can be viewed as a collaborative filter in this sense. User-based matrices (e.g., preference matrices (Zhang et al., 2020)) make this approach technically feasible because users are often seen to have limited preferences. So, the number of items to compare a new item against is limited compared to comparing against a theoretically unlimited number of items in a system-wide perspective.

One of the differences between MAB to RL is the optimizers. Blasco et al. (Blasco and Gunduz, 2014) use combinatorial optimizers – searching for an optimal policy by exploring a finite set of feasible solution sets (typically state-action pairs). It is at least NP-hard to solve. NP-hard problems are both computational and resource intensive, both of which contradict the objectives of maximizing QoS and the cost of ACC.

In this section, we discussed the different approaches for optimal guaranteed adaptive data caching in the literature. In the next section, we discuss related work in adaptive caching using scalable Cloud caches.

IX. SCALABLE CLOUD-BASED CONTEXT CACHING

One of the major challenges to ACC is the large volume of context information about an entity (Perera et al., 2014), which is theoretically larger than the data attributed to that same entity (e.g., the properties). We predominately discussed adaptive caching in limited-sized cache memory in the previous sections, but identified elastic caching, e.g., in the Cloud, as a viable option to overcome several issues (Bao et al., 2020; Choi et al., 2020), e.g., consistency in QoS. In this section, we survey the Cloud caching options, and their feasibility to implement in ACC, also comparing them against limited-sized cache memories.

*A. Methods to handle the scale of caching*

One of the techniques to manage scale efficiently with limited resources is load balancing (Figure 16). Maggs et al. (Bruce M. Maggs and Sitaraman, 2015) group together content provider objects by serial number and then hashed them to the same bucket (e.g., binning). The authors use this hash to redundantly cache a piece of data in multiple distributed cached instances since it is time-inefficient to satisfy all the data requests using a centralized cache memory. Cui et al. (Cui et al., 2020), Liu et al. (Liu and Wang, 2018), and Shuai et al. (Shuai et al., 2017) propose a context-aware load balancing schemes based on cost. In comparison to the literature discussed previously, work in load balancing has focused on content-wise adaptation, e.g., per file. But load balancing (LB) is not excessively scalable, and the cache size is limited which can still cause competition or under utility. Consider an increasing trend in the volume of context queries. Then, the number of cache memory instances is subject to scale, manually or otherwise.

We iterate the fact that context data lack patterns (Medvedev et al., 2017). Query loads can show diurnal variability and unpredictable peaks. Several works (Guo et al., 2013; Kabir and Chiu, 2012) suggest the advantages of using an in-Cloud cache that expands and contracts correspondingly. A similar motive can be found in Dynacache (Cidon et al., 2015).

Another technique is to dynamically re-size the cache memory, which we are interested in this paper. Verma et al. (Verma and Bala, 2021) survey auto-scaling techniques for IoT-based applications on the client side. Two interesting surveys are (Qureshi et al., 2020; Singh and Chana, 2016) which discuss QoS-aware elasticity for near real-time services. Overall, low cost is the main advantage of Cloud computing (Bao et al., 2020), which is in line with the cost-efficiency objective of ACC. The client-side Cloud caching technique investigated in (Banditwattanawong and Uthayopas, 2013) extends these advantages with, (a) scalability and (b) service responsiveness. A significant body of research in cost-efficient Cloud caching strategies bound by SLAs can be found in (Didona et al., 2014; Herodotou et al., 2011) because Cloud resources are leased at a price. Hence, context cache memory scaling is a coherent cost-aware selection problem similar to selective context caching.

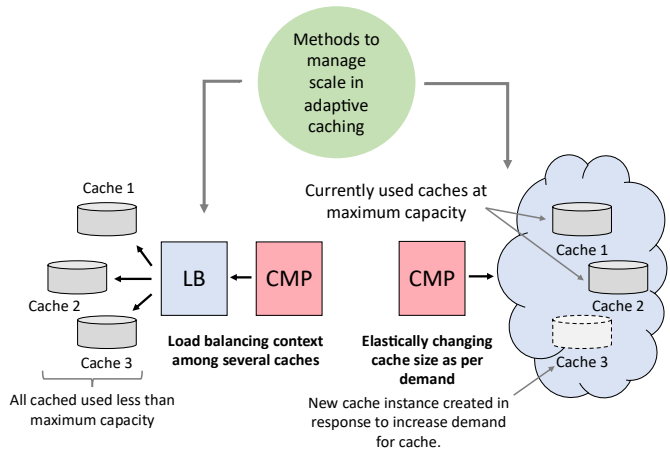

**Fig.16.** Methods to handle scale in Adaptive Context Caching.

Adaptive scaling has been investigated either: (a) virtually (Shen et al., 2011; Verma and Bala, 2021) or (b) physically (Bibal Benifa and Dejey, 2019; Choi et al., 2020; Guo et al., 2013) scale. PaaS products such as those offered by Amazon Web Services (AWS) (Andrade et al., 2019; Kabir and Chiu, 2012; Scouarnec et al., 2014) are used to implement virtual scaling whereas IaaS technologies such as Apache Ignite ("Distributed Database - Apache Ignite," n.d.) are used to implement physical scaling.

There are several metrics authors use to make scaling decisions. Choi et al. (Choi et al., 2020) estimate the data access pattern per tenant to predict the hit rate for estimated cache space. Chockler et al. (Chockler et al., 2011) implement a similar estimator constrained by an SLA. Another group estimates the required cache space size as a function of the popularity of data items (Andrade et al., 2019; Guo et al., 2013; Scouarnec et al., 2014). Resource utility (which is related to *hold up cost*) is a parameter used in RLPAS (Bibal Benifa and Dejey, 2019). Scouarnec et al. (Scouarnec et al., 2014), Kabir et al. (Kabir and Chiu, 2012), and Bao et al. (Bao et al., 2020) use a cost prioritized scaling approach.

*B. Considerations of context caching in the Cloud*

In this survey, we do not discuss specific caching technologies to implement the cache memory for ACC. But we identify several key features in existing technologies that could be used to evaluate the feasibility.

Firstly, cache memories implemented using IaaS options are more desirable to implement elastic caching based on the work of (Bibal Benifa and Dejey, 2019; Choi et al., 2020; Guo et al., 2013; Kabir and Chiu, 2012). But technologies such as Redis ("Redis," n.d.), Memcached ("Memcached - a distributed memory object caching system," n.d.), or EhCache ("Ehcache," n.d.) are stateful. Horizontally scaling on-demand is non-trivial and expensive because caching nodes (e.g., servers, containers) need to be deployed, relinquished, and re-deployed regularly. On the contrary, Caching-as-a-Service products - MemCachier ("MemCachier," n.d.), Elasticache ("Amazon ElastiCache - In-memory datastore and cache," n.d.), and Azure cache ("Azure Cache for Redis | Microsoft Azure," n.d.) are virtually scaled, abstracting this complexity from the ACC developers.

Secondly, the technology of cache memory units is an important aspect when elastic caching is implemented using IaaS. Bao et al. (Bao et al., 2020) signify the cost of using Non-Volatile-Memory (NVM) for write-heavy caching, i.e., cached context is write-heavy due to refreshing. It is a question of priority and balancing between the recoverability of context and cost. NVMs are however low in write endurance.

Thirdly, performance cliffs or *post-scaling degradation* (Hafeez et al., 2018; Verma and Bala, 2021) is a major concern to the profitability of elastic caching. Dynacache (Cidon et al., 2015) attempts to solve this problem using a stack distance based approach per slab class using Memcached ("Memcached - a distributed memory object caching system," n.d.). There are several drawbacks to it. Firstly, stack distance is a notion related to least-recently-used (LRU) items. Secondly, stack distance suffers in sparsely allocated caches. Hence, could be erroneous in scenarios that trigger scaling-down actions. Thirdly, it uses a convex solver. Multi-attribute optimization like Adaptive Context Caching is not guaranteed convex problems but rather concave, e.g., consider the findings in (Weerasinghe et al., 2022b). Pareto optimal methods are suitable candidate strategies instead (Naz et al., 2016).

Finally, elastic caching takes a long restoration time to reach efficiency. ElMem (Hafeez et al., 2018) propose migrating hot data to scaled-up instances from existing ones and from victim instances (scale down) to remaining instances to balance cache utility. The authors depict a significant reduction of 90% in restoration time. The problem of hot context information migration needs to be resolved in relation to query and context information dependencies, which is not yet investigated in the literature. For instance, *goodForWalking* depends on the cached context attributes about the weather. Similarly, these attributes can be a dependency on a different context function or a context query such as *goodForJogging*. Selectively migrating context cache pages based on access frequency and then mandatorily evicting any context information in the victimized instances are not straightforward in ACC. The scale-down, migration, and eviction process could negatively impact the QoS of the CMP due to cache misses when responding to a set of context queries (Hafeez et al., 2018).

Based on the above and our cost-efficiency objective, cache eviction is still a valid problem to maintain a context cache memory at a cost-efficient level of utility, i.e., using the *primary caching resources*. But, eviction in a scalable environment could create distributed data fragments such as in Redis ("Redis," n.d.) which is not fragment aware. Previous work in Cloud data eviction (Banditwattanawong and Uthayopas, 2013) does not provide compelling evidence on solutions to questions such as – (a) how fragmented cached data (sparse cache) are reallocated to maintain affinity and (b) how to selectively migrate cached data of in a victim cache.

TABLE VI
COMPARISON OF CLOUD (SCALABLE) CACHES VERSUS LIMITED CAPACITY CACHES FOR ADAPTIVE CONTEXT CACHING

|  | Limited Size Cache | Scalable Cloud Cache |
|---|---|---|
| **Pros** | <ul><li>Simple to manage.</li><li>Lower seek time especially when the cache space is not virtualized across distributed nodes (benefit of locality of cache memory).</li></ul> | <ul><li>Elastic to demand (Faticanti et al., 2021). Therefore, suitable for developing query load aware adaptive cache management.</li><li>Capacity is theoretically unlimited for scaling unless constrained by monetary budget. So, time-aware cache policies can be implemented to maximize HR of context accessed over a long time.</li><li>Offers high-availability, failover-recovery, and replication due to distributed cache nodes.</li><li>QoS offering is high (low variance in cache performance).</li></ul> |
| **Cons** | <ul><li>Oblivious to size of the query load.</li><li>Competition for cache space. Requires cache replacement which can be non-trivial of a question to optimize for different types of query loads.</li><li>Consistency in QoS offering is low. (Variance in quality metrics are high)</li></ul> | <ul><li>Replication based distributed caching can suffer in performance for each instance the system scales. Similar case is *performance cliffs* (Cidon et al. 2016). This can be problematic in achieving the *graceful* requirement of ACC in Section XIII-B.</li><li>Unconstrainted scaling is expensive.</li><li>Cache manager should be utility aware to relinquish any caches which are no longer required or underutilized to fully-realize load awareness of highly variable context query loads. Some cloud caches can be difficult to scale down.</li><li>Continuous monitoring is required to manage the cache pool under cost constraints and sometimes fragmentation, e.g., using Redis. This is problematic in achieving *affinity* for caching context information which we will discuss in Section XIII-B.</li><li>Data fragmentation across distributed caches can minimize cache utility, and unnecessary scaling.</li><li>Need to manage complex virtualization tasks to provide a single interface for caching, and abstraction of internal complexities e.g., federating multiple physical caches as a single virtual cache unit.</li></ul> |

*C. Adaptive context caching in limited-sized versus scalable cache memory*

Based on our discussions in the previous sub-sections, we summarize the pros and cons of using scalable and limited-sized cache memories for ACC in TABLE VI.

In this section, we discussed the usability of Cloud caching techniques for ACC. In the next section, we briefly summarize all the parameters used in our reviewed literature.

X. PARAMETERS USED IN ADAPTING CACHING DECISIONS

Overall, we identify that there are two classes of parameters that need to be defined in adaptive caching – (a) metrics to make the decision and (b) metrics to evaluate the decision. In this paper, we focus more on the former because the latter is specific to individual domains. Cost, and performance metrics, e.g., response latency, are of primary interest to us in this work. The metrics to make the ACC decision are also interesting to discuss because we observed a plethora of different parameters in the literature. Previous work uses only a single or a small sub-set among these parameters which is a significant drawback.

In the surveys on performance-efficient web data prefetching (Ali et al., 2011; Domènech et al., 2007, 2006), authors list six metrics to evaluate the prefetching decisions: (a) precision, (b) byte precision, (c) recall, (d) byte recall, (e) traffic increase, and (f) latency ratio.

We identify five primary parameters as inputs to the adaptive caching decision: (a) popularity (i.e., frequency of access), (b) size of data, (c) lifetime, (d) locality information (e.g., the origin of the context query), and (e) costs, e.g., network, and cache costs. The three main constraints are (a) cache capacity, (b) monetary budget, and (c) QoC, QoS, or QoE requirements (i.e., based on SLAs).

In Figure 17, we summarize all the relevant parameters for making ACC decisions based on our survey. The key parameters are (a) freshness, (b) latency, (c) popularity, (d) reliability, (e) mismatch cost, (f) retrieval cost, (g) publisher load, (h) processing cost, (i) publishing cost, (j) cloud resource cost, (k) criticality of the context query, e.g., compare $Q_1$ against $Q_7$, and (l) push/pull nature of the query. The arrowed branches point to parameters that the parent depends on. For instance, latency depends on network transmission delays, and processing delays. Accordingly, ACC is a complex decision involving a multitude of parameters that is non-trivial to solve.

In the next section, we discuss several existing CMPs and their cache awareness or cache readiness.

XI. STATE OF THE ART OF CONTEXT CACHING

Research and development in CMPs are in their early stages, with investigations being only a few years old. FIWARE Orion context broker ("FIWARE-Orion," n.d.) and ETSI's Context Information Management (CIM) platform (*Context Information Management (CIM): Application Programming Interface (API)*, 2020) are critical research work apart from the CoaaS platform (Hassani et al., 2018). For instance, the current standardization efforts in CMPs and context query languages are led by the ETSI working group, whereas FIWARE is a commercially available product ("FIWARE-Orion," n.d.). COSMOS (Chabridon et al., 2013) and CA4IoT (Perera et al., 2012) are earlier works with limited context management capabilities. Interestingly, however, none of these CMPs implement caching at the time of this paper. So, we investigate the reasons that enable these CMPs to operate without caching.

First, let us survey the sparingly implemented ACC algorithms in recent literature. There are several works in transient IoT data caching (Zhu et al., 2019), including theoretical frameworks such as (Schwefel et al., 2007). But ACC is uniquely different and poses several challenges not yet resolved under traditional context-aware caching (Weerasinghe et al., 2022c). Medvedev et al. (Medvedev, 2020) develops a cost-efficient proactive cached context refreshing strategy for CoaaS. Figure 18 illustrates the logical context cache structure proposed for CoaaS, but proactive context refreshing is limited to L1. The rest of the logical levels are updated reactively.

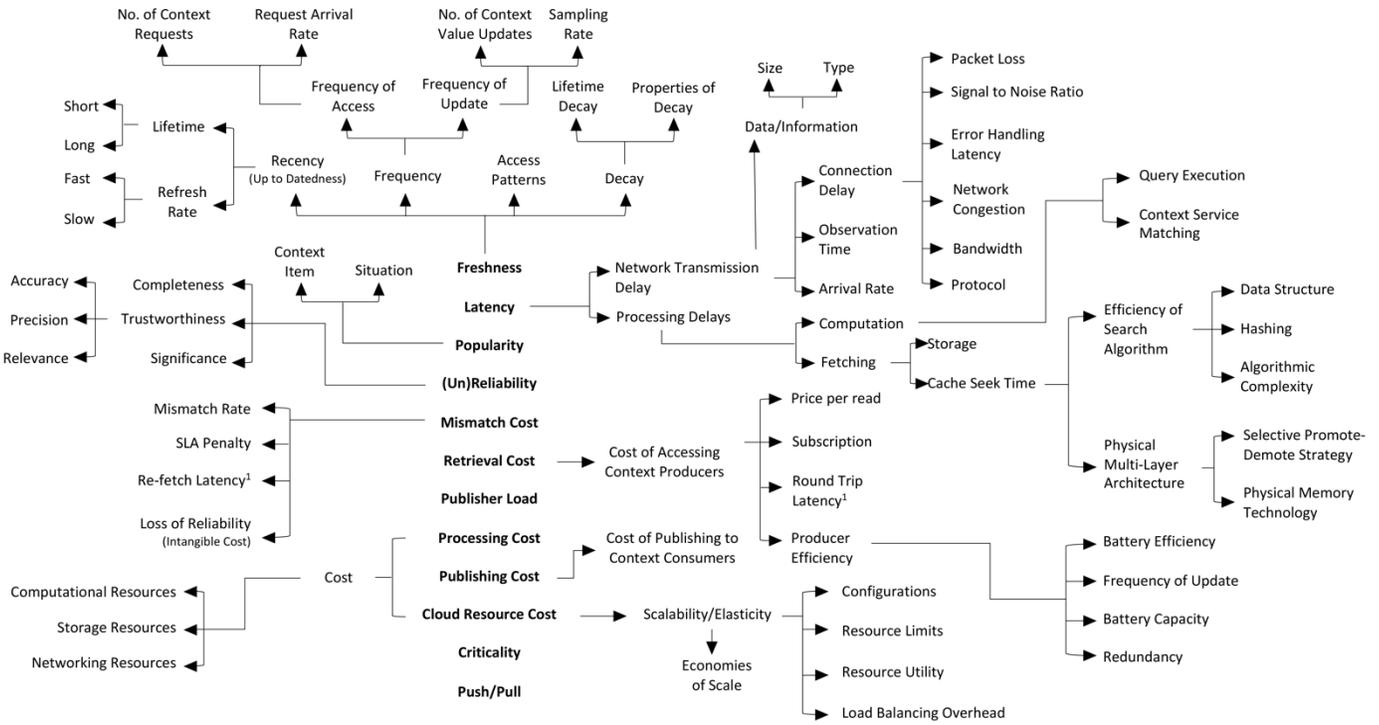

**Fig.17.** Parameters involved in Adaptive Context Caching decisions. [1]Branches out similar to "Latency"

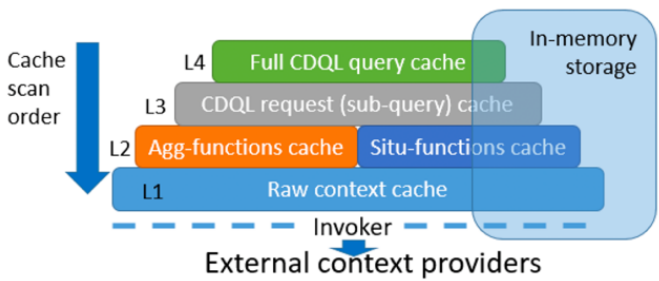

**Fig.18.** Logical context cache in CoaaS (Medvedev, 2020).

Khargahari et al. (Khargharia et al., 2022) investigated an adaptive context caching algorithm that proactively caches context entities based on the estimated popularity of their type (i.e., cluster). Despite the authors indicating a decrease in response latency by 43.8% compared to no-caching (also referred to as the redirector mode), the approach is oblivious to QoC and cost-efficiency. Several other problems need to be resolved when clustering and generalizing measured parameters over all the context information in a cluster. We discussed them in Section VIII-C. Hence, in (Weerasinghe et al., 2023) it is proved that popularity-based policies (using LVF replacement as a benchmark) are not always cost and/or performance efficient for CMPs to manage the context in cache memories. Our algorithm was up to 49.5% cost-efficient while reducing response latency by 54.6%. We also developed a technique to estimate the lifetime of context information

(Weerasinghe et al., 2022b) to alleviate the drawbacks of using a predefined static lifetime in all of the above work. Yet, these recent works are still in their early stages of research and this paper intends to provide further direction in Section XIII.

After careful investigation, we identify five potential reasons for not implementing context caching in the existing state-of-the-art in CMPs. Our focus here is on the Orion ("FIWARE-Orion," n.d.), and CIM platform (*Context Information Management (CIM): Application Programming Interface (API)*, 2020). They are as follows:

- Context queries are executed as a subscribed service. In CoaaS, we refer to them as push-based queries. IoT data are ingested in streams into local storages prior to being processed, i.e., database mode. Then, transiency is no longer an issue as long as the providers reliably push data either, (a) at the rate of sampling or (b) when a different value to the previous measurement is sensed. But we indicated the problems posed by network delays e.g., the asynchronistic of network queues, which still affects transiency. This argument is backed by the notion of tolerance to delays which we tested in (Weerasinghe et al., 2022b).
- Context are managed by the separation of concerns. For instance, Orion ("FIWARE-Orion," n.d.) use JSON for context dissemination, and each *generic enabler* – APIs to access retrieved context, use a different type of context storage, e.g., HDFS, MySQL, MongoDB, etc. Context is accessed via dedicated components optimized for a certain feature, e.g., data type, that is faster than retrieving from the provider.
- In both Orion ("FIWARE-Orion," n.d.), and CIM (*Context Information Management (CIM):*

*Application Programming Interface (API)*, 2020), context queries are developed using NGSI-LD, which can only specify a single entity per query. Complex queries such as $Q_1$-$Q_8$ in our motivating scenario are expressed using multiple context queries each. Several challenges of ACC, e.g., intra-, and inter-context query dependencies are not considered in these CMPs. As a result, there exists no logical structure (such as that depicted in Figure 3 and 18) to context information that can be assumed. So, learning patterns for reuse and/or repurposing is impossible.

- Caching approaches can be classified into three: (a) client caching, (b) proxy caching, and (c) server caching. A CMP operating as a middleware acts both as a proxy and as a server at the same time, e.g., as in CoaaS (Hassani et al., 2018), and Orion ("FIWARE-Orion," n.d.), since it redirects the contextual data from the provider to the consumer, as well as handles the execution of a query. This process is complex and computationally intensive (Medvedev et al., 2019). So, adaptive caching is an additional technical overhead in relation to features like response latency, which is often viewed as a non-functional requirement.
- The massive dimensionality problem involving ACC decisions is clear using Figure 17. According to (Ali et al., 2011; Domènech et al., 2007, 2006), there are at least six different input parameters involved in adaptive data caching – which is a less complex domain compared to ACC. It is a complicated task to design and develop a near real-time algorithm for ACC that incorporates all the dimensions while monitoring each of them in an operational sense.

In the counter-argument, we experimentally found that database mode, e.g., full-coverage strategy in (Weerasinghe et al., 2022b), is not always cost-efficient.

There are also three main reasons to opt for caching in comparison to the database mode. Firstly, the CMP requires considerable processing capacity to derive high-level context for each query, from the low-level context that has been ingested into the persistent storages. Secondly, we find that developing an extensible context caching and storage system is yet to be investigated. Consider generic enablers in Orion ("FIWARE-Orion," n.d.). The objective of the authors is to develop a dedicated interface, optimal for each identified scenario e.g., Orion is focused on five areas: (a) smart agriculture, (b) smart cities, (c) smart energy, (d) smart industry, and (e) smart water. We already find at least nine database technologies in Orion, and CIM uses graph-based technologies (Li et al., 2018) as well. Thirdly, consider $Q_7$ which is only intermittently executed by the CMP for the Context Consumer when the situation is likely to occur (e.g., using the situation monitoring engine in CoaaS (Hassani et al., 2018)). It is then unnecessary to continuously store context persistently compared to proactively retrieving the context information when it could reliably predict the situation (i.e., the context information) would occur.

Context caching is an area that is still in its infancy. There is minimal work to review. We discussed in this section the potential reasons why caching context is not integrated into our benchmarks - FIRWARE Orion ("FIWARE-Orion," n.d.), and ETSI CIM (*Context Information Management (CIM): Application Programming Interface (API)*, 2020). So, in the next section, we elaborate on the concept of adaptive context caching as the basis for identifying features of ACC, and directions for further work in the area.

## XII. ADAPTIVE CONTEXT CACHING

In this section, we compare the concept of adaptive context caching against adaptive data caching which we critically analysed by methods, techniques, policies, and strategies. The main objectives are to, (a) identify a unique area of research in ACC that is different from adaptive data caching and (b) formally define ACC for the benefit of future direction in the research area.

TABLE VII
DIFFERENCE BETWEEN CONTEXT CACHING AND ADAPTIVE DATA CACHING (WEERASINGHE ET AL., 2022)

| Feature | Traditional Data Caching | Context Caching |
|---|---|---|
| Similarity of cached data to original | Same (Copies). | Different considering either or all of structure, form, format, and type, suitable to represent context. |
| Impact of network features on the cache resident time | Only for transient data e.g., IoT data. | Highly vulnerable, e.g., the situation of a location. |
| Need for refreshing | Needed only for transient data. | Necessary. |
| Fault Tolerance (Recoverability) | Recoverable. | Not recoverable (need to re-derive). |
| Quality concerns of cached data | Limited e.g., latency, hit rate. | Multi-variate and complicated. |
| Partially caching | Applicable. | Not applicable. |
| Feasible techniques to implement | Various in literature. | Limited. Not all techniques investigated in traditional data caching are applicable. |
| Priors to make caching decisions | Available. | Unavailable, or limited. |
| Size of cache action space | Pre-definable and limited. | Extensible, and hard to pre-define. |
| Optimizable for transactions | Feasible. | Limited feasibility. |
| Write back to cache (Dirty caches) | Applications write updates back to the cache. | Not performed. Consumers do not write back to the cache. |
| Write heaviness of caching operations | Typically, lighter than context caching. | Heavy due to refreshing. |
| Cost of data management as a function of number of cached items | Theoretically inexpensive than adaptive context caching. E.g., flat files do not incur refreshing costs. | Increases exponentially where the magnitude is subject to complexity of context queries. Therefore, cost-efficiency aware selection of context information to cache is significantly important. |

*A. Adaptive data caching versus ACC*

Differences between traditional data caching and context caching are extensively discussed in our previous work (Weerasinghe et al., 2022c). However, we provide in TABLE VII the summary of those differences for the clarity of the readers.

In the next section, we use this difference to formally define ACC based on the differences and the discussions in this paper.

*B. Definition of Adaptive Context Caching*

In this sub-section, data, and context caching are compared based on several features found in literature in order to establish the uniqueness of ACC.

In this section, we attempt to establish a working definition for the concept of adaptation in context caching (ACC) in a large-scale CMP. The following are three definitions for adaptation, adaptive caching, and ACC in literature:

- definition by Chatterjee et al. (Chatterjee and Misra, 2014) - "Preserving the accuracy of information and conserving the network resources simultaneously against changes in the physical environment",
- definition in (Sadeghi et al., 2019b, 2019a) - Adjusting the (caching) policies upon which the actions are taken through interactive learning in a dynamically evolving environment on which the system operates to achieve long-term cumulative cost minimization,
- definition by Medvedev et al. (Medvedev, 2020) – "Decision about *what* context data to cache and *what* to retrieve from external Context Providers in an ad-hoc manner"

We cannot fully accept any of the definitions above since each encapsulates an application domain, or a predefined structure of deployment e.g., external and internal cache structure in (Chatterjee and Misra, 2014). Chatterjee et al.'s (Chatterjee and Misra, 2014) focus on maintaining the quality of cached data, compromising performance and cost. The definition in (Sadeghi et al., 2019b, 2019a) is relatively adequate for ACC but we discussed several scenarios in which the mechanism needs to proactively and reactively adjust to short-term variations. The definition in (Medvedev, 2020) does not explicitly state the dependence of caching on related components in the CMP when the objective is cost-effectiveness. For instance, what context is cached affects what SLAs apply during refreshing, and the rate of refreshing (e.g., depending on sampling rate) which is investigated in the same work. These parameters are especially tied to how much network is utilized – components external to the cache memory. Further, in the definition of adaptiveness given by Gramacy et al. (Gramacy et al., 2002), "measuring against a comparator" is a key-phrase. In an environment that is dynamic, volatile, and noisy, besides the lack of prior knowledge, it is non-trivial to effectively evaluate any actions taken in near real-time. None of the definitions above underpin this notion.

Accordingly, we define Adaptive Context Caching (ACC) as:

*The decision about what context to cache in near real-time to achieve cost efficiency goal(s) measured by Quality of Service (QoS) and Quality of Context (QoC) parameters. These decisions should be comparable with a benchmark trained with absolute prior knowledge while using the minimal necessary resources and ensuring minimum loss to quality of cached context (QoCaC).*

The minimum necessary resources underpin all computing, caching, storage, and networking resources needed to execute ACC that fully meets the optimization goal(s). The minimum loss to QoCaC refers to the maximum tolerance on QoC and QoS formally agreed upon in SLAs, e.g., freshness. So, ACC is an umbrella term that we use to encompass:

- adaptive selective context caching: the decision of selectively storing context in the cache memory,
- adaptive context refreshing: the process of maintaining adequate QoCaC in cache memory,
- adaptive cache resource scaling: the process of facilitating size, and QoS necessities by scaling, and
- adaptive context eviction: the process of maximizing effective cache memory utility. Effective cache memory is the minimal cache memory required to achieve the optimization goal(s) (we refer to as the *primary cache resources* in Section XIII).

Figure 19 illustrates this definition as a cycle to indicate the cached context lifecycle in ACC.

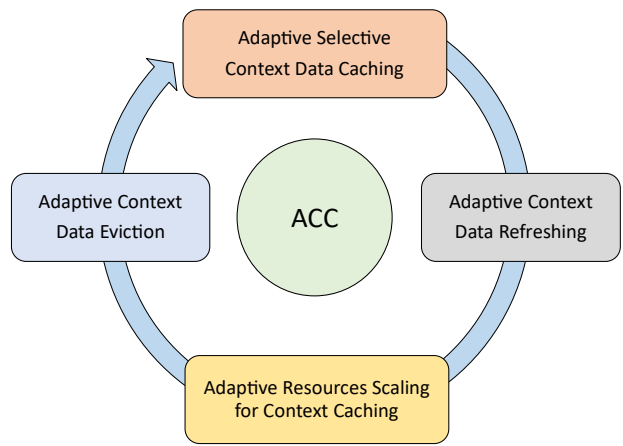

**Fig.19.** Cached Context Lifecycle in ACC.

In this section, we culminated our critical analysis of related literature by first comparing adaptive data caching against adaptive context caching. Then, we used that and the relevant definitions to provide a comprehensive conceptualization and definition for ACC, also indicating significant gaps in the area requiring further research efforts.

XIII. REQUIREMENTS OF ADAPTIVE CONTEXT CACHING

In the previous sections, we laid the groundwork to identify the nature of a well-designed ACC that achieves cost and performance efficiencies. We identified several features and areas that ACC should address in the definition. In this section, we discuss the functional requirements of ACC. We first identify the five problems that ACC should solve to derive an *adaptive context caching plan*. Then, they are translated into technical requirements. Next, the areas of adaptation in ACC are put into the process.

*A. Design Considerations of ACC – The 5Ws*

Adaptation in context caching can be broken down into solving five questions which we call the 5Ws. We summarize them in Figure 20. In a simple sense, system-wide optimization is a matter of adapting to five core-areas in a distributed CMP. We refer to the collective actions in each of these areas as the *Adaptive Context Caching Plan*. We describe it further below.

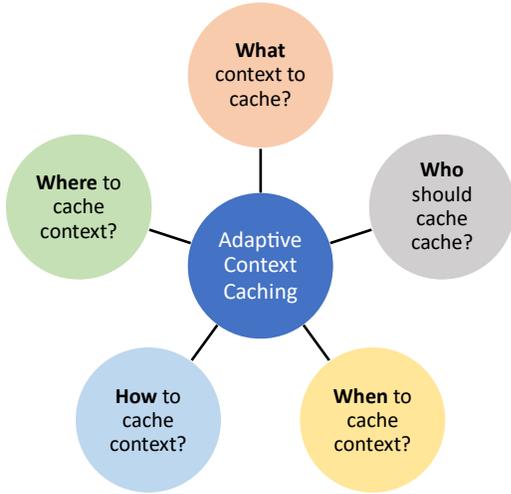

**Fig.20.** Design considerations of ACC – the 5Ws.

1) **What context information to cache?**

This question is related to the temporal decisions of *what* context information to selectively cache (Guo et al., 2013), refresh, and evict. Consider scalable Cloud-based cache memory. Then, the temporal decision on *what* resources to scale e.g., whether to scale the in-memory, or the persistent cache instances, either reactively or proactively with respect to the variations in the context query load and context access patterns is also encompassed in this question.

A significant problem when resolving *what* to cache is the existence of dependencies among context information. To the best of the authors' knowledge, previous literature has not considered this aspect. We describe several of them below in relation to the logical cache structure illustrated previously in Figure 18. Consider the JSON response from a car park for $Q_1$-$Q_4$ in Figure 21.

- The CMP requires retrieving the entire response rather than a partial, incurring the full cost of retrieval (i.e., the price paid to the Context Provider and total memory utilized by the response payload, for each refreshing operation) if at least one of the context attributes of this car park entity is cached, e.g., availability in the first level.
- Consider that the availability in the second level of the car park is also cached. There could exist a discrepancy between refreshing the two attributes because of, (a) different quality requirements, e.g., freshness, for the attributes since they are accessed by different Context Consumers (e.g., different applicable SLAs), (b) cached at different time instances (assuming refreshed independently), or (c) sensed using different types of sensors, e.g., having different sampling rates. It is required to perform two independent retrievals to refresh each item which is cost inefficient. Figure 22 illustrates this case and how it can lead to cost inefficiencies unless these dependencies are considered. Blue and orange arrows indicate accesses to attributes available slots in the first ($a_1$) and second ($a_2$) levels respectively. The green diagonal line indicates the loss of freshness, and the red dotted line indicates the minimum freshness of context ($f_{thr}$) accepted by the consumer.

```
{
    "sucess": true,
    "timestamp": "Thu, 19\/08\/2021 – 12:36",
    "id":"543cbdf4-9b52-4013-8ddb-8721a33bda97",
    "name":"car park 6",
    "location": {
        "street": "HG, 221, Burwood Hwy",
        "suburb": "Burwood",
        "state": "VIC",
        "postal_code": 3125
    },
    "availability": [
        {
            "level": 1,
            "slots": 85,
            "max_slots": 150
        },
        {
            "level": 2,
            "slots": 123,
            "max_slots": 150
        }
    ],
    "meta": {
        "max_height": {
            "value": 8,
            "unit": "meters"
        },
        "capacity": 300,
        "charge": {
            "value": 1.6,
            "unit": "per hour"
        }
    }
}
```

**Fig.21.** JSON response from a car park (a Context Provider).

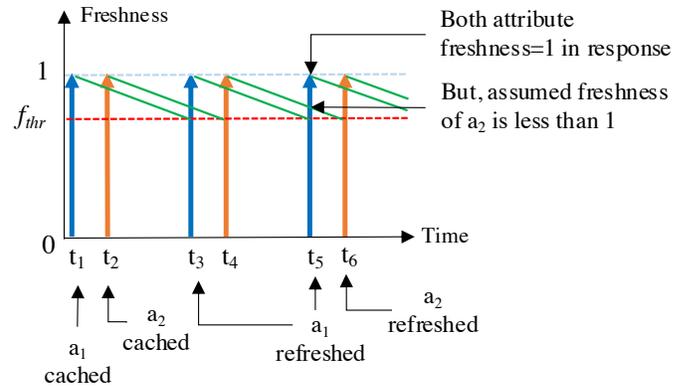

**Fig.22.** Cost inefficiencies arising from discrepancies in caching attributes originating from the same context data response.

- Note the indirect relation between $Q_4$ and $Q_5$ based on the shared context in Figure 4. The caching decision should acknowledge the complex coherencies among context information to improve efficiency by reuse.
- We can consider caching only the context attributes related to *goodForWalking*. Here, we can identify two strategies to re-compute the function, (a) reactively – computing for each context query requesting the item, and (b) proactively – computing each time at least when one of the underlying attributes are refreshed.
- In Section II, we discussed the similarity between *goodForWalking* and *goodForJogging*. Based on our arguments on repurposing, caching one of the two could be a viable cache space-efficient strategy, e.g., *goodForJogging* cached using a pointer to the cache key of *goodForWalking* assuming the context is the same and the function occupies more cache space compared to the pointer.

2) **Where to cache context?**

We refer to deciding the (a) logical, and (b) physical placement of context information in cache memory. It can be explained further as follows:
- Physical assignment of an item
  - **Address space allocation** refers to assigning a piece of context information to the physical level of cache under a multi-level optimization scheme that uses multiple memory and storage classes (Weerasinghe et al., 2022c). For instance, context information used in queries generated by CCs that have a higher tolerance to RT (e.g., when seek time in a slower cache is smaller than the maximum tolerated latency) can be cached in a slower cache, i.e., persistent cache memory.
  - **Byte assignment** - Partially caching a piece of context information is invalid in CMPs because a part of the context can be irrational, e.g., caching only a part of a situation description, unlike in (Faticanti et al., 2021; Hou et al., 2019; Zhang et al., 2020). We, therefore, identified that context caching demand for large coherent physical space assignment, i.e., as contiguous blocks (also referred to as physical affinity) for related context information such as of a context query, e.g., caching logically related context in affinity to minimize cache access latency. It is a research gap, especially when implementing CMPs as a distributed system because of (a) cluttering e.g., caching the *available slots* attribute in a different cache memory instance to where *location* attribute may be cached of the same car park entity and (b) fragmentation across physically separated context cache memory instances after several context eviction cycles. For example, assume the utility of cache memory instances A, B, and C of the same size are 30%, 20%, and 10% respectively (we refer to them being sparsely utilized). They can cause (a) low space utility at a higher resource cost, e.g., consider paying $6 for the above utility versus using a single instance at $2 at 60% utility when the cost of an instance is $2, and (b) higher network cost, e.g., cache seek time across distributed caches. Apart from developing strategies to manage over-scaling due to these issues, maximizing space utility of the *primary cache resources* (the minimal cache requirement at a given time, e.g., a single instance versus three in the example) with the objectives of (a) minimizing the cost of storing one byte of context information in cache and (b) accommodating for scaling down of unnecessary resources, could be identified as requirements.
  - **Assignment to a local caching node** refers to the spatial decision of selecting a caching node to store the context information. For instance, context response to $Q_7$ can be useful to cache for a metropolitan area (as in the motivating scenario), compared to when riding on a suburban trail. We referred to this as Locality Optimization in Section VI.
- **Logical assignment of cache** refers to caching context information with logical affinity. For instance, we illustrated that context information in a response to a context query is logically structured. Then, the logical assignment should ensure the related context of different logical levels are placed closely so that, each logical level data can be accessed quickly. Similarly, when tightly dependent context exists, e.g., car parks and data to derive the status of an access road, it is more logical to place them closer so that cache seeks time in processing a context query is unified.

3) **When to cache a piece of context information?**

The temporal decisions of (a) *when* to cache, refresh, and evict a piece of context information, in addition to (b) *when* to scale, are considered under this question. Our previous work (Weerasinghe et al., 2022b) and Medvedev et al. (Medvedev, 2020), specifically investigated this problem for cached context refreshing. However, caching and evicting at the "correct" point of time has a profound impact on achieving the optimization goals such as cost-efficiency. One of the aspects, in this case, is proactive caching (also referred to by *prefetching*). We have discussed it in Section IV. The other aspect is *when* to cache reactively. Consider an increase in context query load for $Q_1$-$Q_4$ during the high traffic hours during Monday-Friday indicated in Figure 23 (Auditor General of Victoria, 2013) using green in Melbourne, assuming the size of the query load is proportional to the traffic volume. Firstly, ACC should decide, *when* to start caching for $Q_1$-$Q_4$, e.g., when 30% of the car park seeking traffic is observed at 6:00 am. Secondly, the increase in traffic plateaus, e.g., 10:00 am-2:00 pm. Therefore, cached items of $Q_1$-$Q_4$ can be incurring *hold up costs* – accumulated Cloud cache cost proportional to the occupied size and refreshing cost if they are not frequently accessed. Similarly, the traffic drops after 5:30 pm, which is indicative of a traffic outflow and a low requirement for parking spaces. We can assume the popularity of the cached context would then be low and hence, we could selectively evict and scale down the cache memory required for responding to these context queries. These decisions are empirical.

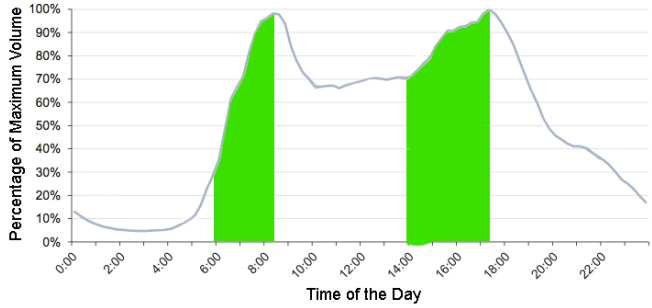

**Fig.23.** Traffic variation in Melbourne from Monday-Friday (Auditor General of Victoria, 2013).

Let us now consider caching for $Q_1$-$Q_4$ close to the peak volume to be observed, say from 8:00 am. Considering the change in traffic volume against time. This decision is late because it is intuitive to commence caching for the queries when the volume of queries of this kind starts to *burst*. Although the definition of burstiness, and whether responding to bursts cost-efficiently is influenced by how the optimization problem is defined, adapting in near real-time, e.g., as quickly *when* a variation in the observed parameter, such as popularity, can maximize the number of context queries responses that utilize the cache.

Caching or not caching a certain piece of context information "now" has several implications on the efficiency of a CMP (Weerasinghe et al., 2023). Context not cached would incur repeated network costs for each subsequent query, whilst an item cached with no subsequent accesses would incur significant holdup costs.

Consider the work of Wu et al. (Wu et al., 2019) in managing short and long-term optimization goals using LSTM and our example questioning the applicability of the optimal caching policy during 8:00-9:00 am to 1:00-4:00 am for $Q_1$-$Q_4$ in Melbourne (Auditor General of Victoria, 2013). Then, the ACC decision has two parts:

- **Short-term adaptation decision** – e.g., applying the most cost-effective caching policy, (a) during 8:00-9:00 am and 1:00-4:00 am, and (b) handling unforeseen variations such as bursts of queries.
- **Long-term adaptation decision** – e.g., cost-effectively for all context queries in a day such as switching between the best-caching policies.

Using this as the basis, we can derive a principal feature related to the *when* question of an efficient adaptive context caching mechanism as – "long-term optimal, but short-term aware".

4) **How to a cache a piece of context?**

This question relates to the structure of the context information for caching and algorithms to access it in the cache. We previously discussed the differences between context and traditional data. Based on that, we can identify three primary requirements for a data structure to cache context information: (a) the ability to represent intra- (e.g., sub-query and entity) and inter- (e.g., co-occurring context) query logical relationships, (b) ability to manage different data formats e.g., the value of a temperature attribute (low-level context) is a real number versus, situational description of a room (high-level context) is JSON, and (c) minimal search complexity (or in other words, time-efficient cache seeking).

It is noteworthy the significance of an efficient data structure and algorithms to perform efficient storage, update, and retrieval of cached context information. Unlike a local limited-sized cache memory, distributed caches pose the unique problem of scanning, and filtering across multiple, often geographically sparse caching nodes that numerically be infinite. The *how* problem is therefore a primary design problem for ACC.

5) **Who should cache context?**

We refer to the question of *who* (e.g., which agents running in the distributed nodes) should: (a) process the adaptive context caching decisions, e.g., making the selective caching decisions of the context information processed for the context query *received* at Node-A (Guo et al., 2013), and (b) executes the decisions e.g., cache placement operation by Node-A, in a distributed system. Consider selective caching as an example and we can identify three different scenarios as follows. We refer to "adequate processing power" to describe nodes in a distributed system that either has dedicated, or remaining resources (e.g., CPU cycles and memory) to run the ACC agents, complying with QoS standards underpinned by an SLA (e.g., remaining size of memory at the caching node should be at least 120% the size of the context information to be cached to accommodate caching space and processing for it).

- Both selective caching decisions are processed and executed centrally, e.g., in the Cloud.
- The selective caching decision is processed centrally and executed at the node, e.g., caching in Node-A.
- The selective caching decision is performed at the nearest node having adequate processing power and executed at the node.
- Both selective caching decision is processed and executed at the nearest node that is cache enabled and contains adequate processing power.
- Both selective caching decision is processed and executed at the receiving node given it is cache-enabled and adequate processing power is available.

A complicated scenario motivated by literature is federation, e.g., different agents executing a portion of the ACC processes in different nodes and federating to produce a single context query result. But we identify it as further development of a distributed CMP, which therefore is out of the scope of this paper. In Figure 24, we illustrate how the 5Ws are related to developing a cohesive adaptation strategy defined as the *adaptive context caching plan.*

In this sub-section, we identified the five questions that an adaptive context caching mechanism should address. In the next sub-section, we translate these into technical requirements that are necessary to efficiently solve the 5Ws.

*B. Nine-fold functional requirements of ACC*

The 5Ws identify all the considerations that the Adaptive Context Caching mechanism needs to address for cost and performance efficiency purposes. In this sub-section, we translate them into nine implementable requirements: (a) fast processing, (b) logical coherency, (c) near real-time-ness, (d) cost-awareness, (e) elasticity, (f) resilience, (g) gracefulness, (h) reliability, and (i) non-intrusiveness. They are discussed below.

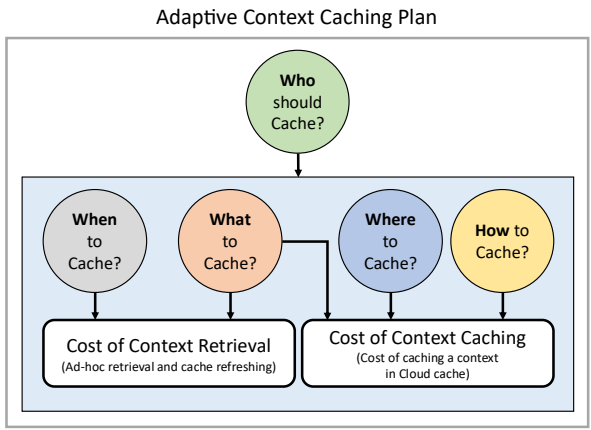

**Fig.24.** Relationships among the 5Ws in a distributed system.

1) **Enable fast query processing**

There are two primary objectives of fast adaptation in ACC: (a) make adaptive caching decisions with minimal latency and (b) converge to an optimal adaptation policy in minimum time with respect to real-time metrics.

The SLAs in (Bibal Benifa and Dejey, 2019; Medvedev, 2020; Weerasinghe et al., 2022b) monetize the time criticality as a penalty for exceeding an acceptable response time. Therefore, apart from the contractual commitment to deliver context fast (which is qualitative), it is essential that the accumulated penalty is minimized for cost-efficiency (a quantitative metric).

Time to converge is a matter of minimizing the time till ACC produces adaptive context caching plans that are comparable with models trained with complete prior knowledge. We emphasized the lack of prior knowledge in making near real-time ACC decisions in Section VIII and, hence fast convergence is a significant requirement to minimize cost-inefficiencies.

2) **Logical coherence in adaptive actions**

We point out that ACC should be a process that attempts to align and orchestrate the optimal performance of each resource (e.g., caches memories, components, and modules) in order to execute efficiently. We propose aligning the objectives of each component for this purpose. For instance, a selective caching process optimized for long-term cost benefit, e.g., caching context information that is accessed over a long period of time, can be competitive with selective eviction optimized to retain context information that has higher recency in access, e.g., using Least Recently Used (LRU) policy. Similarly, Medvedev et al. (Medvedev, 2020) state that the performance of a real IoT system's storage component (or even cache memory in our case) depends not only on the underlying technology but also on the ingestion pipeline, messaging queue, and other components of the platform which take part in data processing or storage. Hence, the components of ACC must operate coherently.

3) **Adapts in near real-time**

The context load is time-variant. We extensively discussed it in relation to the *when* question. It is hence necessary that the adaptations occur in near real time with respect to the observed or expected variations in context and context query loads so that the CMP could leverage to maintain or improve the cost-efficiency, QoC, and QoS.

4) **Cost-awareness in Adaptation**

Commercially viable CMPs need to minimize costs for improved profits. Theoretically, the requirements of fast and near real-time can be best achieved by unconstrained scaling. For instance, context queries could be processed the fastest when the necessary pieces of context information are available in the in-memory caches whilst unlimited computation power is available for processing. However, this approach is expensive. According to (Kabir and Chiu, 2012), it is reasonable in some scenarios to incur a performance penalty to maintain costs at an economically viable level. So, we have indicated throughout our discussions that ACC is at least a tri-objective optimization problem one of which is cost-efficiency. A similar objective can be found in (Kabir and Chiu, 2012).

Accordingly, this requirement constrains the scalability of the resources (e.g., computing, storage, cache) used by the CMP. It, however, clearly aligns with our concerns discussed in solving the *where* question.

5) **Elastic to Context Query Variations**

As we identified above, (a) the high volume and velocity of context queries (Perera et al., 2014) pose the challenge of load balancing, and horizontal scaling, while (b) the variety could impact fast query processing unless the system manages to scale vertically. Therefore, in order to achieve massive parallelism in context query processing (in achieving the *adapting* near *real-time* requirement), it is essential that a CMP is capable of dynamically scaling in relation to demand parameters, e.g., request rate (Choi et al., 2020; Guo et al., 2013; Kabir and Chiu, 2012). We discussed adaptive caching in the Cloud for this matter and established the advantages. However, the Cloud resource consumption must match or approximates the primary cache resources (refer to Section IX) to achieve cost-efficiency. Cache capacity may be relinquished when they are no longer required, e.g., 2:00-4:00 am according to Figure 23.

6) **Resilient to data issues**

One of the key features comparing data caching versus context caching is the lack of prior knowledge for making adaptive caching decisions. Making ACC decisions are trivial under several circumstances, e.g., novel context information. Further, machine learning models such as RL suffer from the *cold start problem*. Minimizing convergence time (which is a feature of *fast processing*) is an existing research gap identified in (Chugh and Hybinette, 2004; Nasehzadeh and Wang, 2020; Zhu et al., 2019). Initial degradation of performance in dynamic models is inevitable as we observed in the literature. But it is important that ACC is resilient and that it can manage the stated circumstances *gracefully*, such that the period of sub-optimal performance is minimized. For instance, we propose to use similarity to previously observed context information such as *goodForWalking* when attempting to make a selective caching decision for a new piece of context information such as *goodForJogging*,

### 7) Gracefulness in Adapting

Resilience underpins *graceful* management of cost and performance under a lack of data scenarios. But, also consider (a) post-scaling degradation (Hafeez et al., 2018), (b) failures in scaled nodes (Verma and Bala, 2021), and (c) post-eviction miss-rate (MR). Context information cached in Cloud cache instances selected for victimizing must be relocated to an available space in a preserving instance. In a similar sense, minimizing the overhead of cache refreshing, replacement (Choi et al., 2020), and eviction demand graceful cost and QoS management as well. For instance, we propose (a) multi-level cache organization, e.g., fast cache such as in-memory that first evicts an item to a slow cache such as a persistent cache before permanently evicting, and (b) a ghost state of context information prior to being evicted when the freshness demand can no longer be met, to minimize MR. We illustrate the latter in the state transition diagram in Figure 25.

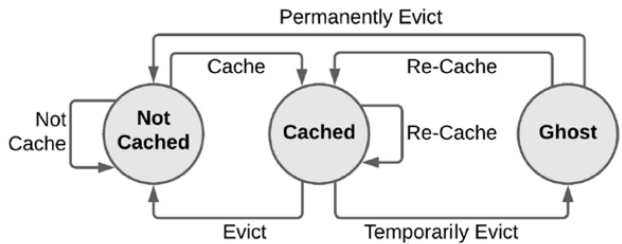

**Fig.25.** State transition diagram of a context.

### 8) Ensure context reliability

QoC is determined by several factors (see Section IV-A). We discussed managing the precision and accuracy of context for $Q_7$ extending our motivating scenario in Figure 9. The completeness factor can be identified using $Q_1$-$Q_4$ where the Context Consumer must receive details about all available parking slots that meet all the contextual criteria. In (Weerasinghe et al., 2022b) we quantified the different impacts of context transiency on ACC, defining notions such as the invalid periods (InvPrd), and invalid rate (IR). It is essential that the adaptive context caching actions, (a) improve the reliability of cached context, e.g., retrieving from a sorted list of most reliable context services, (b) improve the reliability in adhering to the SLAs, and (c) ensure cached low-level contexts are as consistent with the measured values such that the derived context information is based on accurate data, e.g., caching at the highest precision level as demanded by multiple applicable SLAs. Therefore, it further elaborates ACC as a trade-off between the cached lifetime (∝HR) and the number of retrievals, where the former minimizes response time whilst the latter maximizes the reliability of context information (Weerasinghe et al., 2022b).

### 9) Non-intrusive (Automatic) Adaptation

The most significant requirement of ACC is non-intrusiveness. We define self-adaption as automatically adapting context caching based on observations (referred to as reactive context caching) or expectations (referred to as proactive context caching), e.g., bursts in query load, and statistically modelled expectations e.g., expected request rate for $Q_4$ during 8:00-8:30 am.

There are two types of users to a CMP – (a) the end user, e.g., the driver in $Q_1$-$Q_4$, and rider in $Q_6$-$Q_7$, who is a secondary user as they do not interact directly with the CMP, and then (b) the application developer, the primary user who composes context queries. But the developer is still not tightly coupled with the CMP, i.e., our perspective on optimization is system-wide, not specific tenants. Therefore, it is impossible to manually tune for efficiency, unlike in IoT silos, e.g., indexing data. Further, we cannot rely on manually tuning a CMP for a particular use case or a set of use cases. We referred to this feature as the "objectivity" of a CMP in Section IV-B. The variety of queries can be massive, and as these context queries can change over time, manual tuning can be tedious and highly repetitive. ACC should be cost and performance efficient for a wide range of use-cases. So, the NoD-NoR approach requires efficient self-adaptation mechanisms (Medvedev, 2020).

We propose the Monitor-Analyse-Plan-Execute and Knowledge (MAPE-K) model [149] to make near real-time adaptive context caching decisions. As illustrated in Figure 26, it is an iterative model that continuously, (a) monitors QoC, and QoS metrics of the CMP (and adaptive context caching actions), (b) calculate the Cost of Context Caching (and metric analysis, e.g., calculating the reward of an ACC action), (c) develops an adaptive context caching plan to adjust efficiency, and (d) executes ACC actions e.g., selective caching, selective eviction, and scaling cache memory. In this process, we propose a knowledge base developed which includes cache state-action-reward tuples (replay memory), logs, transition probability distributions, etc. It alleviates the lack of prior knowledge problem and increases the efficiency in making caching decisions and the quality of ACC decisions.

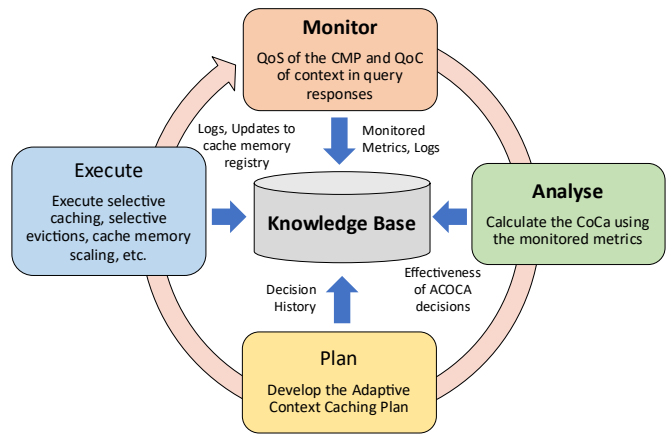

**Fig.26.** MAPE-K process in Adaptive Context Caching.

Figure 27 indicates the dependencies between the identified requirements of ACC. The directed arrows point to the underlying needs a requirement must satisfy in achieving it, e.g., elasticity should meet all (a) cost awareness, (b) gracefulness, and (c) fast response, at the same time. For example, post-scaling degradation (Hafeez et al., 2018) is a result of elasticity without "graceful in adapting".

"Lightweight" can be identified as an additional requirement in a distributed system with heterogeneous resources with limited capacity such as mobile phones. But being lightweight is optional as it depends on the deployed environment.

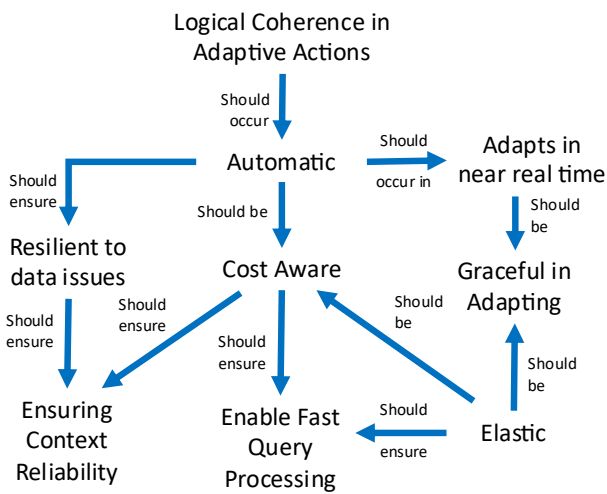

**Fig.27.** Functional requirements of an Adaptive Context Caching mechanism and their relationships.

Accordingly, TABLE VIII indicates the applicability of each requirement in solving the 5Ws. For instance, the intersection of *What* and *Elasticity* can be read as: "what resources to scale?". "X" is used to indicate objectives for ACC in the Cloud (i.e., centralized CMP which is scalable), and "O" to indicate objectives in a massively heterogeneous distributed system (i.e., when ACC decisions are made and executed across many different nodes with different cache, memory, and processing by size, technology, etc.).

TABLE VIII
DESIGN CONSIDERATIONS-FUNCTIONAL REQUIREMENTS OF ADAPTIVE CONTEXT CACHING

| Requirement \ Adaptive Problem | What | When | Where | How | Who |
|---|---|---|---|---|---|
| Enable fast query processing | X | X | X | X | O |
| Logical coherency | X | X | X | X | O |
| Adapts in near real-time | X | X | X | O | O |
| Cost-awareness | X | X | X | X | O |
| Elasticity | X | X | O | X | - |
| Resilience | X | O | O | O | O |
| Gracefulness | X | X | X | - | O |
| Context reliability | X | X | X | X | - |
| Non-intrusiveness (Automatic) | X | X | X | X | O |

In this section, we discussed the areas of adaptation in ACC using the 5Ws model and further elaborated on nine technical requirements based on our critical analysis. So, we defined the requirements of an efficient ACC mechanism.

## XIV. CONCLUSION AND FURTHER WORK

Context Management Platforms are more relevant than ever in making performance-driven smarter decisions, enabling various advanced Internet of Things applications, services, and systems. However, the time-sensitivity of such applications and the sheer demand for context poses a novel challenge in storing, managing, and disseminating context fast, concurrently, and reliably, at scale. This emphasizes a need to investigate and develop innovative solutions such as adaptive context caching. A large number of adaptive solutions exist for caching, however, in different related areas and not all applicable to Adaptive Context Caching (ACC). We critically surveyed and analysed a significant body of research and identified different techniques, methods, approaches, and policies. The discussions were made in relation to a real-world scenario with the goal of objectively evaluating the state-of-the-art against the requirements of adaptive context caching. Then, we also established the unique characteristics and challenges of ACC compared to traditional adaptive data caching. Further, the aspects and technical requirements of a well-designed, objective optimal ACC mechanism are proposed. We propose a comprehensive definition of Adaptive Context Caching as a possible reference framework for future research. Finally, the features of an efficient Adaptive Context Caching mechanism are identified and discussed in the form of design considerations – the 5Ws and the nine functional requirements. The authors are working towards developing an ACC mechanism according to the identified functional requirements and incorporating it into a distributed contextual intelligence system in further work.

ACKNOWLEDGMENTS

Support for this publication from the Australian Research Council (ARC) Discovery Project Grant DP200102299 is thankfully acknowledged.

REFERENCES

Abd-Elmagid, M.A., Pappas, N., Dhillon, H.S., 2019. On the Role of Age of Information in the Internet of Things. IEEE Commun. Mag. 57, 72–77. https://doi.org/10.1109/MCOM.001.1900041

Abowd, G.D., Dey, A.K., Brown, P.J., Davies, N., Smith, M., Steggles, P., 1999. Towards a Better Understanding of Context and Context-Awareness, in: Gellersen, H.-W. (Ed.), Handheld and Ubiquitous Computing, Lecture Notes in Computer Science. Springer Berlin Heidelberg, Berlin, Heidelberg, pp. 304–307. https://doi.org/10.1007/3-540-48157-5_29

Ale, L., Zhang, N., Wu, H., Chen, D., Han, T., 2019. Online Proactive Caching in Mobile Edge Computing Using Bidirectional Deep Recurrent Neural Network. IEEE Internet Things J. 6, 5520–5530. https://doi.org/10.1109/JIOT.2019.2903245

Ali, W., Shamsuddin, S.M., Ismail, A.S., 2011. A Survey of Web Caching and Prefetching A Survey of Web Caching and Prefetching. Int. J. Adv. Soft Comput. Its Appl. 3.

Al-Turjman, F.M., Al-Fagih, A.E., Hassanein, H.S., 2013. A value-based cache replacement approach for Information-Centric Networks, in: 38th Annual IEEE Conference on Local Computer Networks - Workshops. Presented at the 2013 IEEE 38th Conference on Local Computer Networks Workshops (LCN Workshops), IEEE, Sydney, Australia, pp. 874–881. https://doi.org/10.1109/LCNW.2013.6758526

Amazon ElastiCache - In-memory datastore and cache [WWW Document], n.d. URL https://aws.amazon.com/elasticache/

Anandarajah, M., Indulska, J., Robinson, R., 2006. Caching context information in pervasive systems, in: Proceedings of the 3rd International Middleware Doctoral Symposium on - MDS '06. Presented at the the 3rd international Middleware doctoral symposium, ACM Press, Melbourne, Australia, p. 1. https://doi.org/10.1145/1169100.1169101

Andrade, X., Cedeno, J., Boza, E., Aragon, H., Abad, C., Murillo, J., 2019. Optimizing Cloud Caches For Free: A Case for Autonomic Systems with a Serverless Computing Approach, in: 2019 IEEE 4th International Workshops on Foundations and Applications of Self* Systems (FAS*W). Presented at the 2019 IEEE 4th International Workshops on Foundations and Applications of Self* Systems (FAS*W), IEEE, Umea, Sweden, p. 6. https://doi.org/10.1109/FAS-W.2019.00044

Auditor General of Victoria (Ed.), 2013. Managing traffic congestion. Victorian Government Printer, Melbourne, Victoria.

Australia's IoT Opportunity: Driving Future Growth, 2018. . Aust. IoT Oppor. Driv. Future Growth 104.

Azure Cache for Redis | Microsoft Azure [WWW Document], n.d. URL https://azure.microsoft.com/en-us/services/cache/

Baccelli, E., Mehlis, C., Hahm, O., Schmidt, T.C., Wählisch, M., 2014. Information Centric Networking in the IoT: Experiments with NDN in the Wild. Proc. 1st Int. Conf. Inf.-Centric Netw. - INC 14 77–86. https://doi.org/10.1145/2660129.2660144

Baldauf, M., Dustdar, S., Rosenberg, F., 2007. A survey on context-aware systems. Int. J. Ad Hoc Ubiquitous Comput. 2, 263. https://doi.org/10.1504/IJAHUC.2007.014070

Banditwattanawong, T., Uthayopas, P., 2013. Improving cloud scalability, economy and responsiveness with client-side cloud cache, in: 2013 10th International Conference on Electrical Engineering/Electronics, Computer, Telecommunications and Information Technology. Presented at the 2013 10th International Conference on Electrical Engineering/Electronics, Computer, Telecommunications and Information Technology (ECTI-CON 2013), IEEE, Krabi, Thailand, pp. 1–6. https://doi.org/10.1109/ECTICon.2013.6559553

Bao, N., Chai, Y., Zhang, Y., Wang, C., Zhang, D., 2020. More Space may be Cheaper: Multi-Dimensional Resource Allocation for NVM-based Cloud Cache, in: 2020 IEEE 38th International Conference on Computer Design (ICCD). Presented at the 2020 IEEE 38th International Conference on Computer Design (ICCD), IEEE, Hartford, CT, USA, pp. 565–572. https://doi.org/10.1109/ICCD50377.2020.00100

Bauer, M., Becker, C., Rothermel, K., 2002. Location Models from the Perspective of Context-Aware Applications and Mobile Ad Hoc Networks. Pers. Ubiquitous Comput. 6, 322–328. https://doi.org/10.1007/s007790200036

Bibal Benifa, J.V., Dejey, D., 2019. RLPAS: Reinforcement Learning-based Proactive Auto-Scaler for Resource Provisioning in Cloud Environment. Mob. Netw. Appl. 24, 1348–1363. https://doi.org/10.1007/s11036-018-0996-0

Bilal, M., Kang, S.-G., 2014. Time Aware Least Recent Used (TLRU) cache management policy in ICN, in: 16th International Conference on Advanced Communication Technology. Presented at the 2014 16th International Conference on Advanced Communication Technology (ICACT), Global IT Research Institute (GIRI), Pyeongchang, Korea (South), pp. 528–532. https://doi.org/10.1109/ICACT.2014.6779016

Blasco, P., Gunduz, D., 2014. Learning-based optimization of cache content in a small cell base station, in: 2014 IEEE International Conference on Communications (ICC). Presented at the ICC 2014 - 2014 IEEE International Conference on Communications, IEEE, Sydney, NSW, pp. 1897–1903. https://doi.org/10.1109/ICC.2014.6883600

Bøgsted, M., Olsen, R.L., Schwefel, H.-P., 2010. Probabilistic models for access strategies to dynamic information elements. Perform. Eval. 67, 43–60. https://doi.org/10.1016/j.peva.2009.08.015

Boytsov, A., Zaslavsky, A., 2011a. From Sensory Data to Situation Awareness: Enhanced Context Spaces Theory Approach, in: 2011 IEEE Ninth International Conference on Dependable, Autonomic and Secure Computing. Presented at the 2011 IEEE 9th International Conference on Dependable, Autonomic and Secure Computing (DASC), IEEE, Sydney, Australia, pp. 207–214. https://doi.org/10.1109/DASC.2011.55

Boytsov, A., Zaslavsky, A., 2011b. ECSTRA – Distributed Context Reasoning Framework for Pervasive Computing Systems, in: Balandin, S., Koucheryavy, Y., Hu, H. (Eds.), Smart Spaces and Next Generation Wired/Wireless Networking, Lecture Notes in Computer Science. Springer Berlin Heidelberg, Berlin, Heidelberg, pp. 1–13. https://doi.org/10.1007/978-3-642-22875-9_1

Boytsov, A., Zaslavsky, A., 2010. Extending Context Spaces Theory by Proactive Adaptation, in: Balandin, S., Dunaytsev, R., Koucheryavy, Y. (Eds.), Smart Spaces and Next Generation Wired/Wireless Networking, Lecture Notes in Computer Science. Springer Berlin Heidelberg, Berlin, Heidelberg, pp. 1–12. https://doi.org/10.1007/978-3-642-14891-0_1

Buchholz, T., Küpper, A., Schiffers, M., 2016. Quality of Context: What it is and why we need it. https://doi.org/10.13140/RG.2.1.3795.9284

Chabridon, S., Conan, D., Abid, Z., Taconet, C., 2013. Building ubiquitous QoC-aware applications through model-driven software engineering. Sci. Comput. Program. 78, 1912–1929. https://doi.org/10.1016/j.scico.2012.07.019

Chatterjee, S., Misra, S., 2016. Adaptive data caching for provisioning sensors-as-a-service, in: 2016 IEEE International Black Sea Conference on Communications and Networking (BlackSeaCom). Presented at the 2016 IEEE International Black Sea Conference on Communications and Networking (BlackSeaCom), IEEE, Varna, Bulgaria, pp. 1–5. https://doi.org/10.1109/BlackSeaCom.2016.7901549

Chatterjee, S., Misra, S., 2014. Dynamic and adaptive data caching mechanism for virtualization within sensor-cloud, in: 2014 IEEE International Conference on Advanced Networks and Telecommuncations Systems (ANTS). Presented at the 2014 IEEE International Conference on Advanced Networks and Telecommuncations Systems (ANTS), IEEE, New Delhi, India, pp. 1–6. https://doi.org/10.1109/ANTS.2014.7057243

Chen, T., Dong, B., Chen, Y., Du, Y., Li, S., 2020. Multi-Objective Learning for Efficient Content Caching for Mobile Edge Networks, in: 2020 International Conference on Computing, Networking and Communications (ICNC). Presented at the 2020 International Conference on Computing, Networking and Communications (ICNC), IEEE, Big Island, HI, USA, pp. 543–547. https://doi.org/10.1109/ICNC47757.2020.9049682

Cheng, Y., Chen, W., Wang, Z., Yu, X., Xiang, Y., 2015. AMC: an adaptive multi-level cache algorithm in hybrid storage systems: An Adaptive Multi-Level Cache Algorithm. Concurr. Comput. Pract. Exp. 27, 4230–4246. https://doi.org/10.1002/cpe.3530

Chockler, G., Laden, G., Vigfusson, Y., 2011. Design and implementation of caching services in the cloud. IBM J. Res. Dev. 55, 9:1-9:11. https://doi.org/10.1147/JRD.2011.2171649

Chockler, G., Laden, G., Vigfusson, Y., 2010. Data caching as a cloud service, in: Proceedings of the 4th International Workshop on Large Scale Distributed Systems and Middleware - LADIS '10. Presented at the the 4th International Workshop, ACM Press, Zurich, Switzerland, p. 18. https://doi.org/10.1145/1859184.1859190

Choi, J., Gu, Y., Kim, J., 2020. Learning-based dynamic cache management in a cloud. J. Parallel Distrib. Comput. 145, 98–110. https://doi.org/10.1016/j.jpdc.2020.06.013

Chugh, A., Hybinette, M., 2004. Towards Adaptive Caching for Parallel and Discrete Event Simulation, in: Proceedings of the 2004 Winter Simulation Conference, 2004. Presented at the 2004 Winter Simulation Conference, 2004., IEEE, Washington, D.C., pp. 328–336. https://doi.org/10.1109/WSC.2004.1371334

Cidon, A., Eisenman, A., Alizadeh, M., Katti, S., 2015. Dynacache: Dynamic Cloud Caching, in: HotCloud'15: Proceedings of the 7th USENIX Conference on Hot Topics in Cloud Computing. Presented at the HotCloud'15: Proceedings of the 7th USENIX Conference on Hot Topics in Cloud Computing, USENIX Association, Santa Clara CA, p. 6. https://doi.org/10.5555/2827719.2827738

Context Information Management (CIM): Application Programming Interface (API), 2020. . ETSI Industry Specification Group (ISG).

Cui, Z., Zhao, Y., Li, C., Song, Y., Li, W., 2020. Content-Aware Load Balancing in CDN Network, in: 2020 IEEE 6th International Conference on Computer and Communications (ICCC). Presented at the 2020 IEEE 6th International Conference on Computer and Communications (ICCC), IEEE, Chengdu, China, pp. 88–93. https://doi.org/10.1109/ICCC51575.2020.9345240

Dasgupta, A., Kumar, R., Sarlós, T., 2017. Caching with Dual Costs, in: Proceedings of the 26th International Conference on World Wide Web Companion - WWW '17 Companion. Presented at the the 26th International Conference, ACM Press, Perth, Australia, pp. 643–652. https://doi.org/10.1145/3041021.3054187

de Matos, E., Tiburski, R.T., Moratelli, C.R., Johann Filho, S., Amaral, L.A., Ramachandran, G., Krishnamachari, B., Hessel, F., 2020. Context information sharing for the Internet of Things: A survey. Comput. Netw. 166, 106988. https://doi.org/10.1016/j.comnet.2019.106988

Didona, D., Romano, P., Peluso, S., Quaglia, F., 2014. Transactional Auto Scaler: Elastic Scaling of Replicated In-Memory Transactional Data Grids. ACM Trans. Auton. Adapt. Syst. 9, 1–32. https://doi.org/10.1145/2620001

Distributed Database - Apache Ignite [WWW Document], n.d. URL https://ignite.apache.org

Domènech, J., Gil, J.A., Sahuquillo, J., Pont, A., 2006. Web prefetching performance metrics: A survey. Perform. Eval. 63, 988–1004. https://doi.org/10.1016/j.peva.2005.11.001

Domènech, J., Pont, A., Sahuquillo, J., Gil, J.A., 2007. A user-focused evaluation of web prefetching algorithms. Comput. Commun. 30, 2213–2224. https://doi.org/10.1016/j.comcom.2007.05.003

Dulac-Arnold, G., Evans, R., Sunehag, P., Coppin, B., 2015. Reinforcement Learning in Large Discrete Action Spaces 10. https://doi.org/10.48550/arXiv.1512.07679

Ehcache [WWW Document], n.d. URL https://www.ehcache.org

Fanelli, M., Foschini, L., Corradi, A., Boukerche, A., 2011. QoC-Based Context Data Caching for Disaster Area Scenarios, in: 2011 IEEE International Conference on Communications (ICC). Presented at the ICC 2011 - 2011 IEEE International Conference on Communications, IEEE, Kyoto, Japan, pp. 1–5. https://doi.org/10.1109/icc.2011.5963361

Fatale, S., Prakash, R.S., Moharir, S., 2020. Caching Policies for Transient Data. IEEE Trans. Commun. 68, 4411–4422. https://doi.org/10.1109/TCOMM.2020.2987899

Faticanti, F., Maggi, L., Pellegrini, F.D., Santoro, D., Siracusa, D., 2021. Fog Orchestration meets Proactive Caching, in: 2021 IFIP/IEEE International Symposium on Integrated Network Management (IM). Presented at the 2021 IFIP/IEEE International Symposium on Integrated Network Management (IM), IEEE, Bordeaux, France, p. 6.

FIWARE-Orion [WWW Document], n.d. URL https://github.com/telefonicaid/fiware-orion

Gramacy, R.B., Warmuth, M.K.K., Brandt, S.A., Ari, I., 2002. Adaptive Caching by Refetching. NIPS02 Proc. 15th Int. Conf. Neural Inf. Process. Syst. 8. https://doi.org/10.5555/2968618.2968803

Guo, K., Yang, C., 2019. Temporal-Spatial Recommendation for Caching at Base Stations via Deep Reinforcement Learning. IEEE Access 7, 58519–58532. https://doi.org/10.1109/ACCESS.2019.2914500

Guo, Y., Lama, P., Rao, J., Zhou, X., 2013. V-Cache: Towards Flexible Resource Provisioning for Multi-tier Applications in IaaS Clouds, in: 2013 IEEE 27th International Symposium on Parallel and Distributed Processing. Presented at the 2013 IEEE International Symposium on Parallel & Distributed Processing (IPDPS), IEEE, Cambridge, MA, USA, pp. 88–99. https://doi.org/10.1109/IPDPS.2013.12

Hafeez, U.U., Wajahat, M., Gandhi, A., 2018. ElMem: Towards an Elastic Memcached System, in: 2018 IEEE 38th International Conference on Distributed Computing Systems (ICDCS). Presented at the 2018 IEEE 38th International Conference on Distributed Computing Systems (ICDCS), IEEE, Vienna, pp. 278–289. https://doi.org/10.1109/ICDCS.2018.00036

Hail, M.A.M., Amadeo, M., Molinaro, A., Fischer, S., 2015. On the Performance of Caching and Forwarding in Information-Centric Networking for the IoT, in: Aguayo-Torres, M.C., Gómez, G., Poncela, J. (Eds.), Wired/Wireless Internet Communications, Lecture Notes in Computer Science. Springer International Publishing, Cham, pp. 313–326. https://doi.org/10.1007/978-3-319-22572-2_23

Hassani, A., Haghighi, P.D., Jayaraman, P.P., Zaslavsky, A., Ling, S., Medvedev, A., 2016. CDQL: A Generic Context Representation and Querying Approach for Internet of Things Applications, in: Proceedings of the 14th International Conference on Advances in Mobile Computing and Multi Media - MoMM '16. Presented at the the 14th International Conference, ACM Press, Singapore, Singapore, pp. 79–88. https://doi.org/10.1145/3007120.3007137

Hassani, A., Medvedev, A., Delir Haghighi, P., Ling, S., Zaslavsky, A., Prakash Jayaraman, P., 2019. Context Definition and Query Language: Conceptual Specification, Implementation, and Evaluation. Sensors 19, 1478. https://doi.org/10.3390/s19061478

Hassani, A., Medvedev, A., Haghighi, P.D., Ling, S., Indrawan-Santiago, M., Zaslavsky, A., Jayaraman, P.P., 2018. Context-as-a-Service Platform: Exchange and Share Context in an IoT Ecosystem, in: 2018 IEEE International Conference on Pervasive Computing and Communications Workshops (PerCom Workshops). Presented at the 2018 IEEE International Conference on Pervasive Computing and Communications Workshops (PerCom Workshops), IEEE, Athens, pp. 385–390. https://doi.org/10.1109/PERCOMW.2018.8480240

He, X., Wang, K., Xu, W., 2019. QoE-Driven Content-Centric Caching With Deep Reinforcement Learning in Edge-Enabled IoT. IEEE Comput. Intell. Mag. 14, 12–20. https://doi.org/10.1109/MCI.2019.2937608

Henricksen, K., Indulska, J., Rakotonirainy, A., 2002. Modeling Context Information in Pervasive Computing Systems, in: Mattern, F., Naghshineh, M. (Eds.), Pervasive Computing, Lecture Notes in Computer Science. Springer Berlin Heidelberg, Berlin, Heidelberg, pp. 167–180. https://doi.org/10.1007/3-540-45866-2_14

Herodotou, H., Dong, F., Babu, S., 2011. No one (cluster) size fits all: automatic cluster sizing for data-intensive analytics, in: Proceedings of the 2nd ACM Symposium on Cloud Computing - SOCC '11. Presented at the the 2nd ACM Symposium, ACM Press, Cascais, Portugal, pp. 1–14. https://doi.org/10.1145/2038916.2038934

Hou, L., Lei, L., Zheng, K., Wang, X., 2019. A Q-Learning-Based Proactive Caching Strategy for Non-Safety Related Services in Vehicular Networks. IEEE Internet Things J. 6, 4512–4520. https://doi.org/10.1109/JIOT.2018.2883762

Huebscher, M.C., McCann, J.A., 2004. Adaptive middleware for context-aware applications in smart-homes, in: Proceedings of the 2nd Workshop on Middleware for Pervasive and Ad-Hoc Computing -. Presented at the the 2nd workshop, ACM Press, Toronto, Ontario, Canada, pp. 111–116. https://doi.org/10.1145/1028509.1028511

Jagarlamudi, K.S., Zaslavsky, A., Loke, S.W., Hassani, A., Medvedev, A., 2021. Quality and Cost Aware Service Selection in IoT-Context Management Platforms, in: 2021 IEEE International Conferences on Internet of Things (IThings) and IEEE Green Computing & Communications (GreenCom) and IEEE Cyber, Physical & Social Computing (CPSCom) and IEEE Smart Data (SmartData) and IEEE Congress on Cybermatics (Cybermatics). Presented at the 2021 IEEE International Conferences on Internet of Things (iThings) and IEEE Green Computing & Communications (GreenCom) and IEEE Cyber, Physical & Social Computing (CPSCom) and IEEE Smart Data (SmartData) and IEEE Congress on Cybermatics (Cybermatics), IEEE, Melbourne, Australia, pp. 89–98. https://doi.org/10.1109/iThings-GreenCom-CPSCom-SmartData-Cybermatics53846.2021.00028

Jayaraman, P., Yavari, A., Georgakopoulos, D., Morshed, A., Zaslavsky, A., 2016. Internet of Things Platform for Smart Farming: Experiences and Lessons Learnt. Sensors 16, 1884. https://doi.org/10.3390/s16111884

Kabir, F., Chiu, D., 2012. Reconciling Cost and Performance Objectives for Elastic Web Caches, in: 2012 International Conference on Cloud and Service Computing. Presented at the 2012 International Conference on Cloud and Service Computing (CSC), IEEE, Shanghai, China, pp. 88–95. https://doi.org/10.1109/CSC.2012.21

Kangasharju, J., Roberts, J., Ross, K.W., 2002. Object replication strategies in content distribution networks. Comput. Commun. 25, 376–383. https://doi.org/10.1016/S0140-3664(01)00409-1

Kaul, S., Yates, R., Gruteser, M., 2012. Real-time status: How often should one update?, in: 2012 Proceedings IEEE INFOCOM. Presented at the IEEE INFOCOM 2012 - IEEE Conference on Computer Communications, IEEE, Orlando, FL, USA, pp. 2731–2735. https://doi.org/10.1109/INFCOM.2012.6195689

Khargharia, H.S., Jayaraman, P.P., Banerjee, A., Zaslavsky, A., Hassani, A., Abken, A., Kumar, A., 2022. Probabilistic analysis of context caching in Internet of Things applications, in: 2022 IEEE International Conference on Services Computing (SCC). Presented at the 2022 IEEE International Conference on Services Computing (SCC), IEEE, Barcelona, Spain, pp. 93–103. https://doi.org/10.1109/SCC55611.2022.00025

Kiani, S., Anjum, A., Antonopoulos, N., Munir, K., McClatchey, R., 2012. Context caches in the Clouds. J. Cloud Comput. Adv. Syst. Appl. 1, 7. https://doi.org/10.1186/2192-113X-1-7

Kiani, S.L., Anjum, A., Munir, K., McClatchey, R., Antonopoulos, N., 2011. Towards Context Caches in the Clouds, in: 2011 Fourth IEEE International Conference on Utility and Cloud Computing. Presented at the 2011 IEEE 4th International Conference on Utility and Cloud Computing (UCC 2011), IEEE, Victoria, NSW, pp. 403–408. https://doi.org/10.1109/UCC.2011.67

Kirilin, V., Sundarrajan, A., Gorinsky, S., Sitaraman, R.K., 2019. RL-Cache: Learning-Based Cache Admission for Content Delivery, in: Proceedings of the 2019 Workshop on Network Meets AI & ML - NetAI'19. Presented at the the 2019 Workshop, ACM Press, Beijing, China, pp. 57–63. https://doi.org/10.1145/3341216.3342214

Kosta, A., Pappas, N., Ephremides, A., Angelakis, V., 2017. Age and value of information: Non-linear age case, in: 2017 IEEE International Symposium on Information Theory (ISIT). Presented at the 2017 IEEE International Symposium on Information Theory (ISIT), IEEE, Aachen, Germany, pp. 326–330. https://doi.org/10.1109/ISIT.2017.8006543

Kouame, K.-M., Mcheick, H., 2018. Overview of Software Adaptation Techniques; Guide Adaptation Pattern, in: 2018 International Conference on Computer and Applications (ICCA). Presented at the 2018 International Conference on Computer and Applications (ICCA), IEEE, Beirut, pp. 141–148. https://doi.org/10.1109/COMAPP.2018.8460397

Krause, M., Hochstatter, I., 2005. Challenges in Modelling and Using Quality of Context (QoC), in: Magedanz, T., Karmouch, A., Pierre, S., Venieris, I. (Eds.), Mobility Aware Technologies and Applications, Lecture Notes in Computer Science. Springer Berlin Heidelberg, Berlin, Heidelberg, pp. 324–333. https://doi.org/10.1007/11569510_31

Li, W., Privat, G., Cantera, J.M., Bauer, M., Gall, F.L., 2018. Graph-based Semantic Evolution for Context Information Management Platforms, in: 2018 Global Internet of Things Summit (GIoTS). Presented at the 2018 Global Internet of Things Summit (GIoTS), IEEE, Bilbao, pp. 1–6. https://doi.org/10.1109/GIOTS.2018.8534538

Li, X., Eckert, M., Martinez, J.-F., Rubio, G., 2015. Context Aware Middleware Architectures: Survey and Challenges. Sensors 15, 20570–20607. https://doi.org/10.3390/s150820570

Liu, G., Wang, X., 2018. A Modified Round-Robin Load Balancing Algorithm Based on Content of Request, in: 2018 5th International Conference on Information Science and Control Engineering (ICISCE). Presented at the 2018 5th International Conference on Information Science and Control Engineering (ICISCE), IEEE, Zhengzhou, pp. 66–72. https://doi.org/10.1109/ICISCE.2018.00023

Liu, X., Derakhshani, M., Lambotharan, S., 2021. Contextual Learning for Content Caching With Unknown Time-Varying Popularity Profiles via Incremental Clustering. IEEE Trans. Commun. 69, 3011–3024. https://doi.org/10.1109/TCOMM.2021.3059305

Maggs, Bruce M, Sitaraman, R.K., 2015. Algorithmic Nuggets in Content Delivery. ACM SIGCOMM Comput. Commun. Rev. 45, 15.

Maggs, Bruce M, Sitaraman, R.K., 2015. Algorithmic Nuggets in Content Delivery. ACM SIGCOMM Comput. Commun. Rev. 45, 52–66. https://doi.org/10.1145/2805789.2805800

Malawski, M., Figiela, K., Nabrzyski, J., 2013. Cost minimization for computational applications on hybrid cloud infrastructures. Future Gener. Comput. Syst. 29, 1786–1794. https://doi.org/10.1016/j.future.2013.01.004

Meddeb, M., Dhraief, A., Belghith, A., Monteil, T., Drira, K., 2017. How to Cache in ICN-Based IoT Environments?, in: 2017 IEEE/ACS 14th International Conference on Computer Systems and Applications (AICCSA). Presented at the 2017 IEEE/ACS 14th International Conference on Computer Systems and Applications (AICCSA), IEEE, Hammamet, pp. 1117–1124. https://doi.org/10.1109/AICCSA.2017.37

Medvedev, A., 2020. Performance and Cost Driven Data Storage and Processing for IoT Context Management Platforms 204. https://doi.org/10.26180/5ED89F27786D5

Medvedev, A., Hassani, A., Haghighi, P.D., Ling, S., Indrawan-Santiago, M., Zaslavsky, A., Fastenrath, U., Mayer, F., Jayaraman, P.P., Kolbe, N., 2018. Situation Modelling, Representation, and Querying in Context-as-a-Service IoT Platform, in: 2018 Global Internet of Things Summit (GIoTS). Presented at the 2018 Global Internet of Things Summit (GIoTS), IEEE, Bilbao, pp. 1–6. https://doi.org/10.1109/GIOTS.2018.8534571

Medvedev, A., Hassani, A., Zaslavsky, A., Haghighi, P.D., Ling, S., Jayaraman, P.P., 2019. Benchmarking IoT Context Management Platforms: High-level Queries Matter, in: 2019 Global IoT Summit (GIoTS). Presented at the 2019 Global IoT Summit (GIoTS), IEEE, Aarhus, Denmark, pp. 1–6. https://doi.org/10.1109/GIOTS.2019.8766395

Medvedev, A., Indrawan-Santiago, M., Delir Haghighi, P., Hassani, A., Zaslavsky, A., Jayaraman, P.P., 2017. Architecting IoT context storage management for context-as-a-service platform, in: 2017 Global Internet of Things Summit (GIoTS). Presented at the 2017 Global Internet of Things Summit (GIoTS), IEEE, Geneva, Switzerland, pp. 1–6. https://doi.org/10.1109/GIOTS.2017.8016228

Mehrizi, S., Chatzinotas, S., Ottersten, B., 2020. Content Request Prediction with Temporal Trend for Proactive Caching, in: 2020 IEEE 31st Annual International Symposium on Personal, Indoor and Mobile Radio Communications. Presented at the 2020 IEEE 31st Annual International Symposium on Personal, Indoor and Mobile Radio Communications, IEEE, London, United Kingdom, pp. 1–7. https://doi.org/10.1109/PIMRC48278.2020.9217350

Mehrizi, S., Vu, T.X., Chatzinotas, S., Ottersten, B., 2021. Trend-Aware Proactive Caching via Tensor Train Decomposition: A Bayesian Viewpoint. IEEE Open J. Commun. Soc. 2, 975–989. https://doi.org/10.1109/OJCOMS.2021.3075071

Memcached - a distributed memory object caching system [WWW Document], n.d. URL https://memcached.org

MemCachier [WWW Document], n.d. URL https://www.memcachier.com

Menache, I., Singh, M., 2015. Online Caching with Convex Costs: Extended Abstract, in: Proceedings of the 27th ACM Symposium on Parallelism in Algorithms and Architectures. Presented at the SPAA '15: 27th ACM Symposium on Parallelism in Algorithms and Architectures, ACM, Portland Oregon USA, pp. 46–54. https://doi.org/10.1145/2755573.2755585

Ming Li, Ganesan, D., Shenoy, P., 2009. PRESTO: Feedback-Driven Data Management in Sensor Networks. IEEEACM Trans. Netw. 17, 1256–1269. https://doi.org/10.1109/TNET.2008.2006818

Muller, S., Atan, O., van der Schaar, M., Klein, A., 2017. Context-Aware Proactive Content Caching With Service Differentiation in Wireless Networks. IEEE Trans. Wirel. Commun. 16, 1024–1036. https://doi.org/10.1109/TWC.2016.2636139

Narasayya, V., Menache, I., Singh, M., Li, F., Syamala, M., Chaudhuri, S., 2015. Sharing buffer pool memory in multi-tenant relational database-as-a-service. Proc. VLDB Endow. 8, 726–737. https://doi.org/10.14778/2752939.2752942

Nasehzadeh, A., Wang, P., 2020. A Deep Reinforcement Learning-Based Caching Strategy for Internet of Things, in: 2020 IEEE/CIC International Conference on Communications in China (ICCC). Presented at the 2020 IEEE/CIC International Conference on Communications in China (ICCC), IEEE, Chongqing, China, pp. 969–974. https://doi.org/10.1109/ICCC49849.2020.9238811

Naz, S., Rais, R.N.B., Qayyum, A., 2016. Multi-Attribute Caching: Towards efficient cache management in Content-Centric Networks, in: 2016 13th IEEE Annual Consumer Communications & Networking Conference (CCNC). Presented at the 2016 13th IEEE Annual Consumer Communications & Networking Conference (CCNC), IEEE, Las Vegas, NV, USA, pp. 630–633. https://doi.org/10.1109/CCNC.2016.7444852

Olsen, R.L., Schwefel, H.-P., Hansen, M.B., 2006. Quantitative Analysis of Access Strategies to Remote Information in Network Services, in: IEEE Globecom 2006. Presented at the IEEE Globecom 2006, IEEE, San Francisco, CA, USA, pp. 1–6. https://doi.org/10.1109/GLOCOM.2006.422

Orsini, G., Bade, D., Lamersdorf, W., 2016. Generic Context Adaptation for Mobile Cloud Computing Environments. Procedia Comput. Sci. 94, 17–24. https://doi.org/10.1016/j.procs.2016.08.007

Pahl, M.-O., Liebald, S., Wustrich, L., 2019. Machine-Learning based IoT Data Caching, in: 2019 IFIP/IEEE Symposium on Integrated Network and Service Management (IM). Presented at the 2019 IFIP/IEEE Symposium on Integrated Network and Service Management (IM), IEEE, Arlington, VA, USA, p. 4.

Papaefstathiou, V., Katevenis, M.G.H., Nikolopoulos, D.S., Pnevmatikatos, D., 2013. Prefetching and cache management using task lifetimes, in: Proceedings of the 27th International ACM Conference on International Conference on Supercomputing - ICS '13. Presented at the the 27th international ACM conference, ACM Press, Eugene, Oregon, USA, p. 325. https://doi.org/10.1145/2464996.2465443

Perera, C., Zaslavsky, A., Christen, P., Georgakopoulos, D., 2014. Context Aware Computing for The Internet of Things: A Survey. IEEE Commun. Surv. Tutor. 16, 414–454. https://doi.org/10.1109/SURV.2013.042313.00197

Perera, C., Zaslavsky, A., Christen, P., Georgakopoulos, D., 2012. CA4IOT: Context Awareness for Internet of Things, in: 2012 IEEE International Conference on Green Computing and Communications. Presented at the 2012 IEEE International Conference on Green Computing and Communications (GreenCom), IEEE, Besancon, France, pp. 775–782. https://doi.org/10.1109/GreenCom.2012.128

Psaras, I., Chai, W.K., Pavlou, G., 2012. Probabilistic in-network caching for information-centric networks, in: Proceedings of the Second Edition of the ICN Workshop on Information-Centric Networking - ICN '12. Presented at the the second edition of the ICN workshop, ACM Press, Helsinki, Finland, p. 55. https://doi.org/10.1145/2342488.2342501

Qureshi, M.S., Qureshi, M.B., Fayaz, M., Mashwani, W.K., Belhaouari, S.B., Hassan, S., Shah, A., 2020. A comparative analysis of resource allocation schemes for real-time services in high-performance computing systems. Int. J. Distrib. Sens. Netw. 16, 155014772093275. https://doi.org/10.1177/1550147720932750

Raibulet, C., Masciadri, L., 2009. Evaluation of dynamic adaptivity through metrics: an achievable target?, in: 2009 Joint Working IEEE/IFIP Conference on Software Architecture & European Conference on Software Architecture. Presented at the 3rd European Conference on Software Architecture (ECSA), IEEE, Cambridge, United Kingdom, pp. 341–344. https://doi.org/10.1109/WICSA.2009.5290667

Redis [WWW Document], n.d. URL https://redis.io

Ruggeri, G., Amadeo, M., Campolo, C., Molinaro, A., Iera, A., 2021. Caching Popular Transient IoT Contents in an SDN-Based Edge Infrastructure. IEEE Trans. Netw. Serv. Manag. 18, 3432–3447. https://doi.org/10.1109/TNSM.2021.3056891

Sadeghi, A., Sheikholeslami, F., Giannakis, G.B., 2018. Optimal Dynamic Proactive Caching Via Reinforcement Learning, in: 2018 IEEE 19th International Workshop on Signal Processing Advances in Wireless

Communications (SPAWC). Presented at the 2018 IEEE 19th International Workshop on Signal Processing Advances in Wireless Communications (SPAWC), IEEE, Kalamata, pp. 1–5. https://doi.org/10.1109/SPAWC.2018.8445899

Sadeghi, A., Sheikholeslami, F., Marques, A.G., Giannakis, G.B., 2019a. Reinforcement Learning for Adaptive Caching With Dynamic Storage Pricing. IEEE J. Sel. Areas Commun. 37, 2267–2281. https://doi.org/10.1109/JSAC.2019.2933780

Sadeghi, A., Wang, G., Giannakis, G.B., 2019b. Deep Reinforcement Learning for Adaptive Caching in Hierarchical Content Delivery Networks. IEEE Trans. Cogn. Commun. Netw. 5, 1024–1033. https://doi.org/10.1109/TCCN.2019.2936193

Schwefel, H.-P., Hansen, M.B., Olsen, R.L., 2007. Adaptive Caching Strategies for Context Management Systems, in: 2007 IEEE 18th International Symposium on Personal, Indoor and Mobile Radio Communications. Presented at the 2007 IEEE 18th International Symposium on Personal, Indoor and Mobile Radio Communications, IEEE, Athens, Greece, pp. 1–6. https://doi.org/10.1109/PIMRC.2007.4394813

Scouarnec, N.L., Neumann, C., Straub, G., 2014. Cache Policies for Cloud-Based Systems: To Keep or Not to Keep, in: 2014 IEEE 7th International Conference on Cloud Computing. Presented at the 2014 IEEE 7th International Conference on Cloud Computing (CLOUD), IEEE, Anchorage, AK, USA, pp. 1–8. https://doi.org/10.1109/CLOUD.2014.11

Sheikh, R., Kharbutli, M., 2010. Improving cache performance by combining cost-sensitivity and locality principles in cache replacement algorithms, in: 2010 IEEE International Conference on Computer Design. Presented at the 2010 IEEE International Conference on Computer Design (ICCD 2010), IEEE, Amsterdam, Netherlands, pp. 76–83. https://doi.org/10.1109/ICCD.2010.5647594

Shen, Z., Subbiah, S., Gu, X., Wilkes, J., 2011. CloudScale: elastic resource scaling for multi-tenant cloud systems, in: Proceedings of the 2nd ACM Symposium on Cloud Computing - SOCC '11. Presented at the the 2nd ACM Symposium, ACM Press, Cascais, Portugal, pp. 1–14. https://doi.org/10.1145/2038916.2038921

Sheng, S., Chen, P., Chen, Z., Wu, L., Jiang, H., 2020. Edge Caching for IoT Transient Data Using Deep Reinforcement Learning, in: IECON 2020 The 46th Annual Conference of the IEEE Industrial Electronics Society. Presented at the IECON 2020 - 46th Annual Conference of the IEEE Industrial Electronics Society, IEEE, Singapore, Singapore, pp. 4477–4482. https://doi.org/10.1109/IECON43393.2020.9255111

Shuai, Q., Wang, K., Miao, F., Jin, L., 2017. A Cost-Based Distributed Algorithm for Load Balancing in Content Delivery Network, in: 2017 9th International Conference on Intelligent Human-Machine Systems and Cybernetics (IHMSC). Presented at the 2017 9th International Conference on Intelligent Human-Machine Systems and Cybernetics (IHMSC), IEEE, Hangzhou, pp. 11–15. https://doi.org/10.1109/IHMSC.2017.10

Shuja, J., Bilal, K., Alasmary, W., Sinky, H., Alanazi, E., 2021. Applying machine learning techniques for caching in next-generation edge networks: A comprehensive survey. J. Netw. Comput. Appl. 181, 103005. https://doi.org/10.1016/j.jnca.2021.103005

Singh, S., Chana, I., 2016. QoS-Aware Autonomic Resource Management in Cloud Computing: A Systematic Review. ACM Comput. Surv. 48, 1–46. https://doi.org/10.1145/2843889

Somuyiwa, S.O., Gyorgy, A., Gunduz, D., 2018. A Reinforcement-Learning Approach to Proactive Caching in Wireless Networks. IEEE J. Sel. Areas Commun. 36, 1331–1344. https://doi.org/10.1109/JSAC.2018.2844985

Song, H., Ke, F., Guo, W., Lin, Y., Bo, Y., Liu, S., 2020. Loss-aware adaptive caching scheme for device-to-device communications. IET Commun. 14, 2607–2617. https://doi.org/10.1049/iet-com.2019.0727

Srinivasan, S.T., Lebeck, A.R., 1998. Load latency tolerance in dynamically scheduled processors, in: Proceedings. 31st Annual ACM/IEEE International Symposium on Microarchitecture. Presented at the Proceedings. 31st Annual ACM/IEEE International Symposium on Microarchitecture, IEEE Comput. Soc, Dallas, TX, USA, pp. 148–159. https://doi.org/10.1109/MICRO.1998.742777

Sun, Y., Uysal-Biyikoglu, E., Yates, R., Koksal, C.E., Shroff, N.B., 2016. Update or wait: How to keep your data fresh, in: IEEE INFOCOM 2016 - The 35th Annual IEEE International Conference on Computer Communications. Presented at the IEEE INFOCOM 2016 - IEEE Conference on Computer Communications, IEEE, San Francisco, CA, USA, pp. 1–9. https://doi.org/10.1109/INFOCOM.2016.7524524

Sutton, R.S., Barto, A.G., 1998. Reinforcement learning: an introduction, Adaptive computation and machine learning. MIT Press, Cambridge, Mass.

Tadrous, J., Eryilmaz, A., Gamal, H.E., 2013. Proactive Content Distribution for dynamic content, in: 2013 IEEE International Symposium on Information Theory. Presented at the 2013 IEEE International Symposium on Information Theory (ISIT), IEEE, Istanbul, Turkey, pp. 1232–1236. https://doi.org/10.1109/ISIT.2013.6620423

Ullah, I., Kim, J.-B., Han, Y.-H., 2022. Compound Context-Aware Bayesian Inference Scheme for Smart IoT Environment. Sensors 22, 3022. https://doi.org/10.3390/s22083022

ur Rehman, M.H., Yaqoob, I., Salah, K., Imran, M., Jayaraman, P.P., Perera, C., 2019. The role of big data analytics in industrial Internet of Things. Future Gener. Comput. Syst. 99, 247–259. https://doi.org/10.1016/j.future.2019.04.020

Venkataramani, A., Yalagandula, P., Kokku, R., Sharif, S., Dahlin, M., 2002. The potential costs and benefits of long-term prefetching for content distribution. Comput. Commun. 25, 367–375. https://doi.org/10.1016/S0140-3664(01)00408-X

Verma, S., Bala, A., 2021. Auto-scaling techniques for IoT-based cloud applications: a review. Clust. Comput. 24, 2425–2459. https://doi.org/10.1007/s10586-021-03265-9

Villalonga, C., Roggen, D., Lombriser, C., Zappi, P., Tröster, G., 2009. Bringing Quality of Context into Wearable Human Activity Recognition Systems, in: Rothermel, K., Fritsch, D., Blochinger, W., Dürr, F. (Eds.), Quality of Context, Lecture Notes in Computer Science. Springer Berlin Heidelberg, Berlin, Heidelberg, pp. 164–173. https://doi.org/10.1007/978-3-642-04559-2_15

Vural, S., Navaratnam, P., Wang, N., Wang, C., Dong, L., Tafazolli, R., 2014. In-network caching of Internet-of-Things data, in: 2014 IEEE International Conference on Communications (ICC). Presented at the ICC 2014 - 2014 IEEE International Conference on Communications, IEEE, Sydney, NSW, pp. 3185–3190. https://doi.org/10.1109/ICC.2014.6883811

Vural, S., Wang, N., Navaratnam, P., Tafazolli, R., 2017. Caching Transient Data in Internet Content Routers. IEEEACM Trans. Netw. 25, 1048–1061. https://doi.org/10.1109/TNET.2016.2616359

Wang, R., Li, R., Wang, P., Liu, E., 2019. Analysis and Optimization of Caching in Fog Radio Access Networks. IEEE Trans. Veh. Technol. 68, 8279–8283. https://doi.org/10.1109/TVT.2019.2921615

Wang, Y., Friderikos, V., 2020. A Survey of Deep Learning for Data Caching in Edge Network. Informatics 7, 43. https://doi.org/10.3390/informatics7040043

Wang, Y., He, S., Fan, X., Xu, C., Sun, X.-H., 2019. On Cost-Driven Collaborative Data Caching: A New Model Approach. IEEE Trans. Parallel Distrib. Syst. 30, 662–676. https://doi.org/10.1109/TPDS.2018.2868642

Weerasinghe, S., Zaslavsky, A., Loke, S.W., Abken, A., Hassani, A., Medvedev, A., 2023. Adaptive Context Caching for Efficient Distributed Context Management Systems, in: ACM Symposium on Applied Computing. ACM, Tallinn, Estonia, p. 10. https://doi.org/10.1145/3555776.3577602

Weerasinghe, S., Zaslavsky, A., Loke, S.W., Hassani, A., Abken, A., Medvedev, A., 2022a. From Traditional Adaptive Data Caching to Adaptive Context Caching: A Survey. ArXiv221111259 CsHC 35.

Weerasinghe, S., Zaslavsky, A., Loke, S.W., Medvedev, A., Abken, A., 2022b. Estimating the Lifetime of Transient Context for Adaptive Caching in IoT Applications, in: ACM Symposium on Applied Computing. ACM, Brno, Czech Republic, p. 10. https://doi.org/10.1145/3477314.3507075

Weerasinghe, S., Zaslavsky, A., Loke, S.W., Medvedev, A., Abken, A., Hassani, A., 2022c. Context Caching for IoT-based Applications: Opportunities and Challenges.

Weiser, M., 1999. The computer for the 21st century. ACM SIGMOBILE Mob. Comput. Commun. Rev. 3, 3–11. https://doi.org/10.1145/329124.329126

Wu, B., Kshemkalyani, A.D., 2006. Objective-optimal algorithms for long-term Web prefetching. IEEE Trans. Comput. 55, 2–17. https://doi.org/10.1109/TC.2006.12

Wu, H., Luo, Y., Li, C., 2021. Optimization of heat-based cache replacement in edge computing system. J. Supercomput. 77, 2268–2301. https://doi.org/10.1007/s11227-020-03356-1

Wu, P., Li, J., Shi, L., Ding, M., Cai, K., Yang, F., 2019. Dynamic Content Update for Wireless Edge Caching via Deep Reinforcement Learning. IEEE Commun. Lett. 23, 1773–1777. https://doi.org/10.1109/LCOMM.2019.2931688

Wu, X., Li, X., Li, J., Ching, P.C., Leung, V.C.M., Poor, H.V., 2021. Caching Transient Content for IoT Sensing: Multi-Agent Soft Actor-Critic. IEEE Trans. Commun. 69, 5886–5901. https://doi.org/10.1109/TCOMM.2021.3086535

Wyrzykowski, R., Dongarra, J., Karczewski, K., Waśniewski, J. (Eds.), 2014. Parallel Processing and Applied Mathematics: 10th International Conference, PPAM 2013, Warsaw, Poland, September 8-11, 2013,

Revised Selected Papers, Part I, Lecture Notes in Computer Science. Springer Berlin Heidelberg, Berlin, Heidelberg. https://doi.org/10.1007/978-3-642-55224-3

Xin Chen, Xiaodong Zhang, 2003. A popularity-based prediction model for web prefetching. Computer 36, 63–70. https://doi.org/10.1109/MC.2003.1185219

Xu, C., Wang, X., 2019. Transient content caching and updating with modified harmony search for Internet of Things. Digit. Commun. Netw. 5, 24–33. https://doi.org/10.1016/j.dcan.2018.10.002

Yu, Z., Hu, J., Min, G., Xu, H., Mills, J., 2020. Proactive Content Caching for Internet-of-Vehicles based on Peer-to-Peer Federated Learning, in: 2020 IEEE 26th International Conference on Parallel and Distributed Systems (ICPADS). Presented at the 2020 IEEE 26th International Conference on Parallel and Distributed Systems (ICPADS), IEEE, Hong Kong, pp. 601–608. https://doi.org/10.1109/ICPADS51040.2020.00083

Zameel, A., Najmuldeen, M., Gormus, S., 2019. Context-Aware Caching in Wireless IoT Networks, in: 2019 11th International Conference on Electrical and Electronics Engineering (ELECO). Presented at the 2019 11th International Conference on Electrical and Electronics Engineering (ELECO), IEEE, Bursa, Turkey, pp. 712–717. https://doi.org/10.23919/ELECO47770.2019.8990647

Zhang, Z., Lung, C.-H., Lambadaris, I., St-Hilaire, M., 2018. IoT Data Lifetime-Based Cooperative Caching Scheme for ICN-IoT Networks, in: 2018 IEEE International Conference on Communications (ICC). Presented at the 2018 IEEE International Conference on Communications (ICC 2018), IEEE, Kansas City, MO, pp. 1–7. https://doi.org/10.1109/ICC.2018.8422100

Zhang, Z., Lung, C.-H., St-Hilaire, M., Lambadaris, I., 2020. Smart Proactive Caching: Empower the Video Delivery for Autonomous Vehicles in ICN-Based Networks. IEEE Trans. Veh. Technol. 69, 7955–7965. https://doi.org/10.1109/TVT.2020.2994181

Zhong, C., Gursoy, M.C., Velipasalar, S., 2020. Deep Reinforcement Learning-Based Edge Caching in Wireless Networks. IEEE Trans. Cogn. Commun. Netw. 6, 48–61. https://doi.org/10.1109/TCCN.2020.2968326

Zhong, C., Gursoy, M.C., Velipasalar, S., 2018. A deep reinforcement learning-based framework for content caching, in: 2018 52nd Annual Conference on Information Sciences and Systems (CISS). Presented at the 2018 52nd Annual Conference on Information Sciences and Systems (CISS), IEEE, Princeton, NJ, pp. 1–6. https://doi.org/10.1109/CISS.2018.8362276

Zhou, Z., Zhao, D., Xu, X., Du, C., Sun, H., 2015. Periodic Query Optimization Leveraging Popularity-Based Caching in Wireless Sensor Networks for Industrial IoT Applications. Mob. Netw. Appl. 20, 124–136. https://doi.org/10.1007/s11036-014-0545-4

Zhu, H., Cao, Y., Wei, X., Wang, W., Jiang, T., Jin, S., 2019. Caching Transient Data for Internet of Things: A Deep Reinforcement Learning Approach. IEEE Internet Things J. 6, 2074–2083. https://doi.org/10.1109/JIOT.2018.2882583

Zhu, J., Huang, X., Shao, Z., 2020. Learning-Aided Content Placement in Caching-Enabled fog Computing Systems Using Thompson Sampling, in: ICASSP 2020 - 2020 IEEE International Conference on Acoustics, Speech and Signal Processing (ICASSP). Presented at the ICASSP 2020 - 2020 IEEE International Conference on Acoustics, Speech and Signal Processing (ICASSP), IEEE, Barcelona, Spain, pp. 5060–5064. https://doi.org/10.1109/ICASSP40776.2020.9053162